\documentclass[aps,amsmath,amssymb,nofootinbib]{revtex4}

\usepackage{graphicx}
\usepackage{dcolumn}
\usepackage{bm}
\usepackage{color}

\def\bs{\boldsymbol}
\def\t{\tilde}
\def\tm{{\cal T}({\cal M})}
\def\ttm{{\cal T}\left(\t{\cal M}\right)}

\def\cm{{\rm cm}}
\def\dim{{\rm dim}}
\def\eff{{\rm eff}}
\def\emm{{\rm EM}}
\def\g{{\rm g}}
\def\intt{{\rm int}}
\def\ot{{\rm Other}}
\def\P{{\rm P}}

\begin{document}

\title{A New Unified Theory of Electromagnetic and Gravitational Interactions} 

\author{Li-Xin Li}
\email{lxl@pku.edu.cn}
\affiliation{Kavli Institute for Astronomy and Astrophysics, Peking University, Beijing 100871, P. R. China}

\date{\today}

\begin{abstract}
In this paper we present a new unified theory of electromagnetic and gravitational interactions. By considering a four-dimensional spacetime as a hypersurface embedded in a five-dimensional bulk spacetime, we derive the complete set of field equations in the four-dimensional spacetime from the five-dimensional Einstein field equation. Besides the Einstein field equation in the four-dimensional spacetime, an electromagnetic field equation is derived: $\nabla_a F^{ab}-\xi R^b_{\;\,a}A^a=-4\pi J^b$ with $\xi=-2$, where $F^{ab}$ is the antisymmetric electromagnetic field tensor defined by the potential vector $A^a$, $R_{ab}$ is the Ricci curvature tensor of the hypersurface, and $J^a$ is the electric current density vector. The electromagnetic field equation differs from the Einstein-Maxwell equation by a curvature-coupled term $\xi R^b_{\;\,a}A^a$, whose presence addresses the problem of incompatibility of the Einstein-Maxwell equation with a universe containing a uniformly distributed net charge as discussed in a previous paper by the author [L.-X. Li, Gen. Relativ. Gravit. {\bf 48}, 28 (2016)]. Hence, the new unified theory is physically different from the Kaluza-Klein theory and its variants where the Einstein-Maxwell equation is derived. In the four-dimensional Einstein field equation derived in the new theory, the source term includes the stress-energy tensor of electromagnetic fields as well as the stress-energy tensor of other unidentified matter. Under some conditions the unidentified matter can be interpreted as a cosmological constant in the four-dimensional spacetime. We argue that, the electromagnetic field equation and hence the unified theory presented in this paper can be tested in an environment with a high mass density, e.g., inside a neutron star or a white dwarf, and in the early epoch of the universe. 

\vspace{0.4cm}

\end{abstract}

\maketitle

\section{Introduction}
\label{intro}

Since \citet{ein15a,ein15b} discovered the theory of general relativity, many people (including Einstein himself) have attempted to develop a theory which unifies all kinds of fundamental interactions in nature (see \cite{goe04} for a review). The success of general relativity in interpreting gravity as the geometry of spacetime made Einstein enthusiastic about geometrzing electromagnetic fields and combining the electromagnetic interaction with the gravitational interaction (the only long-range forces known to exist in nature) in a unified geometric frame. For this goal, Einstein chose to extend the metric tensor of a four-dimensional spacetime to accommodate the electromagnetic field, e.g., by introducing a complex and Hermitian metric tensor \cite{ein48}, or a real but asymmetric metric tensor \cite{ein55}. Einstein was not alone in pursuing the unified theory for gravitational and electromagnetic interactions. Many other people (some earlier than Einstein) have worked along similar or distinct ways in the frame of a four-dimensional spacetime \cite{goe04}, including \citet{wey18}, \citet{edd21}, and \citet{sch47}. All those efforts have turned out to be unsuccessful, since the derived theory cannot describe the physical reality.

After \citet{nor14}, \citet{kal21} explored the possibility of unifying gravity and electromagnetism in a five-dimensional spacetime: starting from a vacuum Einstein field equation in a five-dimensional spacetime, he was able to derive the Maxwell equation and the Einstein field equation in a four-dimensional spacetime with the stress-energy tensor of an electromagnetic field and an unidentified scalar field as the source. Of the total 15 independent components of the symmetric five-dimensional metric tensor, 10 were interpreted as the components of a four-dimensional spacetime metric, four were interpreted as proportional to the components of an electromagnetic field potential vector in the four-dimensional spacetime, and one was interpreted as an unidentified scalar field. 

In his original work Kaluza adopted the so-called ``cylinder condition'', which assumed that all metric components do not depend on the fifth dimension. \citet{kle26a,kle26b} proposed a quantum interpretation for Kaluza's theory. He introduced the hypothesis that the fifth spacetime dimension is compactified with a very small circumference (e.g., order of the Planck length). Then, all components of the metric were expanded in Fourier series and the $n$-th Fourier mode was interpreted as the $n$-th excited state according to the quantum theory. Because of the extreme smallness of the size of the fifth dimension, all excited states (i.e., all $n\ge 1$ states) would correspond to extreme high energy (e.g., the Planck energy) states so would not be accessible to experiments even in high energy physics. Then, only the ground state with $n=0$ was accessible and Kaluza's cylinder condition was interpreted. For comprehensive reviews on the Kaluza-Klein theory and its variants, see \cite{bai87,ove97}.

Although the Kaluza-Klein theory has not been accepted as the ultimate theory for the unification of electromagnetism and gravity, the idea of extra dimensions has become popular in modern theories attempting to unify all kinds of interactions in nature (the electromagnetic, weak, strong, and gravitational interactions), including the supergravity theory, the superstring theory, and the brane world theory. In both the supergravity theory and the superstring theory it is assumed that the spacetime contains four macroscopic dimensions (one dimension of time and three dimensions of space), and a number of compactified extra space dimensions (seven in the supergravity theory, and six in the superstring theory) that are hypothetically of Planck length scales so that we cannot see the extra dimensions in lab conditions \cite{ove97}. 

Since 1998, people started to realize that large extra dimensions (i.e., extra dimensions of scales much larger than the Planck length) are possible provided that the standard model particles are confined on the macroscopic four-dimensional spacetime (a four-dimensional membrane) and only gravitational interaction can propagate in extra dimensions \cite{ark98,ant98}. The theory of large extra dimensions was proposed to explain the very weakness of gravity relative to the other three kinds of interactions and to address the hierarchy problem in theoretical physics. A more interesting scenario---the so-called brane gravity or brane world---was proposed in 1999 by \citet{ran99a,ran99b}: the four-dimensional spacetime that we live in is in fact a four-dimensional membrane (a hypersurface with a surface stress-energy density) embedded in a five-dimensional anti-de Sitter space. As in the large extra dimension model, it is hypothesized that the standard model particles are confined in the four-dimensional brane so we do not see the fifth dimension. However, in the brane world model, the fifth dimension need not be compactified. It is the curvature of the five-dimensional bulk spacetime that makes the gravity on the four-dimensional bane weak and appear four-dimensional on scales larger than the curvature radius of the bulk spacetime.

In this paper, we propose a new unified theory of electromagnetic and gravitational interactions. Similar to the brane world model, we assume that the four-dimensional spacetime that we live in is a hypersurface embedded in a five-dimensional bulk spacetime. However, we do not assume that the standard model particles are confined in the four-dimensional spacetime {\it a priori}. We also do not attempt to address the hierarchy problem as in the brane world theory. Instead, similar to in the Kaluza-Klein theory, we derive the electromagnetic field equation from the vacuum Einstein field equation in the five-dimensional bulk spacetime. The approach can be outlined as follows: through projection, the metric tensor in the five-dimensional bulk spacetime naturally induces a metric tensor on the four-dimensional spacetime hypersurface embedded in the bulk space. The five-dimensional metric tensor has in total 15 independent components. The four-dimensional metric tensor has 10 independent components, when it is expressed in a coordinate system in the four-dimensional spacetime. Of the rest five independent components of the five-dimensional metric tensor, as in the Kaluza-Klein theory, four are interpreted as proportional to the components of an electromagnetic potential vector in the four-dimensional spacetime, and the rest one as a scalar function. Then, on the four-dimensional spacetime hypersurface, we derive an electromagnetic field equation of the form
\begin{eqnarray}
        \nabla_a F^{ab}-\xi R^b_{\;\,a}A^a=-4\pi J^b \label{meq0}
\end{eqnarray}
with $\xi=-2$, where $F^{ab}$ is the usual antisymmetric electromagnetic field tensor defined by the potential vector $A^a$, $R_{ab}$ is the Ricci curvature tensor of the four-dimensional spacetime, and $J^a$ is the electric current vector.

Technically, the approach adopted in this paper is similar to that adopted in the Hamiltonian formulation of general relativity (see, e.g., \cite{wal84}), where a scalar ``lapse'' function $N$ and a ``shift'' vector $N^a$ tangent to the hypersurface are introduced. However, in the Hamiltonian formulation of general relativity, a spacetime is foliated along a spacelike direction, i.e., the normal to the hypersurface is a timelike vector. For the problem studied in this paper, a five-dimensional spacetime is foliated along a timelike direction, i.e., the normal to the hypersurface is a spacelike vector. In the theory presented in this paper, the ``shift'' vector $N^a$ is interpreted as proportional to the electromagnetic potential vector $A^a$ on the hypersurface: $N^a=2NA^a$ in Planck units. Then, when $N$ is constant on the hypersurface (but can vary with the fifth dimension), the electromagnetic field equation (\ref{meq0}) is derived from the five-dimensional vacuum Einstein field equation, as will be detailed in the paper.

Generally, for an $n$-dimensional spacetime as a hypersurface embedded in an $(n+1)$-dimensional spacetime, the $n$-dimensional spacetime has two curvatures: a Riemann curvature, determined by the metric tensor (also called the first fundamental form in differential geometry) on the hypersurface; and an extrinsic curvature (also called the second fundamental form), determined by the normal to the hypersurface and the metric tensor in the $(n+1)$-dimensional spacetime. The Riemann curvature tensor of the hypersurface depends on the intrinsic geometry of the hypersurface (the metric, and the derivative operator associated with it). The extrinsic curvature tensor depends on the way the hypersurface is embedded in the bulk space \cite{haw73,car03}. In the theory presented in this paper, electromagnetic fields are contained in the extrinsic curvature of the four-dimensional spacetime. As in the standard general relativity, gravitational fields are represented by the Riemann curvature. The electromagnetism in the four-dimensional spacetime arises from the gravity in a five-dimensional spacetime containing the four-dimensional spacetime. So, unlike in the brane world theory, in our theory electromagnetic fields are not assumed to be confined in a four-dimensional membrane {\it a priori}, and the hypersurface representing the four-dimensional spacetime need not be a discontinuous surface layer with a surface stress-energy tensor. 

The electromagnetic field equation (\ref{meq0}) is the most important result of this paper. It differs from the Einstein-Maxwell equation (eq.~\ref{meq1g} in Sec.~\ref{em_general}) by the presence of a term with $A^a$ coupled to the Ricci tensor $R_{ab}$, so is a new electromagnetic field equation. The electromagnetic field equation of the form (\ref{meq0}) with an undetermined dimensionless factor $\xi$ was proposed by \citet{li15} to address the incompatibility of the Einstein-Maxwell equation with a homogeneous and isotropic universe: if a homogeneous and isotropic universe contains a uniformly distributed net charge, $F_{ab}=0$ by the symmetry of the spacetime but $J^b\neq 0$; then equation (\ref{meq0}) is violated if the $\xi R^b_{\;\,a}A^a$ term is not present. In the theory presented in this paper, from a five-dimensional Einstein field equation we can derive an electromagnetic field equation that has exactly the form of equation (\ref{meq0}) as proposed in \cite{li15}, but with a fixed $\xi=-2$.  

The fact that an electromagnetic field equation of the form in equation (\ref{meq0}) is derived also indicates that the theory presented in this paper is distinctly different from the Kaluza-Klein theory. In fact, in the Kaluza-Klein theory, the four-dimensional spacetime metric appearing in the decomposition of a five-dimensional metric is not defined on the spacetime hypersurface mentioned above. The electromagnetic potential vector in the Kaluza-Klein theory is also not a vector tangent to the hypersurface. They are in fact defined on a different hypersurface and are not related to the tensor variables used in this paper by diffeomorphisms, as will be proved in the paper. Hence, the theory presented in this paper is physically distinguishable from the Kaluza-Klein theory.

To some degrees, the present work was motivated by an attempt to ``unify'' the Kaluza-Klein theory and the brane world theory, i.e., to derive an electromagnetic field equation on a spacetime brane. Although the theory presented in this paper cannot be considered as unification or merge of the Kaluza-Klein theory and the brane world theory, it borrows essential ingredients from each of them: derivation of the electromagnetic field equation from the five-dimensional Einstein field equation as in the Kaluza-Klein theory, and derivation of the gravitation field equation in a four-dimensional spacetime by direct projection of the Einstein field equation in a five-dimensional bulk spacetime onto a hypersurface as in the brane world theory.

The paper is organized as follows. In Sec.~\ref{em_general}, we revisit the Maxwell equation in a flat Minkowski spacetime and its generalization to a curved spacetime. We show that, the Maxwell equation can also be expressed in terms of a symmetric tensor instead of the antisymmetric electromagnetic field tensor. When the Maxwell equation expressed in the symmetric tensor is extended to a curved spacetime, the field equation (\ref{meq0}) is naturally obtained, with $\xi=-2$. In Secs.~\ref{4+1}--\ref{field_eqs_II} we describe the theory proposed in this paper in details, which includes the 4+1 decomposition of the five-dimensional gravity described by the vacuum Einstein field equation, the action and the Lagrangian density expressed in terms of the scalar curvature and the extrinsic curvature of the four-dimensional spacetime as a hypersurface, derivation of electromagnetic fields and the electromagnetic field equation, and the Einstein field equation in the four-dimensional spacetime with the stress-energy tensor including electromagnetic fields and other matter as the source. In Sec.~\ref{rel_kk}, the relation and difference between the Kaluza-Klein theory and the theory presented in this paper are discussed. In Sec.~\ref{gauge}, we analyze and discuss gauge and diffeomorphic transformations of electromagnetic fields and gravity.

In Sec.~\ref{cosmo}, we discuss the rest terms in the Lagrangian, i.e., the terms in addition to those representing gravity and electromagnetism. We show that, under certain conditions, the rest terms may represent a cosmological constant. In Sec.~\ref{dis}, we present some discussions on the electromagnetic field equation (\ref{meq0}). In particular, we check under what conditions and in what kind of environments, the field equation (\ref{meq0}) can be tested with experiments and observations. In Sec.~\ref{sum} we summarize the results obtained in this paper.

The paper contains three Appendixes. In Appendix \ref{KK}, which is included as a supplement to Sec.~\ref{rel_kk}, we describe geometric interpretation of the metric decomposition in the Kaluza-Klein theory and derive its relation to the variables used in our theory. In Appendix \ref{decom}, we derive the field equations in the four-dimensional spacetime by projection of the five-dimensional Einstein field equation and discuss their relations to the equations derived from the Lagrangian formulation in Secs.~\ref{field_eqs_I} and \ref{field_eqs_II}. In Appendix \ref{hamilton}, we present a pseudo-Hamiltonian formulation of the new theory, and show that the field equations derived from the pseudo-Hamiltonian formulation confirm the results obtained in Secs.~\ref{field_eqs_I} and \ref{field_eqs_II}.

Throughout the paper we use Planck units with $G=c=\hbar=1$, where $G$ is the gravitational constant, $c$ is the speed of light, and $\hbar$ is the reduced Planck constant. However, in a few places the units are restored to get the magnitudes and dimensions of physical quantities.

\section{Electromagnetic Field Equations and General Relativity}
\label{em_general}

In the paper we adopt the abstract index notation for tensors as used in \cite{wal84}. That is, a tensor will be denoted by a letter followed by lower-case Latin indices, e.g., $v^a$, $T_{ab}$, etc. The component of a tensor in any basis is denoted by a letter followed by greek letter indices (occasionally by lower-case Latin letters from $i$ and onward, and capital Latin letters, as will be manifested in the text), e.g., $v^\mu$, $T_{\mu\nu}$, etc. The summation convention for tensor components is also adopted: an index appearing in both subscripts and superscripts is summed over all dimensions represented by the index.

\subsection{Revisit of Maxwell's equations in a flat spacetime}
\label{max_flat}

In terms of the electric field ${\bf E}$ and the magnetic field ${\bf B}$, in a global inertial frame in a flat spacetime Maxwell's equations can be written as
\begin{eqnarray}
        \nabla\cdot{\bf E} &=& 4\pi\rho_e \;, \label{meq10}\\
        \nabla\times{\bf B}-\frac{\partial{\bf E}}{\partial t} &=& 4\pi{\bf J} \;, \label{meq20}\\
        \nabla\cdot{\bf B} &=& 0 \;, \label{meq30}\\
        \nabla\times{\bf E}+\frac{\partial{\bf B}}{\partial t} &=& 0 \;, \label{meq40}
\end{eqnarray}
where $\rho_e$ is the charge density, and ${\bf J}$ is the current density vector.

In the theory of special relativity, the electric field ${\bf E}$ and the magnetic field ${\bf B}$ are combined to form an antisymmetric electromagnetic field tensor $F_{ab}$, with components in Cartesian coordinates
\begin{eqnarray}
        F_{\mu\nu}=\left(\begin{array}{cccc}
          0 & -E_x & -E_y & -E_z \\
          E_x & 0 & B_z & -B_y \\
          E_y & -B_z & 0 & B_x \\
          E_z & B_y & -B_x & 0\\
          \end{array}\right) \;.
\end{eqnarray}
The charge density and the current density vector are combined to form a four-dimensional current density vector $J^a$, with components $(\rho_e, J_x, J_y, J_z)$.

With the antisymmetric field tensor $F_{ab}$, the inhomogeneous Maxwell equations (\ref{meq10}) and (\ref{meq20}) are equivalent to the equation
\begin{eqnarray}
        \partial_aF^{ab}=-4\pi J^b \;, \label{meq1}
\end{eqnarray}
where $\partial_a$ is the ordinary derivative operator of the global inertial coordinates. The homogeneous Maxwell equations (\ref{meq30}) and (\ref{meq40}) are equivalent to the equation
\begin{eqnarray}
        \partial_aF_{bc}+\partial_bF_{ca}+\partial_cF_{ab}=0 \;. \label{meq2}
\end{eqnarray}
Note, $\partial_a$ is associated with the Minkowski metric tensor $\eta_{ab}$ (i.e., $\partial_a\eta_{bc}=0$).

The action of $\partial_b$ on the Maxwell equation (\ref{meq1}) leads to the equation for charge conservation
\begin{eqnarray}
        \partial_aJ^a=0=\frac{\partial\rho_e}{\partial t}+\nabla\cdot{\bf J} \;. \label{J_con0}
\end{eqnarray}

By the converse of the Poincar\'e lemma, equation (\ref{meq2}) indicates that there must exist a four-dimensional potential vector $A^a$ so that (see, e.g., \cite{wal84})
\begin{eqnarray}
        F_{ab}=\partial_aA_b-\partial_bA_a \;. \label{F_A0}
\end{eqnarray}
If $A^a$ is taken to be the fundamental variable, equation (\ref{meq2}) is automatically satisfied and hence trivial. Then equation (\ref{meq1}) is the only equation determining the evolution of electromagnetic fields.

It is well known that Maxwell's equations are invariant under the gauge transformation. That is, under the gauge transformation $A_a\rightarrow A_a+\partial_a\chi$, where $\chi$ is any scalar function, $F_{ab}$ is unchanged and hence Maxwell's equations are unchanged.

Adopting $A^a$ as the fundamental variable, the Maxwell equation (\ref{meq1}) can also be expressed with a symmetric tensor $H_{ab}$ instead of the antisymmetric tensor $F_{ab}$, where $H_{ab}$ is defined by \cite{li15}
\begin{eqnarray}
        H_{ab}\equiv\partial_aA_b+\partial_bA_a \;. \label{H_A0}
\end{eqnarray}
By the definitions of $F_{ab}$ and $H_{ab}$, we have
\begin{eqnarray}
        F_{ab}=H_{ab}-2\partial_bA_a \;.
\end{eqnarray}
Hence,
\begin{eqnarray}
        \partial_aF^{ab}=\partial_aH^{ab}-\partial^bH \;,
\end{eqnarray}
where $H \equiv \eta_{ab}H^{ab}=2\partial_a A^a$. The Maxwell equation (\ref{meq1}) can then be written as
\begin{eqnarray}
        \partial_aH^{ab}-\partial^bH=-4\pi J^b \;. \label{meq3}
\end{eqnarray}

Although in a flat spacetime equations (\ref{meq1}) and (\ref{meq3}) are equivalent, in the next section we will see that they lead to different equations for the electromagnetic field in a curved spacetime. In particular, later we will see that writing the electromagnetic field equation in terms  of a symmetric tensor makes it easier to derive electromagnetism from five-dimensional gravity.

\subsection{Electromagnetic field equations in a curved spacetime}
\label{em_curv}

In the theory of general relativity, a spacetime is defined by a manifold ${\cal M}$ with a symmetric metric tensor $g_{ab}$ on it. With a derivative operator $\nabla_a$ associated with the metric (i.e., $\nabla_ag_{bc}=0$), the Riemann curvature of the spacetime is defined, hence the Ricci tensor $R_{ab}$. By Einstein's field equation, $R_{ab}$ is related to the stress-energy tensor of matter, $T_{ab}$, by
\begin{eqnarray}
        G_{ab}\equiv R_{ab}-\frac{1}{2}Rg_{ab}=8\pi T_{ab} \;, \label{ein_eq}
\end{eqnarray}
where the Ricci scalar $R\equiv g^{ab}R_{ab}$. By the Bianchi identity $\nabla_aG^{ab}=0$, the Einstein field equation leads to the equation for the conservation of stress-energy
\begin{eqnarray}
        \nabla_aT^{ab}=0 \;. \label{dT0}
\end{eqnarray}

An equation of physics in a flat spacetime is transplanted into a general curved spacetime usually by the ``minimal substitution rule'' (not applied to the equation of gravity, of course), i.e., by replacing the Minkowski metric tensor $\eta_{ab}$ appearing in the equation by a general spacetime metric tensor $g_{ab}$, and the derivative operator $\partial_a$ associated with $\eta_{ab}$ by the derivative operator $\nabla_a$ associated with $g_{ab}$ \cite{wal84}. This is essentially reflection of the equivalence principle \cite{wei72,mis73}: at any point in a curved spacetime, it is possible to choose a local inertial frame so that within a sufficiently small region around that point the laws of nature take the forms as in a flat Minkowski spacetime. The critical point is that the region must be sufficiently small (much smaller than the radius of the spacetime curvature) so that any term of matter coupled to the spacetime curvature can be ignored. So, with the ``minimal substitution rule'', it is not possible to recover terms of matter coupled to the spacetime curvature, if those terms exist in the laws.

Applying the ``minimal substitution rule'' to equations (\ref{F_A0}) and (\ref{H_A0}), we get the definitions of $F_{ab}$ and $H_{ab}$ in a curved spacetime
\begin{eqnarray}
        F_{ab}=\nabla_aA_b-\nabla_bA_a \;, \label{F_A}
\end{eqnarray}
and
\begin{eqnarray}
        H_{ab}=\nabla_aA_b+\nabla_bA_a \;. \label{H_A}
\end{eqnarray}
The definition of $F_{ab}$ in equation (\ref{F_A}) leads to
\begin{eqnarray}
        \nabla_aF_{bc}+\nabla_bF_{ca}+\nabla_cF_{ab}=0 \;, \label{meq2g}
\end{eqnarray}
which is identical to the equation obtained by application of the ``minimal substitution rule'' to equation (\ref{meq2}). 

Application of the ``minimal substitution rule'' to equation (\ref{meq1}) leads to an inhomogeneous electromagnetic field equation in a general curved spacetime
\begin{eqnarray}
        \nabla_aF^{ab}=-4\pi J^b \;, \label{meq1g}
\end{eqnarray}
which implies the conservation of charge in a curved spacetime
\begin{eqnarray}
        \nabla_aJ^a = 0 \;, \label{J_con}
\end{eqnarray}
since $\nabla_a\nabla_bF^{ab}=0$. The equation (\ref{meq1g}) was originally proposed by Einstein \cite{ein55,ein16}. Since then it was widely accepted as the standard generalization of the Maxwell equation in a curved spacetime and sometimes called the Einstein-Maxwell equation.

However, if we apply the ``minimal substitution rule'' to equation (\ref{meq3}), we get an electromagnetic field equation in a curved spacetime
\begin{eqnarray}
        \nabla_aH^{ab}-\nabla^bH=-4\pi J^b \;, \label{meq4}
\end{eqnarray}
where
\begin{eqnarray}
        H=g^{ab}H_{ab}=2\nabla_aA^a \;.
\end{eqnarray}
By the identity
\begin{eqnarray}
        \nabla_aH^{ab}-\nabla^bH = \nabla_aF^{ab}+2R^b_{\;\,a}A^a \;,
\end{eqnarray}
we find that equation (\ref{meq4}) is equivalent to
\begin{eqnarray}
        \nabla_aF^{ab}+2R^b_{\;\,a}A^a=-4\pi J^b \;. \label{meq5}
\end{eqnarray}
Equation (\ref{meq5}) differs from equation (\ref{meq1g}) by a curvature-coupled term $2R^b_{\;\;a}A^a$ on the left-hand side, but is identical to the equation (\ref{meq0}) with $\xi=-2$.

The action of $\nabla_b$ on equation (\ref{meq5}) leads to $\nabla_aJ_\eff^a=0$, where
\begin{eqnarray}
        J_\eff^a\equiv J^a+\frac{1}{2\pi}R^a_{\;\,b}A^b \label{Jeff}
\end{eqnarray}
can be interpreted as an effective current density vector. The usual equation of charge conservation, $\nabla_aJ^a=0$, is preserved if and only if
\begin{eqnarray}
        \nabla^a\left(R_{ab}A^b\right)=0 \;, \label{gac1}
\end{eqnarray}
which is equivalent to
\begin{eqnarray}
        R_{ab}H^{ab}+A^b\nabla_bR=0 \;, \label{gac2}
\end{eqnarray}
after application of the Bianchi identity.

Both equations (\ref{meq1g}) and (\ref{meq5}) are generally covariant, and return to the Maxwell equation (\ref{meq1}) in a flat spacetime. We have seen that, starting from the same equation in the flat spacetime, we can get different corresponding equations in a curved spacetime with the ``minimal substitution rule''. This is easily understood since the Maxwell equation (\ref{meq1}) contains second-order derivatives of $A^a$ and the order of derivatives of a vector matters in a curved spacetime. That is, $\partial_a\partial^bA^a$ becomes $\nabla_a\nabla^bA^a$ in a curved spacetime, but $\partial^b\partial_aA^a$ becomes $\nabla^b\nabla_aA^a=\nabla_a\nabla^bA^a-R^b_{\;\;a}A^a$, although $\partial_a\partial^bA^a=\partial^b\partial_aA^a$. So, the difference between equations (\ref{meq1g}) and (\ref{meq5}) shows the ambiguity in writing down an equation of physics in a curved spacetime according to the corresponding equation in a flat spacetime. Although it is not possible to decide which equation is correct {\it a priori}, in \cite{li15} it has been shown that equation (\ref{meq1g}) is not compatible with a universe with a uniformly distributed charge, but equation (\ref{meq5}) is.

Equations (\ref{meq1g}) and (\ref{meq5}) are not the only possible equations extended to a curved spacetime of the Maxwell equation, even when the ``minimal substitution rule'' is forced. Since $\partial_a\partial^bA^a=(1+\xi)\partial_a\partial^bA^a-\xi\partial^b\partial_aA^a$ where $\xi$ is any number, $\partial_aF^{ab}=\partial_a\partial^aA^b-\partial_a\partial^bA^a$ can be substituted by $\nabla_a\nabla^aA^b-(1+\xi)\nabla_a\nabla^bA^a+\xi\nabla^b\nabla_aA^a =\nabla_aF^{ab}-\xi R^b_{\;\;a}A^a$ in a curved spacetime, according to the ``minimal substitution rule''. Then equation (\ref{meq0}) is obtained. Since $\xi$ is arbitrary (it can even be a function), there are an infinite number of possible equations in a curved spacetime corresponding to the Maxwell equation in a flat spacetime. It is worth to point out that these equations do not violate the equivalence principle, since in a region of size much smaller than the spacetime curvature radius, the curvature-coupled term in the equation is negligibly small for an electromagnetic field with a coherent space scale smaller than or comparable to the size of the region.

The electromagnetic field equation (\ref{meq4}) can be recast in a neater form
\begin{eqnarray}
        \nabla_a\Theta^{ab}=-4\pi J^b \;, \label{meq4t}
\end{eqnarray}
where
\begin{eqnarray}
        \Theta_{ab}=\Theta_{ba}\equiv H_{ab}-Hg_{ab} \;. \label{Theta_ab}
\end{eqnarray}
Then, the conservation of charge demands that
\begin{eqnarray}
        \nabla_a\nabla_b\Theta^{ab}=0 \;. \label{gac3}
\end{eqnarray}

Because of the presence of the curvature-coupled term, in a spacetime with a nonvanishing $R_{ab}$ the field equation (\ref{meq5}) (identically, eqs.~\ref{meq4} and \ref{meq4t}) is not invariant under gauge transformations. This issue will be discussed in details in Sec.~\ref{gauge}.

\subsection{The stress-energy tensor of electromagnetic fields}
\label{stress_em}

The Einstein field equation and the electromagnetic field equations written out in previous subsections hold in a spacetime of any integer number dimensions, although the representation of electromagnetic fields in terms of the vectors ${\bf E}$ and ${\bf B}$ is applicable only to a four-dimensional spacetime. Here we write down the stress-energy tensor of electromagnetic fields in an $n$-dimensional spacetime, where $n$ is any integer $\ge 4$ (then we have $g^{ab}g_{ab}=n$).

It can be checked that, the electromagnetic field equation (\ref{meq4t}) (identical to eqs.~\ref{meq4} and \ref{meq5}) can be derived from an action 
\begin{eqnarray}
        S_\emm = \int {\cal L}_\emm(g_{ab},A_a){\bf e} \label{S_em0}
\end{eqnarray}
by variation with respect to $A_a$, where ${\bf e}$ is a fixed volume element, and 
\begin{eqnarray}
        {\cal L}_\emm(g_{ab},A_a) &=& \sqrt{-g}\left(-\frac{1}{4}F_{ab}F^{ab}+R_{ab}A^aA^b+4\pi A_aJ^a\right)=\sqrt{-g}\left[-\frac{1}{4}\left(H_{ab}H^{ab}-H^2\right)+4\pi A_aJ^a\right] \\
        &=& \sqrt{-g}\left[-\frac{1}{4}\left(\Theta_{ab}\Theta^{ab}-\frac{1}{n-1}\Theta^2\right)+4\pi A_aJ^a\right] \label{L_em}
\end{eqnarray}
is the Lagrangian density of electromagnetic fields. Here, $g$ is the determinant of the component matrix of $g_{ab}$ in the coordinate system associated with ${\bf e}$, and
\begin{eqnarray}
        \Theta\equiv g^{ab}\Theta_{ab}=-(n-1)H \;. \label{Theta_H}
\end{eqnarray}

The stress-energy tensor of electromagnetic fields is derived from the action by variation with respect to $g^{ab}$ (after ignoring the source term $4\pi A_aJ^a$ in the Lagrangian density) \cite{wal84}
\begin{eqnarray}
        T_{\emm,ab}=-\frac{1}{2\pi\sqrt{-g}}\frac{\delta}{\delta g^{ab}}S_\emm(J^a=0) \;.
\end{eqnarray}
Then, for the action defined by equations (\ref{S_em0}) and (\ref{L_em}), we get
\begin{eqnarray}
        T_{\emm,ab} &=& \frac{1}{4\pi}\left(F_{ac}F_b^{\;\;c}-\frac{1}{4}g_{ab}F_{cd}F^{cd}\right) \nonumber\\
        &&-\frac{1}{4\pi}\left\{\nabla^c\nabla_c(A_aA_b)-2\nabla^c\nabla_{(a}(A_{b)}A_c)+4A^cR_{c(a}A_{b)}+g_{ab}\left[\nabla_c\nabla_d(A^cA^d)-R_{cd}A^cA^d\right]\right\} \;; \label{Tem_ab2}
\end{eqnarray}
or, equivalently,
\begin{eqnarray}
	T_{\emm,ab} = \frac{1}{4\pi}\left[\Theta_{ac}\Theta_b^{\;\;c}-\frac{1}{n-1}\Theta\Theta_{ab} -\frac{1}{4}\left(\Theta_{cd}\Theta^{cd}-\frac{1}{n-1}\Theta^2\right)g_{ab} -\nabla^c\left(2A_{(a}\Theta_{b)c}-A_c\Theta_{ab}\right)\right] \;. \label{Tem_ab}
\end{eqnarray}
In equations (\ref{Tem_ab2}) and (\ref{Tem_ab}), the parentheses in the indexes of a tensor denote symmetrization of the tensor.

The term in the first line on the right-hand side of equation (\ref{Tem_ab2}) is just the usual stress-energy tensor of electromagnetic fields described by the Einstein-Maxwell equation (\ref{meq1g}). The terms in the second line are derived from the curvature term in equation (\ref{L_em}), $R_{ab}A^aA^b$, which was discussed in \cite{li15} in details. The divergence of $T_{\emm,ab}$, i.e., $\nabla^aT_{\emm,ab}$, was also calculated and discussed in \cite{li15} (see also Sec.~\ref{dis} in this paper).

\section{4+1 Decomposition of 5D Gravity}
\label{4+1}

By considering the 4+1 decomposition of the Einstein field equation in a five-dimensional (5D) spacetime, \citet{kal21} and \citet{kle26a,kle26b} were able to derive the four-dimensional Einstein-Maxwell equation and the four-dimensional Einstein field equation from the five-dimensional gravity. In other words, electromagnetic fields and gravity were unified into five-dimensional gravity in the Kaluza-Klein theory, at least formally. 

The strategy of Kaluza and Klein was to, without loss of generality, take a five-dimensional coordinate system $\{x^0,x^1,x^2,x^3,x^4=w\}$ and write a five-dimensional spacetime metric tensor $\t{g}_{AB}$ ($A,B=0,1,2,3,4$) in the form
\begin{eqnarray}
	\t{g}_{AB} = \left(\begin{array}{cc}
          \!~_kg_{\mu\nu} +\phi^2A_\mu A_\nu & \phi^2 A_\mu\\[2mm]
          \phi^2 A_\nu & \phi^2
          \end{array}\right)
        \;, \label{ktg_ab}
\end{eqnarray}
where $\mu,\nu=0,1,2,3$. Kaluza and Klein interpreted $\!~_kg_{\mu\nu}$ as the spacetime metric in the four-dimensional spacetime defined by the coordinates $\{x^0,x^1,x^2,x^3\}$, $A_\mu$ as the electromagnetic field potential vector in the four-dimensional spacetime, and $\phi$ as an unidentified scalar field. By the assumption that $\!~_kg_{\mu\nu}$ and $A_\mu$ do not depend on the $w$ coordinate and $\phi=\mbox{const}$, Kaluza and Klein have shown that the vacuum five-dimensional Einstein field equation is equivalent to the combination of a four-dimensional Einstein-Maxwell equation and a four-dimensional Einstein field equation with the stress-energy tensor of electromagnetic fields as the source. This is the basis of the Kaluza-Klein theory.

In this paper, we consider a different 4+1 decomposition of a five-dimensional spacetime, which is constructed by direct projection of the five-dimensional metric onto a timelike hypersurface embedded in the five-dimensional spacetime. The scheme we take is much like that in the Hamiltonian formulation of general relativity and canonical quantization of gravity (see, e.g., \cite{wal84,mis73,arn62}), except that in the latter case the hypersurface is spacelike, i.e., has a timelike normal but here the hypersurface is timelike, i.e., has a spacelike normal. Assuming that the bulk spacetime is described by the vacuum five-dimensional Einstein field equation, we investigate the field equations induced on the spacetime hypersurface by projection. We will find that, in addition to a four-dimensional Einstein field equation, an electromagnetic field equation in the form of equation (\ref{meq5}) is derived. The electromagnetic field equation is equivalent to the equation (\ref{meq5}), but different from the Einstein-Maxwell equation (\ref{meq1g}). As will be explained in details in Sec.~\ref{rel_kk} and Appendix~\ref{KK}, our procedure is distinctly different from that of Kaluza and Klein, and, as a result, the derived field equations are also different.

For generality, we consider an $(n+1)$-dimensional spacetime $(\t{\cal M},\t{g}_{ab})$ that is (locally at least) covered by a coordinate system $\{x^0,...,x^{n-1},x^n=w\}$, and a hypersurface ${\cal M}$ in it defined by $w=\mbox{const}$.\footnote{According to the Campbell theorem, any analytic $n$-dimensional Riemannian space can be locally embedded in an $(n+1)$-dimensional Ricci-flat space \cite{cam26,rom96}. Hence, consideration of an $n$-dimensional spacetime embedded in an $(n+1)$-dimensional spacetime does not seem to put much constraint on the properties of the $n$-dimensional spacetime.} We denote the unit normal to ${\cal M}$ by $n^a$, which is a spacelike vector and hence satisfies the condition
\begin{eqnarray}
	n^an_a=1 \;.
\end{eqnarray}
Then, the tangent vector of the $w$-coordinate line, $w^a=(\partial/\partial w)^a$, can be decomposed as
\begin{eqnarray}
	w^a=N n^a+N^a \;, \label{w^a}
\end{eqnarray}
where $N$ is a scalar function, and the vector $N^a$ is tangent to ${\cal M}$, i.e.,
\begin{eqnarray}
	n^aN_a=0 \;.
\end{eqnarray}
In the Hamiltonian formulation of general relativity, $N$ is called the lapse function, and $N^a$ is called the shift vector.

Given $\t{g}_{ab}$ and $n^a$, an $n$-dimensional metric tensor $g_{ab}$ is naturally induced on ${\cal M}$
\begin{eqnarray}
	g_{ab}\equiv \t{g}_{ab}-n_an_b \;, \label{g_tg}
\end{eqnarray}
which satisfies
\begin{eqnarray}
	g_{ab}n^b=0 \;.
\end{eqnarray}
Then, $({\cal M},g_{ab})$ forms an $n$-dimensional spacetime. By definition, $N^a$ is a vector field on ${\cal M}$, i.e., $N^a\in\tm$.\footnote{We denote a tensor space on a manifold ${\cal M}$ generally by $\tm$, regardless of the type of the tensor (a scalar, a vector and a dual vector, a tensor of any type).} The inverse metric tensor on ${\cal M}$ is
\begin{eqnarray}
	g^{ab}=\t{g}^{ac}\t{g}^{bd}g_{cd}=\t{g}^{ab}-n^an^b \;. \label{bg^ab}
\end{eqnarray}
It can be checked that
\begin{eqnarray}
	g_a^{\;\;b}=\t{g}^{bc}g_{ac}=g_{ac}g^{bc}=\t{\delta}_a^{\;\;b}-n_an^b \;,
\end{eqnarray}
where $\t{\delta}_a^{\;\;b}$ is the identity operator on $\t{\cal M}$. The $g_a^{\;\;b}=\delta_a^{\;\;b}$ defined above is the identity operator on ${\cal M}$, i.e., for any vector $v^a\in\tm$, we have $g_a^{\;\;b}v^a=v^b$.

Note, for any tensor $\in\ttm$, the index can be raised and lowered by $\t{g}^{ab}$ and $\t{g}_{ab}$. For any tensor $\in\tm$, the index can be raised and lowered by $\t{g}^{ab}$ and $\t{g}_{ab}$, or equivalently by $g^{ab}$ and $g_{ab}$. The metric tensor $g_{ab}$ can be used as a projection operator to project a tensor on $\t{\cal M}$ onto ${\cal M}$.

With the metric tensor $g_{ab}$ defined above, the matrix representation of the $(n+1)$-dimensional metric tensor $\t{g}_{ab}$ in the coordinate system $\{x^0,...,x^{n-1},x^n=w\}$ can be written as
\begin{eqnarray}
	\t{g}_{AB} = \left(\begin{array}{cc}
          g_{\mu\nu} & N_\mu\\[2mm]
          N_\nu & N^2+N_\rho N^\rho
          \end{array}\right)
        \;, \label{tg_ab2}
\end{eqnarray}
where $\mu,\nu,\rho=0,1,...,n-1$, and $A,B=0,1,...,n$. By comparison with equation (\ref{ktg_ab}), we can see the difference between our decomposition of the metric $\t{g}_{AB}$ and that of Kaluza and Klein. The $g_{\mu\nu}$ is the metric tensor on ${\cal M}$ by projection of the bulk metric $\t{g}_{AB}$, but the $\!~_kg_{\mu\nu}$ in equation (\ref{ktg_ab}) is not.

Given the spacetime metric tensor $\t{g}_{ab}$, a derivative operator associated with $\t{g}_{ab}$ is defined, which is denoted by $\t{\nabla}_a$ (i.e., $\t{\nabla}_a\t{g}_{bc}=0$). Since ${\cal M}$ is considered as a hypersurface embedded in an $(n+1)$-dimensional spacetime $\t{\cal M}$, we can define the extrinsic curvature tensor of ${\cal M}$ (see, e.g., \cite{wal84})
\begin{eqnarray}
	K_{ab}\equiv g_a^{\;\;c}\t{\nabla}_cn_b=\frac{1}{2}\t{\pounds}_ng_{ab} =K_{ba} \;, \label{Kab}
\end{eqnarray}
where $\t{\pounds}_n$ denotes the Lie derivative with respect to the vector $n^a$. The extrinsic curvature $K_{ab}$ defines how ${\cal M}$ is embedded in $\t{\cal M}$, and is tangent to ${\cal M}$, i.e., $n^aK_{ab}=0$.

By application of the Frobenius' theorem \cite{wal84}, it can be derived that
\begin{eqnarray}
	\t{\nabla}_{a}n_{b}=K_{ab}+n_{a}a_{b} \;, \label{nab_n}
\end{eqnarray}
where
\begin{eqnarray}
	a^b\equiv n^a\t{\nabla}_an^b \;, \hspace{1cm}a^bn_b=0 \;. \label{aa_def}
\end{eqnarray}
The acceleration vector $a^a\in\tm$ describes the curvature of a curve orthogonally intersecting the hypersurface ${\cal M}$ \cite{mis73}. If the curve is a geodesics, we have $a^a=0$.

The derivative operator associated with the $g_{ab}$ on ${\cal M}$, denoted by $\nabla_a$ (i.e.,$\nabla_ag_{bc}=0$), is defined by \cite{wal84}
\begin{eqnarray}
	\nabla_c T^{a_1...a_k}_{\;\;\;\;\;\;\;\;\;\;\;b_1...b_l}\equiv g^{a_1}_{\;\;\;d_1}...g_{b_l}^{\;\;\;e_l}g_c^{\;\;f}\t{\nabla}_fT^{d_1...d_k}_{\;\;\;\;\;\;\;\;\;\;\;e_1...e_l} \;, \label{der_M}
\end{eqnarray}
for any tensor $T^{a_1...a_k}_{\;\;\;\;\;\;\;\;\;\;\;b_1...b_l}\in\tm$.

With the derivative operator $\t{\nabla}_a$, the Riemann curvature tensor $\t{R}_{abc}^{\;\;\;\;\;\;d}$ on $\t{\cal M}$ is defined, then the Ricci tensor $\t{R}_{ab}$ and the Ricci scalar $\t{R}$. With the derivative operator $\nabla_a$, the Riemann curvature tensor $R_{abc}^{\;\;\;\;\;\;d}$ on ${\cal M}$ is defined, then the Ricci tensor $R_{ab}$ and the Ricci scalar $R$. It can be derived that $\t{R}$ and $R$ are related by (e.g., \cite{wal84,mis73})
\begin{eqnarray}
	\t{R}=R-K_{ab}K^{ab}+K^2-2\t{\nabla}_av^a \;, \label{R_R2}
\end{eqnarray} 
where
\begin{eqnarray}
	K\equiv g^{ab}K_{ab}=\t{\nabla}_an^a \;,
\end{eqnarray}
and
\begin{eqnarray}
	v^a\equiv n^a\t{\nabla}_cn^c-n^c\t{\nabla}_cn^a =Kn^a-a^a \;. \label{v^a}
\end{eqnarray}

The Einstein-Hilbert action of gravity in the $(n+1)$-dimensional spacetime $(\t{\cal M},\t{g}_{ab})$ is
\begin{eqnarray}
	\t{S}_G = \int \sqrt{-\t{g}}\,\t{R}\,\t{\bf e} = \int \sqrt{-\t{g}}\left(R-K_{ab}K^{ab}+K^2\right)\t{\bf e} \;, \label{SG}
\end{eqnarray}
where $\t{\bf e}$ is a fixed volume element associated with the coordinate system on $\t{\cal M}$, and $\t{g}$ is the determinant of the component matrix of $\t{g}_{ab}$. In equation (\ref{SG}) we have substituted equation (\ref{R_R2}) and ignored the term $2\t{\nabla}_av^a$ since it does not contribute to the action integral with suitable boundary conditions. The field equation derived from $\delta\t{S}_G/\delta\t{g}^{ab}=0$ is just the vacuum Einstein field equation in the $(n+1)$-dimensional spacetime
\begin{eqnarray}
	\t{G}_{ab}\equiv\t{R}_{ab}-\frac{1}{2}\t{R}\t{g}_{ab}=0 \;, \label{ein_eq_n+1}
\end{eqnarray}
which is equivalent to
\begin{eqnarray}
	\t{R}_{ab}=0 \;. \label{ein_eq2_n+1}
\end{eqnarray}

Substituting equation (\ref{w^a}) into equation (\ref{Kab}), we can express $K_{ab}$ in terms of $N$, $N_a$, and $g_{ab}$
\begin{eqnarray}
	K_{ab} = \frac{1}{2}N^{-1}\left(\dot{g}_{ab}-M_{ab}\right) \;, \label{Kab_M}
\end{eqnarray}
where
\begin{eqnarray}
	\dot{g}_{ab}\equiv\frac{\partial}{\partial w}g_{ab}\equiv g_a^{\;\;c}g_b^{\;\;d}\t{\pounds}_wg_{cd} \;, \label{dot_g}
\end{eqnarray}
and
\begin{eqnarray}
	M_{ab}\equiv\nabla_aN_b+\nabla_bN_a \;. \label{Mab}
\end{eqnarray}
Both $\dot{g}_{ab}$ and $M_{ab}$ are symmetric tensors tangent to ${\cal M}$. The trace of $K_{ab}$ is
\begin{eqnarray}
	K = \frac{1}{2}N^{-1}\left(g^{ab}\dot{g}_{ab}-M\right) \;, \label{K_M}
\end{eqnarray}
where
\begin{eqnarray}
	M\equiv g^{ab}M_{ab}=2\nabla_aN^a \;. \label{M}
\end{eqnarray}

Substituting equations (\ref{Kab_M}), (\ref{K_M}), and $\sqrt{-\t{g}}=N\sqrt{-g}$ into equation (\ref{SG}), we get
\begin{eqnarray}
	\t{S}_G=\int \t{\cal L}_G(N,N_a,g_{ab})\t{\bf e} \;, \label{SG2}
\end{eqnarray}
where the Lagrangian density $\t{\cal L}_G$ is defined by
\begin{eqnarray}
	\t{\cal L}_G(N,N_a,g_{ab}) = \sqrt{-g}N\left[R-\frac{1}{4}N^{-2}\left(M_{ab}M^{ab}-M^2\right)\right] -\frac{1}{4}\sqrt{-g}N^{-1}\left(g^{ac}g^{bd}-g^{ab}g^{cd}\right) \left(\dot{g}_{ab}\dot{g}_{cd}-2\dot{g}_{ab}M_{cd}\right) \;. \label{LG}
\end{eqnarray}

In the following sections we will investigate the field equations derived from the action defined by equations (\ref{SG2}) and (\ref{LG}) by variation with respect to $N$, $N_a$, and $g_{ab}$. We will show that electromagnetic fields are contained in $M_{ab}$ and $M$. The variation of $\t{S}_G$ with respect to $N$, $N_a$, and $g_{ab}$ leads to a scalar constraint equation, the electromagnetic field equation, and the Einstein field equation on ${\cal M}$, respectively.

\section{Inspection of the Lagrangian}
\label{inspec}

The Lagrangian density in equation (\ref{LG}) can be separated into several parts
\begin{eqnarray}
	\t{\cal L}_G ={\cal L}_G+{\cal L}_\emm+{\cal L}_\ot \;, \label{Ltot}
\end{eqnarray}
where
\begin{eqnarray}
	{\cal L}_G &\equiv& \sqrt{-g}NR \;, \label{LG0}
\end{eqnarray}
\begin{eqnarray}
	{\cal L}_\emm &\equiv& -\frac{1}{4}\sqrt{-g}N^{-1}\left(M_{ab}M^{ab}-M^2\right) \;, \label{Lem0}
\end{eqnarray}
and
\begin{eqnarray}
	{\cal L}_\ot &\equiv& -\frac{1}{4}\sqrt{-g}N^{-1}\left(g^{ac}g^{bd}-g^{ab}g^{cd}\right) \left(\dot{g}_{ab}\dot{g}_{cd}-2\dot{g}_{ab}M_{cd}\right) \;. \label{Lother}
\end{eqnarray}

The ${\cal L}_G$ in equation (\ref{LG0}) will be interpreted as the Lagrangian density of gravity on ${\cal M}$, since it is proportional to the $n$-dimensional Ricci scalar $R$. The ${\cal L}_\emm$ in equation (\ref{Lem0}) will be interpreted as the Lagrangian density of electromagnetic fields on ${\cal M}$, which will be explained in details below. The ${\cal L}_\ot$ in equation (\ref{Lother}) contains all other terms in the total Lagrangian density, so will be interpreted as the Lagrangian density of other matter fields on ${\cal M}$ and their interaction with the electromagnetic field, although the nature of the matter cannot be determined. 

Recall that the total Lagrangian density of gravity and electromagnetic fields in an $n$-dimensional spacetime is (see eq.~\ref{L_em} and ref.~\cite{wal84})
\begin{eqnarray}
	{\cal L} =\sqrt{-g}\left[l_\P^{-n+2}R-\left(H_{ab}H^{ab}-H^2\right)\right] \;, \label{LG_em}
\end{eqnarray}
where $l_\P$ is the Planck length in the $n$-dimensional spacetime, which is related to the $n$-dimensional gravitational constant by $l_\P=G_n^{1/(n-2)}$.

If we assume that $\nabla_aN=0$ and let
\begin{eqnarray}
	N_a=2Nl_\P^{n/2-1} A_a \;, \label{N_A}
\end{eqnarray}
the ${\cal L}_\emm$ in equation (\ref{Lem0}) can be written as
\begin{eqnarray}
	{\cal L}_\emm= -\sqrt{-g}Nl_\P^{n-2}\left(H_{ab}H^{ab}-H^2\right) \;. \label{Lem0a}
\end{eqnarray}
Then we have
\begin{eqnarray}
	{\cal L}_G+{\cal L}_\emm &=& \sqrt{-g}Nl_\P^{n-2} \left[l_\P^{-n+2}R-\left(H_{ab}H^{ab}-H^2\right)\right] \;. \label{LG_em2}
\end{eqnarray}

Comparing equations (\ref{LG_em}) and (\ref{LG_em2}), we see that ${\cal L}_G+{\cal L}_\emm$ is identical to the Lagrangian density of electromagnetic fields and gravity, up to a constant multiplier in the total Lagrangian. This fact indicates that ${\cal L}_\emm$ can be interpreted as the Lagrangian density of electromagnetic fields on ${\cal M}$. The ${\cal L}_\ot$ in equation (\ref{Lother}) contains a term $\propto M_{cd}\propto H_{cd}$, which can be interpreted as representing the interaction of electromagnetic fields with other matter. Variation of the total action with respect to $A_a$ will then give rise to an electromagnetic field equation in the form of equation (\ref{meq5}), at least in the case of $\nabla_aN=0$ (see next section).  Hence, we see that, in a spacetime as a hypersurface embedded in a higher-dimensional spacetime, electromagnetism is contained in the extrinsic curvature tensor of the hypersurface.

Because the total Lagrangian can be defined up to a constant multiplier, the gravitational constant on ${\cal M}$ cannot be determined from the ${\cal L}_G+{\cal L}_\emm$ in equation (\ref{LG_em2}). However, this does not affect the field equations on ${\cal M}$ derived from the action. With the ${\cal L}_G$, ${\cal L}_\emm$, and ${\cal L}_\ot$ defined above, variation of the total action with respect to $g^{ab}$ (the details will be given in Sec.~\ref{field_eqs_II}) leads to the Einstein field equation on ${\cal M}$
\begin{eqnarray}
	G_{ab}=8\pi l_\P^{n-2}\left(T_{\emm,ab}+T_{\ot,ab}\right) \;, \label{ein_eq_n2}
\end{eqnarray}
where $T_{\emm,ab}$ is the stress-energy tensor of electromagnetic fields given by equation (\ref{Tem_ab}) (equivalent to eq.~\ref{Tem_ab2}), and
\begin{eqnarray}
	T_{\ot,ab}=-\frac{1}{8\pi Nl_\P^{n-2}\sqrt{-g}}\frac{\delta}{\delta g^{ab}}\int {\cal L}_\ot\,\t{\bf e} 
\end{eqnarray}
is the stress-energy tensor of other matter fields, including their interaction with electromagnetic fields.

Finally, we have some comments on the condition $\nabla_aN=0$. By the definition of $n^a$, we have
\begin{eqnarray}
	n_a=Ndw_a=N\nabla_aw \;. \label{n_a}
\end{eqnarray}
Hence, we have
\begin{eqnarray}
	\tilde{\nabla}_bn_a=N\t{\nabla}_a\t{\nabla_b}w+\t{\nabla}_aw\t{\nabla}_bN \;.
\end{eqnarray}
Since $n_an^a=1$, we get
\begin{eqnarray}
	0 = n^a\tilde{\nabla}_bn_a=Nn^a\t{\nabla}_a\t{\nabla}_bw+n^a\t{\nabla}_aw\t{\nabla}_bN \;. \label{ndn}
\end{eqnarray}
From equation (\ref{n_a}) we have $\nabla_aw=N^{-1}n_a$. Substituting it into equation (\ref{ndn}), we get
\begin{eqnarray}
	0 = n^a\t{\nabla}_an_b+\frac{1}{N}\left(\t{\nabla}_bN-n_bn^a\t{\nabla}_aN\right) = n^a\t{\nabla}_an_b+\frac{1}{N}\nabla_bN \;.
\end{eqnarray}
Hence, by the definition of $a_a$ (eq.~\ref{aa_def}), we get
\begin{eqnarray}
	a_a=-\nabla_a\ln N \;. \label{a_dN}
\end{eqnarray}

Therefore, the condition of $\nabla_aN=0$ (i.e., $N$ is constant on ${\cal M}$) is equivalent to the condition of $a_a=0$, i.e., ${\cal M}$ is orthogonal to a congruence of spacelike geodesics. This condition must also be fulfilled if we require that the ${\cal L}_G$ in equation (\ref{LG0}) is equal to $R$ multiplied by a constant on ${\cal M}$.

\section{Derivation of the Electromagnetic Field Equation}
\label{field_eqs_I}

The Lagrangian density $\t{\cal L}_G$ in equation (\ref{LG}), which contains three independent variables $N$, $N_a$, and $g_{ab}$, can be divided into several parts according to equations (\ref{Ltot})--(\ref{Lother}). As discussed in Sec.~\ref{inspec}, ${\cal L}_G$ is interpreted as the Lagrangian density of gravity on ${\cal M}$, ${\cal L}_\emm$ is interpreted as the Lagrangian density of electromagnetic fields, and ${\cal L}_\ot$ is interpreted as the Lagrangian density of other matter fields and their interaction with electromagnetic fields. The field equations on ${\cal M}$ are derived by variation of the action $\t{S}_G$ with respect to $N$, $N_a$, and $g_{ab}$, respectively.

The ${\cal L}_\ot$ can be further separated into two parts,
\begin{eqnarray}
	L_\ot={\cal L}_m+{\cal L}_\intt \;,
\end{eqnarray}
where
\begin{eqnarray}
	{\cal L}_m \equiv -\frac{1}{4}\sqrt{-g}N^{-1}\left(g^{ac}g^{bd}-g^{ab}g^{cd}\right)\dot{g}_{ab}\dot{g}_{cd} \;, \label{Lm_M}
\end{eqnarray}
and
\begin{eqnarray}
	{\cal L}_\intt \equiv \frac{1}{2}\sqrt{-g}N^{-1}\left(g^{ac}g^{bd}-g^{ab}g^{cd}\right)\dot{g}_{ab}M_{cd} \;. \label{Lint_M}
\end{eqnarray}
The ${\cal L}_m$ does not contain $N_a$ and is interpreted as the Lagrangian density of matter fields. The $\dot{g}_{ab}$ is interpreted as matter fields, although we do not know the nature of the matter (normal matter, dark matter, or dark energy that have been observed in cosmology). The ${\cal L}_\intt$ contains both $\dot{g}_{ab}$ and $N_a$, hence is interpreted as representing the interaction between the electromagnetic field and the matter field.\footnote{The Lagrangian in equation (\ref{LG}) does not contain any derivatives of $N$ so we do not interpret $N$ as a matter field.}

Then, the total Lagrangian density can be written as
\begin{eqnarray}
	\t{\cal L}_G = {\cal L}_G+{\cal L}_\emm+{\cal L}_m+{\cal L}_\intt \;. \label{LG2a}
\end{eqnarray}
Accordingly, the total action can be written as
\begin{eqnarray}
	\t{S}_G = S_G+S_\emm+S_m+S_\intt \;, \label{SG2x}
\end{eqnarray}
where $S_G$, $S_\emm$, $S_m$, and $S_\intt$ are the integration of corresponding Lagrangian densities.

To simplify the derived equations, we define variables
\begin{eqnarray}
	\Psi_{ab}\equiv \frac{1}{2}N^{-1}\left(M_{ab}-Mg_{ab}\right) \;, \hspace{1cm} \Psi\equiv g^{ab}\Psi_{ab}= -\frac{1}{2}(n-1)N^{-1}M \;, \label{Psi_ab}
\end{eqnarray}
and
\begin{eqnarray}
	\Phi_{ab}\equiv -\frac{1}{2}N^{-1}\left(g_a^{\;\;c}g_b^{\;\;d}-g_{ab}g^{cd}\right)\dot{g}_{cd} \;, \hspace{1cm} \Phi\equiv g^{ab}\Phi_{ab}= \frac{1}{2}(n-1)N^{-1}g^{ab}\dot{g}_{ab} \;. \label{Phi_ab}
\end{eqnarray}
The variable $\Psi_{ab}$ is used to replace $M_{ab}$ and hence represents the electromagnetic field, and $\Phi_{ab}$ is used to replace $\dot{g}_{ab}$ and hence represents the presumed matter field.

Then, ${\cal L}_\emm$, ${\cal L}_m$, and ${\cal L}_\intt$, can be rewritten as
\begin{eqnarray}
	{\cal L}_\emm = -\sqrt{-g}N\left(\Psi_{ab}\Psi^{ab}-\frac{1}{n-1}\Psi^2\right) \;, \label{Lem_M2}
\end{eqnarray}
\begin{eqnarray}
	{\cal L}_m = -\sqrt{-g}N\left(\Phi_{ab}\Phi^{ab}-\frac{1}{n-1}\Phi^2\right) \;, \label{Lm_M2}
\end{eqnarray}
and
\begin{eqnarray}
	{\cal L}_\intt = -2\sqrt{-g}N\left(\Psi_{ab}\Phi^{ab}-\frac{1}{n-1}\Psi\Phi\right) \;, \label{Lint_M2}
\end{eqnarray}
respectively.

The $N$ appears in the Lagrangian density only as a multiplication parameter. Variation of the Lagrangian density $\t{\cal L}_G$ with respect to $N$ leads to\footnote{Note, all $\Psi_{ab}\Psi^{ab}$, $\Psi^2$, $\Phi_{ab}\Phi^{ab}$, $\Phi^2$, $\Psi_{ab}\Phi^{ab}$, and $\Psi\Phi$ are $\propto N^{-2}$.}
\begin{eqnarray}
	\delta\t{\cal L}_G = \sqrt{-g}\left(R+\Pi_{ab}\Pi^{ab}-\frac{1}{n-1}\Pi^2\right)\delta N \;,
\end{eqnarray}
where
\begin{eqnarray}
	\Pi_{ab}\equiv \Psi_{ab}+\Phi_{ab} \;, \hspace{1cm} \Pi\equiv g^{ab}\Pi_{ab}=\Psi+\Phi \;. \label{Pi_ab}
\end{eqnarray}
Hence, $\delta\t{S}_G/\delta N=0$ leads to a scalar constraint equation
\begin{eqnarray}
	 R+\Pi_{ab}\Pi^{ab}-\frac{1}{n-1}\Pi^2 = 0 \;. \label{scalar_eq_L}
\end{eqnarray}

By equation (\ref{Kab_M}), we have
\begin{eqnarray}
	\Pi_{ab}=-K_{ab}+Kg_{ab} \;. \label{Pi_ab_K}
\end{eqnarray}
Hence, equation (\ref{scalar_eq_L}) is equivalent to
\begin{eqnarray}
	R+K_{ab}K^{ab}-K^2 = 0 \;, \label{scalar_eq_LK} 
\end{eqnarray}
which is in fact just $\t{G}_{ab}n^an^b=0$ (see Appendix \ref{decom}).

The electromagnetic field equation is obtained by variation of the action with respect to $N_a$, which is contained in $M_{ab}$ and $\Psi_{ab}$. By equations (\ref{Mab}) and (\ref{M}), we get
\begin{eqnarray}
	\delta M_{ab}=\nabla_a\delta N_b+\nabla_b\delta N_a \;, \hspace{1cm}
        \delta M=2\nabla^c\delta N_c \;.
\end{eqnarray}
Then, by equation (\ref{Psi_ab}), we get
\begin{eqnarray}
	\delta\Psi_{ab}=\frac{1}{2}N^{-1}\left(\nabla_a\delta N_b+\nabla_b\delta N_a-2g_{ab}\nabla^c\delta N_c\right) \;,
\end{eqnarray}
and
\begin{eqnarray}
	\delta\Psi=g^{ab}\delta\Psi_{ab}=-(n-1)N^{-1}\nabla^c\delta N_c \;.
\end{eqnarray}
Hence,
\begin{eqnarray}
	\delta \left(N\Psi_{ab}\Psi^{ab}\right) = 2\Psi^{ab}\nabla_a\delta N_b-2\Psi\nabla^c\delta N_c \;, \hspace{1cm} \delta\left(\frac{N}{n-1}\Psi^2\right) =-2\Psi\nabla^c\delta N_c \;,
\end{eqnarray}
and
\begin{eqnarray}
	\delta \left(N\Psi_{ab}\Phi^{ab}\right) = \Phi^{ab}\nabla_a\delta N_b-\Phi\nabla^c\delta N_c \;, \hspace{1cm} \delta\left(\frac{N}{n-1}\Psi\Phi\right) =-\Phi\nabla^c\delta N_c \;.
\end{eqnarray}

Therefore, we get
\begin{eqnarray}
	\delta{\cal L}_\emm = -2\sqrt{-g}\Psi^{ab}\nabla_a\delta N_b = \sqrt{-g}\nabla_a[...]^a+2\sqrt{-g}\left(\nabla_a\Psi^{ab}\right)\delta N_b \;, \label{dlem}
\end{eqnarray}
and
\begin{eqnarray}
	\delta{\cal L}_\intt = -2\sqrt{-g}\Phi^{ab}\nabla_a\delta N_b = \sqrt{-g}\nabla_a[...]^a+2\sqrt{-g}\left(\nabla_a\Phi^{ab}\right)\delta N_b \;. \label{dlint}
\end{eqnarray}
In equations (\ref{dlem}) and (\ref{dlint}), the terms in $[...]$ are not written out since $\nabla_a[...]^a$ does not contribute to the action integral. Since ${\cal L}_G$ and ${\cal L}_m$ do not contain $N_a$, we have $\delta{\cal L}_G=\delta{\cal L}_m=0$. Then, in terms of $\Psi_{ab}$ and $\Phi_{ab}$, $\delta\t{\cal L}_G$ can be written as
\begin{eqnarray}
	\delta\t{\cal L}_G \doteq 2\sqrt{-g}\left(\nabla_a\Pi^{ab}\right)\delta N_b\;,
\end{eqnarray}
where $\doteq$ means ``equal up to a divergence term that does not contribute to the action integral''.

Hence, $\delta\t{S}_G/\delta N_a=0$ leads to a vector equation
\begin{eqnarray}
	\nabla_a\Pi^{ab}=0 \;, \label{dPi_ab_0}
\end{eqnarray}
or, equivalently,
\begin{eqnarray}
	\nabla_a\Psi^{ab}=-\nabla_a\Phi^{ab} \;. \label{vector_eq_L}
\end{eqnarray}
Since $\Psi_{ab}$ is interpreted as the electromagnetic field, the right-hand side of equation (\ref{vector_eq_L}) can be interpreted as the charge current density vector. If we define a current vector $J^b$ by
\begin{eqnarray}
        J^b\equiv\frac{1}{4\pi}\nabla_a\Phi^{ab} \;, \label{cal_J}
\end{eqnarray}
the vector equation (\ref{vector_eq_L}) can be written as
\begin{eqnarray}
	\nabla_a\Psi^{ab}=-4\pi J^b \;. \label{emfield_eq}
\end{eqnarray}
This equation will be interpreted as the electromagnetic field equation according to the reasons given bellow.

According to equation (\ref{N_A}), when $\nabla_aN=0$ is satisfied we have $M_{ab}=2Nl_\P^{n/2-1} H_{ab}$ and $M=2Nl_\P^{n/2-1} H$, where $H_{ab}$ is defined by equation (\ref{H_A}). Hence, when $\nabla_aN=0$, by equation (\ref{Psi_ab}) we have
\begin{eqnarray}
	\Psi_{ab}=l_\P^{n/2-1} \Theta_{ab} \;,
\end{eqnarray}
where $\Theta_{ab}$ is defined by equation (\ref{Theta_ab}). Therefore, if we adopt Planck units by setting $l_\P=1$, $\Psi_{ab}$ is identical to $\Theta_{ab}$, and equation (\ref{emfield_eq}) is identical to the electromagnetic field equation (\ref{meq4t}), which is equivalent to the equation (\ref{meq5}).

By equation (\ref{Pi_ab_K}), equation (\ref{vector_eq_L}) is equivalent to
\begin{eqnarray}
	\nabla_aK^{ab}-\nabla^bK= 0 \;, \label{emfield_eqK}
\end{eqnarray}
which is in fact just $g^{bc}\t{R}_{cd}n^d=0$ (equivalent to $g^{bc}\t{G}_{cd}n^d=0$; see Appendix \ref{decom}).

\section{Derivation of the Gravitational Field Equation}
\label{field_eqs_II}

The gravitational field equation on ${\cal M}$ are obtained by variation of $\t{S}_G$ with respect to $g^{ab}$. For variation with respect to $g^{ab}$, $N$ and $N_a$ are treated as invariant quantities, but
\begin{eqnarray}
	\delta N^a=N_b\delta g^{ab} \;, \hspace{1cm}
        \delta g_{ab}=-g_{ac}g_{bd}\delta g^{cd} \;,  \label{dg_ab}
\end{eqnarray}
and
\begin{eqnarray}
	\delta\sqrt{-g}=-\frac{1}{2}\sqrt{-g}g_{ab}\delta g^{ab} \;. \label{dsqrtg}
\end{eqnarray}

The variation of $S_G$ with respect to $g^{ab}$ is simple to evaluate. The result is
\begin{eqnarray}
	\frac{1}{\sqrt{-g}}\frac{\delta S_G}{\delta g^{ab}} = NG_{ab} -\nabla_a\nabla_bN +g_{ab}\nabla_c\nabla^c N \;.
        \label{dSG_M}
\end{eqnarray}

We denote the results of variation of $S_\emm$, $S_m$, and $S_\intt$ with respect to $g^{ab}$ by $\kappa T_{\emm,ab}$, $\kappa T_{m,ab}$, and $\kappa T_{\intt,ab}$, respectively, where $\kappa=8\pi$ in Planck units. That is, we define
\begin{eqnarray}
	\kappa T_{\emm,ab}\equiv -\frac{1}{N\sqrt{-g}}\frac{\delta S_\emm}{\delta g^{ab}} \;, \label{dSem_M}
\end{eqnarray}
\begin{eqnarray}
        \kappa T_{m,ab}\equiv -\frac{1}{N\sqrt{-g}}\frac{\delta S_m}{\delta g^{ab}} \;, \label{dSm_M}
\end{eqnarray}
and
\begin{eqnarray}
	\kappa T_{\intt,ab}\equiv -\frac{1}{N\sqrt{-g}}\frac{\delta S_\intt}{\delta g^{ab}} \;. \label{dSint_M}
\end{eqnarray}

Then, the gravitational field field equation on ${\cal M}$ is
\begin{eqnarray}
	G_{ab} = \kappa\left(T_{\emm,ab}+T_{m,ab}+T_{\intt,ab}\right) +\frac{1}{N}\left(\nabla_a\nabla_bN -g_{ab}\nabla_c\nabla^c N\right) \;. \label{Ein_eq_n}
\end{eqnarray}
Since $\nabla_aG^{ab}=0$, the divergence of right-hand side of equation (\ref{Ein_eq_n}) vanishes. Hence, the right-hand side can be interpreted as the stress-energy tensor of matter. Then, equation (\ref{Ein_eq_n}) is just the Einstein field equation in an $n$-dimensional spacetime
\begin{eqnarray}
	G_{ab} = \kappa T_{ab} \;, \label{ein_eq_n}
\end{eqnarray}
if we define the total stress-energy tensor
\begin{eqnarray}
	T_{ab} \equiv T_{\emm,ab}+T_{m,ab}+T_{\intt,ab} +\frac{1}{\kappa N}\left(\nabla_a\nabla_bN -g_{ab}\nabla_c\nabla^c N\right) \;. \label{T_hT}
\end{eqnarray}
When $\nabla_a N=0$, the terms in the parentheses of equation (\ref{T_hT}) vanish.

\subsection{Derivation of the $T_{\emm,ab}$}

By the definition of $M_{ab}$, for variation with respect to $g^{ab}$ we have
\begin{eqnarray}
	\delta M_{ab}=-2N_c\delta\Gamma^c_{\;\;ab} \;, \hspace{1cm} \delta M = -2N_cg^{de}\delta\Gamma^c_{\;\;de}+M_{cd}\delta g^{cd} \;,
\end{eqnarray}
where $\Gamma^c_{\;\;ab}$ is the Christoffel symbol. Then, by the definition of $\Psi_{ab}$, we have
\begin{eqnarray}
	\delta \Psi_{ab} = N^{-1}\left(-N_c\delta\Gamma^c_{\;\;ab}+g_{ab}N_cg^{de}\delta\Gamma^c_{\;\;de}\right) -\frac{1}{2}N^{-1}\left(g_{ab}M_{cd}-g_{ac}g_{bd}M\right)\delta g^{cd} \;, \label{dPsi_ab}
\end{eqnarray}
and
\begin{eqnarray}
	\delta \Psi = N^{-1}(n-1)\left(N_cg^{de}\delta\Gamma^c_{\;\;de}-\frac{1}{2}M_{cd}\delta g^{cd}\right) \;. \label{dPsi}
\end{eqnarray}

From equations (\ref{dPsi_ab}) and (\ref{dPsi}) we get
\begin{eqnarray}
	\delta\left(\Psi_{ab}\Psi^{ab}-\frac{1}{n-1}\Psi^2\right) =-2N^{-1}N_c\Psi^{ab}\delta\Gamma^c_{\;\;ab} +2\Psi_{ac}\Psi_b^{\;\;c}\delta g^{ab}+N^{-1}M\Psi_{ab}\delta g^{ab} \;. 
\end{eqnarray}
The variation of $\Gamma^c_{\;\;ab}$ is given by \cite{wal84}
\begin{eqnarray}
	\delta\Gamma^c_{\;\;ab}=\frac{1}{2}g^{cd}\left(\nabla_a\delta g_{bd}+\nabla_b\delta g_{ad}-\nabla_d\delta g_{ab}\right) \;. \label{del_Gamma}
\end{eqnarray}
Then, by integration by parts and substitution of equation (\ref{Psi_ab}) for $M$, we get
\begin{eqnarray}
	\delta \left[N\left(\Psi_{ab}\Psi^{ab}-\frac{1}{n-1}\Psi^2\right)\right] \doteq -\nabla^c\left(N_a\Psi_{bc}+N_b\Psi_{ac}-N_c\Psi_{ab}\right)\delta g^{ab} +2N\left(\Psi_{ac}\Psi_b^{\;\;c}-\frac{1}{n-1}\Psi\Psi_{ab}\right)\delta g^{ab} \;. 
\end{eqnarray}

Hence, by equations (\ref{dsqrtg}), (\ref{dSem_M}) and the expression for ${\cal L}_\emm$ in equation (\ref{Lem_M2}), we get the stress-energy tensor of electromagnetic fields
\begin{eqnarray}
	T_{\emm,ab} = \frac{2}{\kappa}\left(\Psi_{ac}\Psi_b^{\;\;c}-\frac{1}{n-1}\Psi\Psi_{ab}\right) -\frac{1}{2\kappa}\left(\Psi_{cd}\Psi^{cd}-\frac{1}{n-1}\Psi^2\right)g_{ab} -\frac{1}{\kappa N}\nabla^c\left(2N_{(a}\Psi_{b)c}-N_c\Psi_{ab}\right) \;, \label{Tab_em_M}
\end{eqnarray}
which agrees with equation (\ref{Tem_ab}) when $\nabla_aN=0$.
The trace of $T_{\emm,ab}$ is
\begin{eqnarray}
	T_\emm = \frac{2}{\kappa}\left(1-\frac{1}{4}n\right)\left(\Psi_{cd}\Psi^{cd}-\frac{1}{n-1}\Psi^2\right) -\frac{1}{\kappa N}\nabla^c\left(2N^a\Psi_{ac}-N_c\Psi\right) \;. \label{T_em_M}
\end{eqnarray}
When $n=4$, we have
\begin{eqnarray}
	T_\emm = -\frac{1}{\kappa N}\nabla^c\left(2N^a\Psi_{ac}-N_c\Psi\right) \;. \label{T_em_M4}
\end{eqnarray}

\subsection{Derivation of the $T_{m,ab}$}

By the definition of $\Phi_{ab}$, we have
\begin{eqnarray}
	\delta\Phi^{ab}= -\frac{1}{2}N^{-1}\left(g^{ac}g^{bd}-g^{ab}g^{cd}\right)\delta\dot{g}_{cd}-\frac{1}{2}N^{-1}\dot{g}_{cd} \left(g^{ac}\delta g^{bd}+g^{bd}\delta g^{ac}-g^{ab}\delta g^{cd}-g^{cd}\delta g^{ab}\right) \;. \label{dPhi_ab}
\end{eqnarray}
Hence,
\begin{eqnarray}
	\delta\left(\Phi_{ab}\Phi^{ab}\right) = -N^{-1}\left(\Phi^{ab}-\Phi g^{ab}\right)\delta\dot{g}_{ab}-2\Phi_{ac}\Phi_b^{\;\;c}\delta g^{ab} -N^{-1}\Phi_{ab}\dot{g}_{cd}\left(2g^{ac}\delta g^{bd}-g^{ab}\delta g^{cd}-g^{cd}\delta g^{ab}\right) \;. \label{dPhiPhix}
\end{eqnarray}

By equation (\ref{Phi_ab}) we have
\begin{eqnarray}
	\dot{g}_{ab}=-2N\left(\Phi_{ab}-\frac{1}{n-1}\Phi g_{ab}\right) \;. \label{dotg_Phi}
\end{eqnarray}
Substituting equation (\ref{dotg_Phi}) into equation (\ref{dPhiPhix}), we get
\begin{eqnarray}
	\delta\left(\Phi_{ab}\Phi^{ab}\right) = -N^{-1}\left(\Phi^{ab}-\Phi g^{ab}\right)\delta\dot{g}_{ab}+2\Phi_{ac}\Phi_b^{\;\;c}\delta g^{ab} -\frac{2}{n-1}\left(n\Phi\Phi_{ab}-\Phi^2g_{ab}\right)\delta g^{ab} \;. 
\end{eqnarray}

By equations (\ref{dotg_Phi}) and (\ref{dPhi_ab}), we get
\begin{eqnarray}
	\delta \Phi^2 = (n-1)\left(N^{-1}\Phi g^{ab}\delta\dot{g}_{ab}-2\Phi\Phi_{ab}\delta g^{ab}\right) +2\Phi^2g_{ab}\delta g^{ab} \;.
\end{eqnarray}
Hence,
\begin{eqnarray}
	\delta\left(\Phi_{ab}\Phi^{ab}-\frac{1}{n-1}\Phi^2\right) =-N^{-1}\Phi^{ab}\delta\dot{g}_{ab} +2\left(\Phi_{ac}\Phi_b^{\;\;c}-\frac{1}{n-1}\Phi\Phi_{ab}\right)\delta g^{ab} \;. \label{dPhiPhiab}
\end{eqnarray}

Since
\begin{eqnarray}
	\delta\dot{g}_{ab}=\delta\frac{\partial}{\partial w}g_{ab}=\frac{\partial}{\partial w}\delta g_{ab} \;, \label{ddotg}
\end{eqnarray}
by equations (\ref{Lm_M2}) and (\ref{dPhiPhiab}) we get
\begin{eqnarray}
	\delta{\cal L}_m \doteq -\frac{\partial}{\partial w}\left(\sqrt{-g}\Phi^{ab}\right)\delta g_{ab} -2\sqrt{-g}N\left(\Phi_{ac}\Phi_b^{\;\;c}-\frac{1}{n-1}\Phi\Phi_{ab}\right)\delta g^{ab} -N\left(\Phi_{ab}\Phi^{ab}-\frac{1}{n-1}\Phi^2\right)\delta\sqrt{-g} \;.
\end{eqnarray}
Then, by equations (\ref{dg_ab}) and (\ref{dsqrtg}), we get
\begin{eqnarray}
	\delta{\cal L}_m &\doteq& g_{ac}g_{bd}\frac{\partial}{\partial w}\left(\sqrt{-g}\Phi^{cd}\right)\delta g^{ab} -2\sqrt{-g}N\left(\Phi_{ac}\Phi_b^{\;\;c}-\frac{1}{n-1}\Phi\Phi_{ab}\right)\delta g^{ab} \nonumber\\
        &&+\frac{1}{2}\sqrt{-g}N\left(\Phi_{cd}\Phi^{cd}-\frac{1}{n-1}\Phi^2\right)g_{ab}\delta g^{ab} \;.  
\end{eqnarray}

Then, by equation (\ref{dSm_M}), we get the stress-energy tensor of the matter field
\begin{eqnarray}
	T_{m,ab} = -g_{ac}g_{bd}\frac{1}{\kappa N\sqrt{-g}}\frac{\partial}{\partial w}\left(\sqrt{-g}\Phi^{cd}\right) +\frac{2}{\kappa}\left(\Phi_{ac}\Phi_b^{\;\;c}-\frac{1}{n-1}\Phi\Phi_{ab}\right) -\frac{1}{2\kappa}\left(\Phi_{cd}\Phi^{cd}-\frac{1}{n-1}\Phi^2\right)g_{ab} \;. \label{Tab_m_M}
\end{eqnarray}
The trace of $T_{m,ab}$ is
\begin{eqnarray}
	T_m = -g_{cd}\frac{1}{\kappa N\sqrt{-g}}\frac{\partial}{\partial w}\left(\sqrt{-g}\Phi^{cd}\right) +\frac{2}{\kappa}\left(1-\frac{1}{4}n\right)\left(\Phi_{cd}\Phi^{cd}-\frac{1}{n-1}\Phi^2\right) \;. \label{T_m_M}
\end{eqnarray}
When $n=4$, we have
\begin{eqnarray}
	T_m = -g_{cd}\frac{1}{\kappa N\sqrt{-g}}\frac{\partial}{\partial w}\left(\sqrt{-g}\Phi^{cd}\right) \;. \label{T_m_M4}
\end{eqnarray}

\subsection{Derivation of the $T_{\intt,ab}$}

By equations (\ref{dPhi_ab}) and (\ref{dotg_Phi}), we get
\begin{eqnarray}
	\Psi_{ab}\delta\Phi^{ab}=-\frac{1}{2}N^{-1}\left(\Psi^{ab}-\Psi g^{ab}\right)\delta\dot{g}_{ab} -\Psi \Phi_{ab}\delta g^{ab} +\left[2\Phi_{c(a}\Psi_{b)}^{\;\;c}-\frac{1}{n-1}\Phi\left(\Psi_{ab}-\Psi g_{ab}\right)\right]\delta g^{ab} \;,
\end{eqnarray}
and
\begin{eqnarray}
	\Psi\delta\Phi = \frac{1}{2}(n-1)\left(N^{-1}\Psi g^{ab}\delta\dot{g}_{ab}-2\Psi\Phi_{ab}\delta g^{ab}\right) +\Phi\Psi g_{ab}\delta g^{ab} \;.
\end{eqnarray}
Hence,
\begin{eqnarray}
	\Psi_{ab}\delta\Phi^{ab}-\frac{1}{n-1}\Psi\delta\Phi=-\frac{1}{2}N^{-1}\Psi^{ab}\delta\dot{g}_{ab} +\left[2\Phi_{c(a}\Psi_{b)}^{\;\;c}-\frac{1}{n-1}\Phi\Psi_{ab}\right]\delta g^{ab} \;. \label{dPsiPhi1}
\end{eqnarray}

By equations (\ref{dPsi_ab}) and (\ref{dPsi}), we get
\begin{eqnarray}
	\Phi^{ab}\delta \Psi_{ab}-\frac{1}{n-1}\Phi\delta\Psi= -N^{-1}N_c\Phi^{ab}\delta\Gamma^c_{\;\;ab} +\frac{1}{2}N^{-1}M\Phi_{ab}\delta g^{ab} \;. \label{dPsiPhi2}
\end{eqnarray}
Combination of equations (\ref{dPsiPhi1}) and (\ref{dPsiPhi2}) leads to
\begin{eqnarray}
	\delta\left(\Psi_{ab}\Phi^{ab}-\frac{1}{n-1}\Psi\Phi\right)= \left[2\Phi_{c(a}\Psi_{b)}^{\;\;c}-\frac{1}{n-1}(\Phi\Psi_{ab}+\Psi\Phi_{ab})\right]\delta g^{ab} -\frac{1}{2}N^{-1}\Psi^{ab}\delta\dot{g}_{ab}-N^{-1}N_c\Phi^{ab}\delta\Gamma^c_{\;\;ab} \;,
\end{eqnarray}
where equation (\ref{Psi_ab}) has been used to substitute for $M$.

Then, by equation (\ref{Lint_M2}), we have
\begin{eqnarray}
	\delta{\cal L}_\intt &=& \sqrt{-g}\Psi^{ab}\delta\dot{g}_{ab}+2\sqrt{-g}N_c\Phi^{ab}\delta\Gamma^c_{\;\;ab} -2\sqrt{-g}N\left[2\Phi_{c(a}\Psi_{b)}^{\;\;c}-\frac{1}{n-1}(\Phi\Psi_{ab}+\Psi\Phi_{ab})\right]\delta g^{ab} \nonumber\\
        &&-2N\left(\Psi_{cd}\Phi^{cd}-\frac{1}{n-1}\Psi\Phi\right)\delta\sqrt{-g} \;.
\end{eqnarray}
By equations (\ref{dg_ab}) and (\ref{ddotg}), we get
\begin{eqnarray}
	\sqrt{-g}\Psi^{ab}\delta\dot{g}_{ab} = \frac{\partial}{\partial w}\left(\sqrt{-g}\Psi^{ab}\delta g_{ab}\right)-\frac{\partial}{\partial w}\left(\sqrt{-g}\Psi^{ab}\right)\delta g_{ab} \doteq g_{ac}g_{bd}\frac{\partial}{\partial w}\left(\sqrt{-g}\Psi^{cd}\right)\delta g^{ab} \;.
\end{eqnarray}
By equation (\ref{del_Gamma}), we get
\begin{eqnarray}
	2\sqrt{-g}N_c\Phi^{ab}\delta\Gamma^c_{\;\;ab} \doteq \sqrt{-g}\nabla^c\left[2N_{(a}\Phi_{b)c}-N_c\Phi_{ab}\right]\delta g^{ab} \;. 
\end{eqnarray}
Hence, we have
\begin{eqnarray}
	\delta{\cal L}_\intt &\doteq& g_{ac}g_{bd}\frac{\partial}{\partial w}\left(\sqrt{-g}\Psi^{cd}\right)\delta g^{ab}+ \sqrt{-g}\nabla^c\left[2N_{(a}\Phi_{b)c}-N_c\Phi_{ab}\right]\delta g^{ab} \nonumber\\
        &&-2\sqrt{-g}N\left[2\Phi_{c(a}\Psi_{b)}^{\;\;c}-\frac{1}{n-1}(\Phi\Psi_{ab}+\Psi\Phi_{ab})\right]\delta g^{ab} +\sqrt{-g}N\left(\Psi_{cd}\Phi^{cd}-\frac{1}{n-1}\Psi\Phi\right)g_{ab}\delta g^{ab} \;.
\end{eqnarray}
where equation (\ref{dsqrtg}) has been used.

Finally, by equation (\ref{dSint_M}), we get the stress-energy tensor of the interaction
\begin{eqnarray}
	T_{\intt,ab} &=& -g_{ac}g_{bd}\frac{1}{\kappa N\sqrt{-g}}\frac{\partial}{\partial w}\left(\sqrt{-g}\Psi^{cd}\right) +\frac{2}{\kappa}\left[2\Phi_{c(a}\Psi_{b)}^{\;\;c}-\frac{1}{n-1}\left(\Phi\Psi_{ab}+\Psi\Phi_{ab}\right)\right] \nonumber\\
        &&-\frac{1}{\kappa}\left(\Phi_{cd}\Psi^{cd}-\frac{1}{n-1}\Phi\Psi\right)g_{ab} -\frac{1}{\kappa N}\nabla^c\left[2N_{(a}\Phi_{b)c}-N_c\Phi_{ab}\right] \;. \label{Tab_int_M}
\end{eqnarray}
The traces of $T_{\intt,ab}$ is
\begin{eqnarray}
	T_\intt = -g_{cd}\frac{1}{\kappa N\sqrt{-g}}\frac{\partial}{\partial w}\left(\sqrt{-g}\Psi^{cd}\right) +\frac{4}{\kappa}\left(1-\frac{1}{4}n\right)\left(\Phi_{cd}\Psi^{cd}-\frac{1}{n-1}\Phi\Psi\right) -\frac{1}{\kappa N}\nabla^c\left(2N^a\Phi_{ac}-N_c\Phi\right) \;. \label{T_int_M}
\end{eqnarray}
When $n=4$, we have
\begin{eqnarray}
	T_\intt = -g_{cd}\frac{1}{\kappa N\sqrt{-g}}\frac{\partial}{\partial w}\left(\sqrt{-g}\Psi^{cd}\right) -\frac{1}{\kappa N}\nabla^c\left(2N^a\Phi_{ac}-N_c\Phi\right) \;. \label{T_int_M4}
\end{eqnarray}

\subsection{The total stress-energy tensor}

Substituting equations (\ref{Tab_em_M}), (\ref{Tab_m_M}), and (\ref{Tab_int_M}) into equation (\ref{T_hT}), we get the total stress-energy tensor
\begin{eqnarray}
	T_{ab}&=&-g_{ac}g_{bd}\frac{1}{\kappa N\sqrt{-g}}\frac{\partial}{\partial w}\left(\sqrt{-g}\Pi^{cd}\right) +\frac{2}{\kappa}\left(\Pi_{ac}\Pi^c_{\;\;b}-\frac{1}{n-1}\Pi\Pi_{ab}\right) -\frac{1}{2\kappa}\left(\Pi_{cd}\Pi^{cd}-\frac{1}{n-1}\Pi^2\right)g_{ab} \nonumber\\ 
        &&+\frac{1}{\kappa N}\left(\nabla_a\nabla_bN -g_{ab}\nabla_c\nabla^c N\right) -\frac{1}{\kappa N}\nabla^c\left(2N_{(a}\Pi_{b)c}-N_c\Pi_{ab}\right) \;. \label{Tm+Tint+Tem}
\end{eqnarray}
The trace of $T_{ab}$ is
\begin{eqnarray}
	T &=& -g_{cd}\frac{1}{\kappa N\sqrt{-g}}\frac{\partial}{\partial w}\left(\sqrt{-g}\Pi^{cd}\right)-\frac{n-1}{\kappa N}\nabla_c\nabla^c N +\frac{2}{\kappa}\left(1-\frac{1}{4}n\right)\left(\Pi_{cd}\Pi^{cd}-\frac{1}{n-1}\Pi^2\right) \nonumber\\
        && -\frac{1}{\kappa N}\nabla^c\left(2N^a\Pi_{ac}-N_c\Pi\right) \;. \label{T_em+T_m+T_int}
\end{eqnarray}

By the identities
\begin{eqnarray}
	\frac{1}{\sqrt{-g}}\frac{\partial}{\partial w}\sqrt{-g}=\frac{1}{2}g^{ab}\dot{g}_{ab}=\frac{1}{n-1}N\Phi \label{dw_sqrtg}
\end{eqnarray}
and
\begin{eqnarray}
	\frac{\partial}{\partial w}\Pi^{ab} &=& Ng^a_{\;\;c}g^b_{\;\;d}\t{\pounds}_n\Pi^{cd} +N^c\nabla_c\Pi^{ab}-2\Pi^{c(a}\nabla_cN^{b)} \;, \label{pu_Piab}
\end{eqnarray}
we get
\begin{eqnarray}
	\kappa T_{ab}&=&-g_{ac}g_{bd}\t{\pounds}_n\Pi^{cd} +2\Pi_{ac}\Pi^c_{\;\;b}-\frac{3}{n-1}\Pi\Pi_{ab} -\frac{1}{2}\left(\Pi_{cd}\Pi^{cd}-\frac{1}{n-1}\Pi^2\right)g_{ab}-\frac{2}{N}N_{(a}\nabla^c\Pi_{b)c} \nonumber\\ 
        &&+\frac{1}{N}\left(\nabla_a\nabla_bN -g_{ab}\nabla_c\nabla^c N\right) \;. \label{Tm+Tint+Tem2}
\end{eqnarray}

By the vector field equation (\ref{dPi_ab_0}), the term $N_{(a}\nabla^c\Pi_{b)c}$ in equation (\ref{Tm+Tint+Tem2}) vanishes. Hence we have
\begin{eqnarray}
	\kappa T_{ab}&=&-g_{ac}g_{bd}\t{\pounds}_n\Pi^{cd} +\left(2\Pi_{ac}\Pi^c_{\;\;b}-\frac{3}{n-1}\Pi\Pi_{ab}\right) -\frac{1}{2}\left(\Pi_{cd}\Pi^{cd}-\frac{1}{n-1}\Pi^2\right)g_{ab} \nonumber\\ 
        &&+\frac{1}{N}\left(\nabla_a\nabla_bN -g_{ab}\nabla_c\nabla^c N\right) \;. \label{Tm+Tint+Tem2a}
\end{eqnarray}

Substituting equation (\ref{Pi_ab_K}) into equation (\ref{Tm+Tint+Tem2a}) and making use of the relation
\begin{eqnarray}
	g_{ac}g_{bd}\t{\pounds}_n\Pi^{cd} = -\left(g_a^{\;\;c}g_b^{\;\;d}-g_{ab}g^{cd}\right)\t{\pounds}_nK_{cd}+4K_{ac}K_b^{\;\;c} -2KK_{ab}-2K_{cd}K^{cd}g_{ab} \;, \label{Ln_Piab}
\end{eqnarray}
we get
\begin{eqnarray}
	\kappa T_{ab}&=&-2K_{ac}K_b^{\;\;c}+KK_{ab}+\frac{1}{2}\left(3K_{cd}K^{cd}-K^2\right)g_{ab} +\left(g_a^{\;\;c}g_b^{\;\;d}-g_{ab}g^{cd}\right)\t{\pounds}_nK_{cd} \nonumber\\
        &&+\frac{1}{N}\left(\nabla_a\nabla_bN -g_{ab}\nabla_c\nabla^c N\right) \;. \label{Tm+Tint+Tem3}
\end{eqnarray}

By equation (\ref{a_dN}), we have
\begin{eqnarray}
	\nabla_aa_b-a_aa_b=-\frac{1}{N}\nabla_a\nabla_bN \;,
\end{eqnarray}
and
\begin{eqnarray}
	\nabla_ca^c-a_ca^c=-\frac{1}{N}\nabla_c\nabla^cN \;.
\end{eqnarray}
Hence, equation (\ref{Tm+Tint+Tem3}) is equivalent to
\begin{eqnarray}
	\kappa T_{ab}&=&-2K_{ac}K_b^{\;\;c}+KK_{ab}+\frac{1}{2}\left(3K_{cd}K^{cd}-K^2\right)g_{ab} +\left(g_a^{\;\;c}g_b^{\;\;d}-g_{ab}g^{cd}\right)\t{\pounds}_nK_{cd} \nonumber\\
        &&-\nabla_aa_b+a_aa_b+\left(\nabla_ca^c-a_ca^c\right)g_{ab} \;. \label{Tm+Tint+Tem4}
\end{eqnarray}
The $\kappa T_{ab}$ in equation (\ref{Tm+Tint+Tem4}) agrees with the right-hand side of equation (\ref{tensor_eq}) in Appendix \ref{decom}. Hence, the $n$-dimensional Einstein field equation (\ref{ein_eq_n}) derived from the Lagrangian formulation is equal to the full projection of the $(n+1)$-dimensional vacuum Einstein field equation onto ${\cal M}$, up to addition of a term proportional to $g_b^{\;\;c}\t{R}_{cd}n^d$.

Equation (\ref{Tm+Tint+Tem4}) can be further simplified. By equation (\ref{Tm+Tint+Tem4}), the trace of $\kappa T_{ab}$ is
\begin{eqnarray}
	\kappa T = \left(\frac{3}{2}n-2\right)K_{cd}K^{cd}-\left(\frac{n}{2}-1\right)K^2 -(n-1)\left(g^{cd}\t{\pounds}_nK_{cd}-\nabla_ca^c+a_ca^c\right) \;. \label{TT}
\end{eqnarray}
The trace of equation (\ref{ein_eq_n}) gives rise to
\begin{eqnarray}
	\kappa T=-\left(\frac{n}{2}-1\right)R \;. \label{TR}
\end{eqnarray}
Substituting equation (\ref{scalar_eq_LK}) into equation (\ref{TR}), we get
\begin{eqnarray}
	\kappa T=\left(\frac{n}{2}-1\right)\left(K_{cd}K^{cd}-K^2\right) \;. \label{TR2}
\end{eqnarray}
Then, eliminating $\kappa T$ from equations (\ref{TT}) and (\ref{TR2}), we get
\begin{eqnarray}
	g^{cd}\t{\pounds}_nK_{cd}-K_{cd}K^{cd}-\nabla_ca^c+a_ca^c=0 \;. \label{scalar_eq_LKb}
\end{eqnarray}
Substituting equation (\ref{scalar_eq_LKb}) into equation (\ref{Tm+Tint+Tem4}), we get
\begin{eqnarray}
	\kappa T_{ab}=-2K_{ac}K_b^{\;\;c}+KK_{ab}+\frac{1}{2}\left(K_{cd}K^{cd}-K^2\right)g_{ab} +g_a^{\;\;c}g_b^{\;\;d}\t{\pounds}_nK_{cd}-\nabla_aa_b+a_aa_b\;. \label{Tm+Tint+Tem5s}
\end{eqnarray}

Equation (\ref{scalar_eq_LKb}) is another scalar constraint equation on ${\cal M}$, which can be used to replace equation (\ref{scalar_eq_LK}). In fact, with the $T_{ab}$ given by equation (\ref{Tm+Tint+Tem5s}), the $n$-dimensional Einstein field equation (\ref{ein_eq_n}) implies equation (\ref{scalar_eq_LK}). By equation (\ref{Rnn_K}), the scalar equation (\ref{scalar_eq_LKb}) corresponds to $\t{R}_{ab}n^an^b=0$.

It can be checked that the scalar equation (\ref{scalar_eq_LKb}) is equivalent to
\begin{eqnarray}
	g_{cd}\frac{1}{\sqrt{-g}}\frac{\partial}{\partial w}\left(\sqrt{-g}\Pi^{cd}\right) =(3-n)N\left(\Pi_{cd}\Pi^{cd}-\frac{1}{n-1}\Pi^2\right)-(n-1)\nabla_c\nabla^c N -\nabla^c\left(2N^a\Pi_{ac}-N_c\Pi\right) \;;  \label{scalar_eq_La}
\end{eqnarray}
i.e.,
\begin{eqnarray}
	g_{cd}\t{\cal L}_n\Pi^{cd} = -(n-3)\Pi_{cd}\Pi^{cd}+\frac{n-4}{n-1}\Pi^2 -(n-1)\frac{1}{N}\nabla_c\nabla^c N \;,  \label{scalar_eq_Lb}
\end{eqnarray}
after substitution of equation (\ref{dPi_ab_0}).

With the supplement of the scalar constraint equation (\ref{scalar_eq_Lb}) (or, equivalently, eq.~\ref{scalar_eq_LKb}), the electromagnetic field equation (\ref{emfield_eq}) (or, equivalently, eq~\ref{emfield_eqK}) and the Einstein field equation (\ref{ein_eq_n}) with the $T_{ab}$ given by equation (\ref{Tm+Tint+Tem2a}) (or, equivalently, eq~\ref{Tm+Tint+Tem5s}) form a complete system of field equations on ${\cal M}$.

\section{Relation to the Kaluza-Klein Theory}
\label{rel_kk}

\begin{figure}
\vspace{3pt}
\begin{center}\includegraphics[angle=0,scale=0.9]{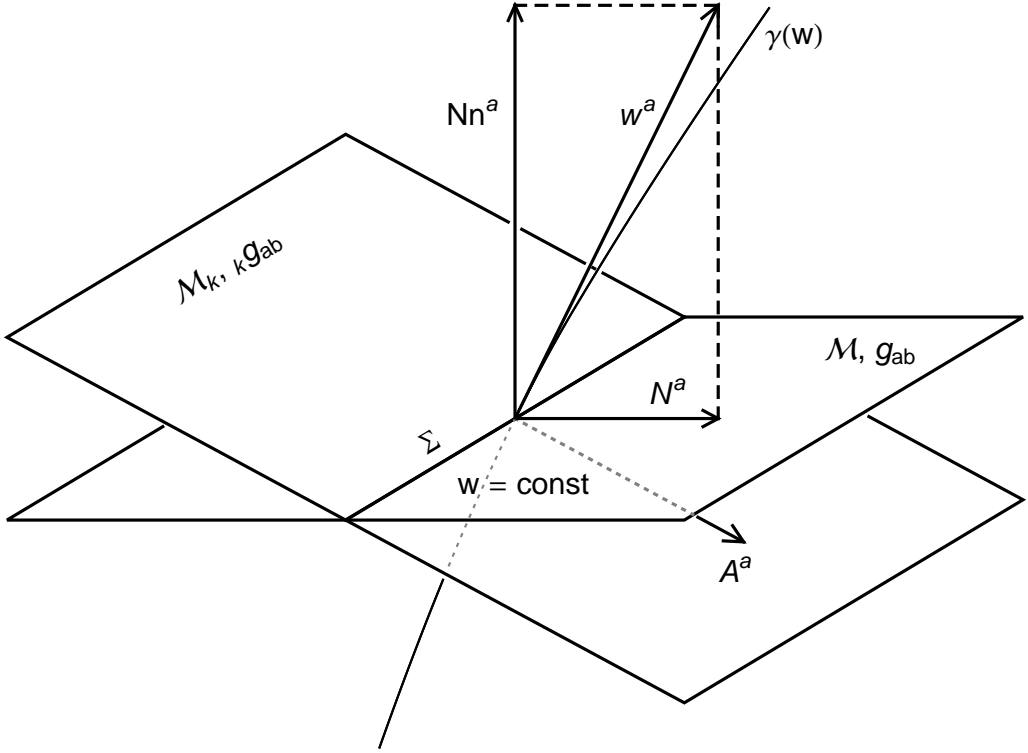}\end{center}
\caption{Geometric relation between the Kaluza-Klein theory and the new unified theory presented in this paper. In coordinate system $\{x^0,x^1,x^2,x^3,w\}$ in a five-dimensional spacetime, a hypersurface ${\cal M}$ is defined by $w=\mbox{const}$. If $w$-coordinate lines, $\gamma(w)$, are hypersurface orthogonal, there exists a hypersurface ${\cal M}_k$ orthogonal to the vector $w^a$ tangent to $\gamma(w)$. The ${\cal M}_k$ and ${\cal M}$ intersect at a three-dimensional manifold $\Sigma$. Projection of the five-dimensional metric tensor $\t{g}_{ab}$ onto ${\cal M}$ gives rise to a four-dimensional metric tensor $g_{ab}$ on ${\cal M}$. Projection of $\t{g}_{ab}$ onto ${\cal M}_k$ gives rise to a four-dimensional metric tensor $\!\!\!~_kg_{ab}$ on ${\cal M}_k$. Projection of $w^a$ onto ${\cal M}$ gives rise to a four-dimensional vector field $N^a$ on ${\cal M}$. Projection of $-N^{-1}n^a$ (normal to ${\cal M}$) onto ${\cal M}_k$ gives rise to a four-dimensional vector field $A^a$ on ${\cal M}_k$ (eq.~\ref{gab_n_k}). The theory presented in this paper is defined by $\left({\cal M},g_{ab},N^a,N\right)$, where $N=\t{g}_{ab}w^an^b$. The Kaluza-Klein theory is defined by $\left({\cal M}_k,\!\!~_kg_{ab},A^a,\phi\right)$, where $\phi^2=\t{g}_{ab}w^aw^b$. As discussed in Sec.~\ref{rel_kk}, the two theories are not related by diffeomorphisms hence they represent different physics in a four-dimensional spacetime.}
\label{kk2}
\end{figure}

Since both the theory presented in this paper and the Kaluza-Klein theory are constructed from the Einstein theory of gravity in a five-dimensional bulk spacetime, and in both theories electromagnetic field equations are derived from the five-dimensional Einstein field equation, the relation and difference between the two theories should be clarified. 

The geometric interpretation of the Kaluza-Klein decomposition of a five-dimensional metric is described in Appendix~\ref{KK}. It is shown that the metric tensor constructed from $\!\!\!~_kg_{\mu\nu}$, i.e., $\!\!\!~_kg_{ab}=\!\!~_kg_{\mu\nu} dx^\mu_adx^\nu_b$ where $\mu,\nu=0,1,2,3$, is a tensor $\in{\cal T}({\cal M}_k)$, where ${\cal M}_k$ is a hypersurface orthogonal to the integral curves of $w^a=(\partial/\partial w)^a$. Similarly, the $A_a=A_\mu dx^\mu_a$ is a vector $\in{\cal T}({\cal M}_k)$. Since when $N^a\neq 0$ the hypersurface ${\cal M}$ defined by $w=\mbox{const}$ is not orthogonal to $w^a$, ${\cal M}_k$ and ${\cal M}$ are two hypersurfaces intersecting at a three-dimensional manifold $\Sigma$ (Fig.~\ref{kk2}).

By equation (\ref{ktg_ab}), in the Kaluza-Klein representation the line element in the five-dimensional spacetime is
\begin{eqnarray}
	d\t{s}^2 = \!~_kg_{\mu\nu}dx^\mu dx^\nu+\phi^2\left(A_\mu dx^\mu+dw\right)\left(A_\nu dx^\nu+dw\right) \;, \label{ds_5d}
\end{eqnarray}
from which we derive the line element on the hypersurface ${\cal M}$ defined by $w=\mbox{const}$
\begin{eqnarray}
	ds^2 = \left(\!~_kg_{\mu\nu}+\phi^2A_\mu A_\nu\right)dx^\mu dx^\nu \;.  \label{ds_4d}
\end{eqnarray}
Hence, in the Kaluza-Klein theory, the metric tensor on ${\cal M}$ is $\!~_kg_{\mu\nu}+\phi^2A_\mu A_\nu$, not $\!~_kg_{\mu\nu}$. This indicates that the Kaluza-Klein theory is not a unified theory of the electromagnetism and gravity in the four-dimensional spacetime spanned by the coordinates $\{x^0,x^1,x^2,x^3\}$.

By equation (\ref{kgab_uu}), $\!\!~_kg_{ab}$ is a projection operator onto ${\cal M}_k$. In contrast, $g_{ab}$ is a projection operator onto ${\cal M}$. They are related by equation (\ref{g_g}). By equation (\ref{gab_n_k}), $A^a$ is related to the normal to the hypersurface ${\cal M}$, i.e., defined by the projection of $-N^{-1}n^a$ onto ${\cal M}_k$. In contrast, $N^a$ is the projection of $w^a$ onto ${\cal M}$. $A^a$ and $N^a$ are related by equation (\ref{Aa_Na}).

Since the variables $\!\!~_kg_{ab}$ and $A^a$ are defined on ${\cal M}_k$, the Kaluza-Klein theory is inherently defined on the hypersurface ${\cal M}_k$. In contrast, the theory presented in this paper is defined on ${\cal M}$. The existence of ${\cal M}_k$ orthogonal to the vector field $w^a$ requires that $w^a$ is hypersurface orthogonal, i.e., $w^a$ must satisfy the condition $w_{[a}\t{\nabla}_bw_{c]}=0$ \cite{wal84}. Hence, the new unified theory described in this paper is more general than the Kaluza-Klein theory, since it does not require that $w^a$ is hypersurface orthogonal.

The unit normal to ${\cal M}$ is the $n^a$ defined by equation (\ref{w^a}), i.e.,
\begin{eqnarray}
        n^a=\frac{1}{N}(w^a-N^a) \;. \label{na_wa}
\end{eqnarray}

By equation (\ref{phi}), the unit normal to ${\cal M}_k$ is
\begin{eqnarray}
        \!\!\!~_kn^a=\hat{w}^a=\frac{1}{\phi}w^a \;. \label{kna_wa}
\end{eqnarray}
Assume that the hypersurface ${\cal M}_k$ is defined by $w=f(x^\mu)$, i.e., by $F(x^\mu,w)=f(x^\mu)-w=0$. Then we have
\begin{eqnarray}
	\!\!\!~_kn^a\propto\t{\nabla}^aF=\t{g}^{ab}\left(\frac{\partial f}{\partial x^\mu}dx^\mu_b-dw_b\right)=\left(\!\!\!~_kg^{\mu\nu}\frac{\partial f}{\partial x^\mu}+A^\nu\right)\left(\frac{\partial}{\partial x^\nu}\right)^a-\left(A^\mu\frac{\partial f}{\partial x^\mu}+\frac{1}{\phi^2}+A_\rho A^\rho\right)\left(\frac{\partial}{\partial w}\right)^a \;, \label{k_na}
\end{eqnarray}
where equation (\ref{ktg^ab1}) has been applied. Therefore, we get the equation defining the function $f(x^\mu)$
\begin{eqnarray}
	\frac{\partial f}{\partial x^\mu}=-A_\mu \;. \label{pf_pxm}
\end{eqnarray} 

From equations (\ref{na_wa}) and (\ref{kna_wa}) we get
\begin{eqnarray}
        \t{g}_{ab}n^a\!\!\!~_kn^b=\frac{N}{\sqrt{N^2+N_c N^c}}=\frac{1}{\sqrt{1+\phi^2A_c A^c}} \;. \label{nakna}
\end{eqnarray}
When $N^a$ and $A^a$ are timelike, we have $\t{g}_{ab}n^a\!\!\!~_kn^b>1$ and the inclination between the two hypersurfaces ${\cal M}$ and ${\cal M}_k$ can be interpreted as relative motion between them, with a relative velocity $\beta=\phi\sqrt{-A_c A^c}=N^{-1}\sqrt{-N_c N^c}$. When $N^a$ and $A^a$ are spacelike, we have $\t{g}_{ab}n^a\!\!\!~_kn^b<1$ and the inclination between ${\cal M}$ and ${\cal M}_k$ can be interpreted as relative spatial rotation, with the rotation angle $\alpha=\arccos\left(\t{g}_{ab}n^a\!\!\!~_kn^b\right)$. When $N^a$ and $A^a$ are null, we have $\t{g}_{ab}n^a\!\!\!~_kn^b=1$, both $N^a$ and $A^a$ are tangent to $\Sigma$ (Proposition 12 in Appendix~\ref{KK}). The ${\cal M}$ and ${\cal M}_k$ are still two different hypersurfaces.

The fundamental variables on ${\cal M}$ are $g_{ab}$, $N^a$, and $N$. The fundamental variables on ${\cal M}_k$ are $\!\!~_kg_{ab}$, $A^a$, and $\phi$. From the results in Appendix~\ref{KK}, we can derive the relations between the two groups of variables, which are given by
\begin{eqnarray}
        g_{ab}=\!\!\!~_kg_{ab}+\frac{1}{1+\phi^2A_c A^c}\left(A_cA^cw_aw_b+A_aw_b+A_bw_a-\phi^2A_aA_b\right) \;, \label{g_gk1}
\end{eqnarray}
and
\begin{eqnarray}
        N^a=\frac{\phi^2}{1+\phi^2A_cA^c}\left(A^a+A_cA^cw^a\right) \;, \hspace{1cm} N=\frac{\phi}{\sqrt{1+\phi^2A_cA^c}} \;. \label{g_gk2}
\end{eqnarray}
Clearly, the transformations described by equations (\ref{g_gk1}) and (\ref{g_gk2}) are not diffeomorphic transformations. Diffeomorphisms map tensors by linear transformations (see, e.g., \cite{wal84,car03}), but the transformations in (\ref{g_gk1}) and (\ref{g_gk2}) are nonlinear.

Therefore, the new theory presented in this paper is physically different from the Kaluza-Klein theory, although both aim at unifying the electromagnetic and gravitational interactions in the framework of general relativity in a five-dimensional spacetime. The two theories are defined on two different hypersurfaces in a five-dimensional spacetime, and are not related by diffeomorphisms. In general relativity, theories defined by tensor fields on manifolds are physically identical only if they are related by diffeomorphic transformations (\cite{wal84}, and next section of this paper). Since both theories are derived from the Einstein theory of gravity in a five-dimensional spacetime, mathematically they are related by the transformations in equations (\ref{g_gk1}) and (\ref{g_gk2}). However, the two theories are physically distinguishable. This is similar to the case of conformal transformations: two spacetimes related by conformal transformations are usually not identical in physics, unless the conformal factor is a constant. For example, the Robertson-Walker spacetime metric of a universe is related to the flat Minkowski spacetime metric by a conformal transformation, but the physics in a curved universe is different from that in a flat spacetime.
 
The physical difference between the two theories is manifested by the fact that in the Kaluza-Klein theory the standard Maxwell equation is derived, but in our theory the derived electromagnetic field equation contains a term coupled to the spacetime curvature. The curvature-coupled term cannot be eliminated by diffeomorphic transformations. Only in a Ricci-flat four-dimensional spacetime the two field equations are identical. In addition, the Kaluza-Klein theory contains an unidentified scalar field, and the total five physical degrees of freedom of gravitons in the five-dimensional bulk spacetime are shared by gravitons (two), photons (two), and the scalar field (one) in the four-dimensional spacetime. In our theory, the scalar field does not exist. The total five physical degrees of freedom of gravitons in the bulk spacetime are shared by gravitons (two) and photons (three) in the four-dimensional spacetime. In our theory, photons have three degrees of freedom since the curvature-coupled term in the field equation breaks the gauge symmetry and causes photons to acquire an effective mass. 

In the two theories, the derived gravitational field equations in a four-dimensional spacetime are also different, since the stress-energy tensors on the right-hand side of the field equation are different. This can be directly verified by comparison of the stress-energy tensor derived in this paper (eqs.~\ref{T_hT}, \ref{Tab_em_M}, \ref{Tab_m_M}, and \ref{Tab_int_M}) with that in the Kaluza-Klein theory \cite{bai87,ove97}.

Finally, our theory and the procedure adopted in this paper are more general than the Kaluza-Klein theory and the procedure used in it. As already mentioned above, our theory does not require that the vector field $w^a$ is hypersurface orthogonal, but the Kaluza-Klein theory does. In addition, the notation of ``normal'' or ``orthogonality'' requires the existence of a prespecified metric tensor, but in both theories the metric tensor needs to be solved from field equations. This is also the reason why in the Hamiltonian formulation of general relativity a Gaussian normal coordinate system cannot be used to simplify the problem.

\section{Diffeomorphism and Gauge Symmetry}
\label{gauge}

It is well known that the Maxwell theory of electromagnetic fields is invariant under the gauge transformation 
\begin{eqnarray}
	A_a\rightarrow A_a+\nabla_a\chi \;, \label{gauge0}
\end{eqnarray}
where $\chi$ is any scalar function. Under this gauge transformation, the antisymmetric electromagnetic field tensor $F_{ab}$ defined by equation (\ref{F_A}) is unchanged, and hence the Einstein-Maxwell equation (\ref{meq1g}) is unchanged. In a Ricci-flat spacetime, the electromagnetic field equation (\ref{meq0}) reduces to the Einstein-Maxwell equation (\ref{meq1g}) and so is invariant under the gauge transformation. However, when the spacetime is not Ricci-flat, the electromagnetic field equation (\ref{meq0}) is not invariant under the gauge transformation. So, in the theory presented in this paper, it appears that the presence of Ricci curvature leads to gauge symmetry breaking to the electromagnetic field equation.

The above observation reminds us the Proca equation, which is generalization of the Maxwell equation by introducing a photon mass term. The presence of a photon mass term makes the Proca equation not invariant under the gauge transformation. However, it is possible to restore the gauge symmetry by introducing a complex scalar field interacting with the electromagnetic field \cite{stu57,pes95}. In high energy state, the scalar field has a true vacuum at a zero value and photons remain massless. In low energy state, the scalar field has a true vacuum at a nonzero value through spontaneous symmetry breaking, and photons acquire a mass through the Higgs-like mechanism. Hence, a theory with a broken gauge symmetry can be fitted into an underlying and more fundamental theory with gauge symmetry. This is a well-known fact in quantum field theory.

For the theory presented in this paper we have a similar situation. Despite the fact that the derived electromagnetic field equation in a four-dimensional spacetime is not invariant under the gauge transformation defined by equation (\ref{gauge0}), the underlying theory---the Einstein theory of gravity in a five-dimensional bulk spacetime from which the electromagnetic field equation is derived---is invariant under the gauge transformation defined by diffeomorphisms in the bulk spacetime. The apparent breakdown of gauge invariance for the electromagnetic field equation (\ref{meq0}) derived from the Einstein field equation in the bulk spacetime originates from slicing of the bulk spacetime with timelike hypersurfaces.

One may wonder how the gauge transformation in equation (\ref{gauge0}) is related to diffeomorphisms in a spacetime. In this section, we show that the gauge transformation (\ref{gauge0}) can be interpreted as a diffeomorphic transformation in the background spacetime and hence can be regarded as arising from diffeomorphisms of spacetime. Since the electromagnetic field equation (\ref{meq0}) is invariant under any diffeomorphism, it is invariant under the gauge transformation (\ref{gauge0}) provided that the corresponding diffeomorphism is applied to all variables appearing in the field equation.  

As a geometric theory, general relativity is known to be invariant under diffeomorphic transformations. In fact, all field equations expressed in tensors (including scalar functions and vectors as special cases) defined on a manifold are invariant under diffeomorphism. Hence, diffeomorphisms comprise the gauge freedom of any theory formulated in terms of tensor fields, including general relativity itself \cite{wal84,haw73,car03}. Assume that $({\cal M}, g_{ab}, \psi)$ defines a spacetime $({\cal M}, g_{ab})$ and matter fields $\psi$ on it, where $\psi$ can be scalar, vector, and tensor fields. Under a diffeomorphism $\phi: {\cal M}\rightarrow{\cal M}$, with the map $\phi$ and its inverse $\phi^{-1}$ any tensor field $T$ is transformed to $T^\prime=\phi_*T$. Then, $({\cal M}, g_{ab}, \psi)$ is transformed to $({\cal M}, \phi_*g_{ab}, \phi_*\psi)$. According to general relativity, $({\cal M}, g_{ab}, \psi)$ and $({\cal M}, \phi_*g_{ab}, \phi_*\psi)$ belong to the same equivalent class under diffeomorphisms and represent the same physics. 

Before going to show that the gauge transformation (\ref{gauge0}) can be generated by a diffeomorphism on the manifold where electromagnetic fields are defined, we present some Lemmas which ensure that any theory defined on a manifold in terms of tensors is invariant under diffeomorphic transformations.

{\sc Lemma 1}. On a manifold ${\cal M}$, any action defined by
\begin{eqnarray}
	S=\int {\cal L}\,{\bs \epsilon} \label{action_g}
\end{eqnarray}
is invariant under diffeomorphisms, where the Lagrangian density ${\cal L}$ is a scalar function, and ${\bs \epsilon}$ is a volume element.

Lemma 1 is a direct consequence of the fact that integral of any $n$-form on an $n$-dimensional manifold is invariant under diffeomorphisms \cite{haw73}. It can also be proved as follows. Let ${\bs \epsilon}=\sqrt{-g}\,{\bf e}$, where ${\bf e}$ is a fixed volume on ${\cal M}$. Under an infinitesimal diffeomorphism generated by a vector field $v^a=(\partial/\partial\tau)^a$, we have
\begin{eqnarray}
	\delta g_{ab}=-(\pounds_vg_{ab})\delta\tau=-2\nabla_{(a}v_{b)}\delta\tau \;, \label{pgab1}
\end{eqnarray}
where $\pounds$ is the Lie derivative operator on ${\cal M}$, and $\nabla_a$ is the derivative operator associated with the metric tensor $g_{ab}$. Then, by equation (\ref{dsqrtg}) we get $\delta\sqrt{-{g}}=-\sqrt{-{g}}({\nabla}_av^a)\delta\tau$. On the other hand, we have $\delta{\cal L}= -({\pounds}_v {\cal L})\delta\tau=-(v^a{\nabla}_a{\cal L})\delta\tau$. Hence, we get
\begin{eqnarray}
	\delta\left(\sqrt{-{g}}\,{\cal L}\right) = -\sqrt{-{g}}\,{\nabla}_a({\cal L}v^a\delta\tau) \;. \label{sqLp}
\end{eqnarray}
The term $\nabla_a({\cal L}v^a\delta\tau)$ in equation (\ref{sqLp}) is a boundary term which has no contribution to the action integral by Gauss's theorem, if we set $v^a=0$ on the boundary. Hence, we get $\delta S=0$ under the variation generated by a diffeomorphism.

{\sc Lemma 2}. For any tensor field $T^{a_1...a_k}_{\;\;\;\;\;\;\;\;\;\;\;b_1...b_l}$ on ${\cal M}$, 
\begin{eqnarray}
	\phi_*\left(\nabla_cT^{a_1...a_k}_{\;\;\;\;\;\;\;\;\;\;\;b_1...b_l}\right)=\nabla_c^\prime\left(\phi_*T^{a_1...a_k}_{\;\;\;\;\;\;\;\;\;\;\;b_1...b_l}\right) \;,
\end{eqnarray}
where $\nabla_a$ is the derivative operator associated with $g_{ab}$, and $\nabla^\prime_a$ is the derivative operator associated with $\phi_*g_{ab}$ (see, e.g, \cite{lia06}).

{\em Proof}. Since $\phi_*(\nabla T)=\phi_*(\nabla(\phi^{-1}_*\phi_*T))\equiv f(\phi_*T)$, $\phi_*(\nabla T)$ is a map of $\phi_*T$. It can be checked that $f(\phi_*T)$ satisfies the definition for a derivative operator on ${\cal M}$, due to the fact that $\phi$ is a linear map preserving tensor type and the relations in the tensor algebra, and $\nabla$ is a derivative operator. For example, the Leibnitz rule can be verified as follows: $f(\phi_*T_1\bigotimes\phi_*T_2)=f(\phi_*(T_1\bigotimes T_2))=\phi_*(\nabla(T_1\bigotimes T_2))=\phi_*(\nabla T_1\bigotimes T_2+T_1\bigotimes\nabla T_2)=\phi_*(\nabla T_1)\bigotimes\phi_*T_2+\phi_*T_1\bigotimes\phi_*(\nabla T_2)=f(\phi_*T_1)\bigotimes\phi_*T_2+\phi_*T_1\bigotimes f(\phi_*T_2)$. Hence, we can write $\phi_*(\nabla T)=\nabla^\prime(\phi_*T)$, where $\nabla^\prime$ is a derivative operator. Since $\nabla g=0$ everywhere, we get $\nabla^\prime(\phi_*g)=\phi_*(\nabla g)=0$. {\em End of Proof}.

Hence, the action of diffeomorphisms preserves both algebraic and derivative relations in tensors, if the derivative operator is associated with the metric tensor. As an example, diffeomorphic transformation of a Riemann curvature tensor defined by a metric and the derivative operator associated with it, is equivalent to the Riemann curvature tensor defined by the diffeomorphic transformation of the metric tensor and the derivative operator associated with it. That is, $\phi_*\left(R_{abc}^{\;\;\;\;\;\;d}\right)=R_{abc}^{\prime\;\;\;\;\;d}$, where $R_{abc}^{\prime\;\;\;\;\;d}$ is defined by $\phi_*(g_{ab})$ and $\nabla_a^\prime$, $\nabla_a^\prime (\phi_*g_{bc})=0$. Hence we have:

{\sc Lemma 3}. In a spacetime physical laws expressed by tensors and their derivatives defined by the derivative operator associated with the metric tensor are invariant under diffeomorphisms.

The above discussions and results apply to any spacetime of any dimensions, including both the $(n+1)$-dimensional $(\t{\cal M}, \t{g}_{ab})$, and the $n$-dimensional $({\cal M}, g_{ab})$ as a hypersurface embedded in $(\t{\cal M}, \t{g}_{ab})$. The Lie derivative operator $\pounds$ on ${\cal M}$ is related to the Lie derivative operator $\t{\pounds}$ on $\t{\cal M}$ by
\begin{eqnarray}
	\pounds_vT^{a_1...a_k}_{\;\;\;\;\;\;\;\;\;\;\;b_1...b_l}=g^{a_1}_{\;\;\;c_1}...g_{b_l}^{\;\;\;d_l}\t{\pounds}_vT^{c_1...c_k}_{\;\;\;\;\;\;\;\;\;\;\;d_1...d_l} \;, \label{pound_M}
\end{eqnarray}
for any $v^a$ and $T^{a_1...a_k}_{\;\;\;\;\;\;\;\;\;\;\;b_1...b_l}\in\tm$.

Since the theory presented in this paper is base on an $(n+1)$-dimensional Einstein field equation on $\t{\cal M}$, Lemmas 1--3 guarantee that all the field equations derived in previous sections are invariant under diffeomorphisms on $\t{\cal M}$. In particular, the derived electromagnetic field equation (\ref{emfield_eq}) is invariant under the gauge transformation defined by all diffeomorphisms on $\t{\cal M}$.

Here we consider diffeomorphisms restricted to the $n$-dimensional ${\cal M}$, which are a subset of the diffeomorphisms on $\t{\cal M}$. Let $\phi_\tau$ be a one-parameter group of diffeomorphisms, which is generated by a vector field $v^a=(\partial/\partial\tau)^a$ on ${\cal M}$. Then, by equation (\ref{pgab1}), under an infinitesimal transformation the metric tensor $g_{ab}$ is transformed to $g^\prime_{ab}=g_{ab}-2\nabla_{(a}v_{b)}\delta\tau$. Any vector $N_a\in\tm$ is transformed to $N^\prime_a=N_a-(\pounds_vN_a)\delta\tau$, with
\begin{eqnarray}
	\pounds_v N_a = \nabla_a\left(v^cN_c\right)+v^c{\cal F}_{ca} \;, \label{Na_Nap2}
\end{eqnarray}
where the antisymmetric tensor
\begin{eqnarray}
	{\cal F}_{ca}\equiv \nabla_cN_a-\nabla_aN_c \;. \label{cF_ab}
\end{eqnarray}

{\sc Lemma 4}. For any vector field $N_a\in\tm$ and any smooth function $\chi$ on ${\cal M}$, there exists a vector field $v^a\in\tm$ so that
\begin{eqnarray}
	\pounds_vN_a=\nabla_a\chi \;. \label{pNa}
\end{eqnarray}

{\em Proof.} When ${\cal F}_{ca}$ is nondegenerate, i.e., $\det {\cal F}_{ca}\neq 0$, there exists an inverse antisymmetric tensor $\hat{\cal F}^{ab}$ defined by ${\cal F}_{ca}\hat{\cal F}^{ab}=\delta_c^{\;\;b}$. Then, for any smooth function $\chi^\prime$, the linear algebraic equation $v^c{\cal F}_{ca}=\nabla_a\chi^\prime$ has a unique solution $v^a=\hat{\cal F}^{ca}\nabla_c\chi^\prime$. Substituting it into equation (\ref{Na_Nap2}), we get equation (\ref{pNa}) with
\begin{eqnarray}
	\chi=\chi^\prime-N_c\hat{\cal F}^{ca}\nabla_a\chi^\prime \;. \label{chi}
\end{eqnarray}
Given any function $\chi(x^0,...,x^{n-1})$ on ${\cal M}$, equation (\ref{chi}) can be solved for $\chi^\prime(x^0,...,x^{n-1})$ as follows. Let $x^\mu=x^\mu(s)$ ($\mu=0,...,n-1$) define the integral curves of the vector field $k^a\equiv -N_c\hat{\cal F}^{ca}$. Then, $k^\mu=dx^\mu/ds$. By $d\chi^\prime/ds=(\partial_\mu\chi^\prime) dx^\mu/ds$, equation (\ref{chi}) is equivalent to a first-order ordinary differential equation
\begin{eqnarray}
	e^{-s}\frac{d}{ds}\left(e^s\chi^\prime(s)\right)=\chi\left(x^0(s),...,x^{n-1}(s)\right) \;,
\end{eqnarray}
whose solution is given by the integral
\begin{eqnarray}
	\chi^\prime(s)=e^{-s}\int e^s\chi\left(x^0(s),...,x^{n-1}(s)\right)ds \;.
\end{eqnarray}

When ${\cal F}_{ca}$ is degenerate, i.e., $\det {\cal F}_{ca}= 0$, the linear algebraic equation $v^c{\cal F}_{ca}=0$ of $v^c$ has an infinite number of solutions. Let $v_1^c$ be a solution. Then, $v^c=\chi^\prime v_1^c$ must also be a solution, where $\chi^\prime$ is any function. Hence, $v^c{\cal F}_{ca}=0$, and equation (\ref{Na_Nap2}) becomes equation (\ref{pNa}) with $\chi=\chi^\prime v_1^cN_c$. {\em End of Proof.}

{\sc Theorem 1}. For any vector field $N_a\in\tm$ and any smooth function $\chi$ on ${\cal M}$, there exists a diffeomorphism $\phi$ so that 
\begin{eqnarray}
	\phi_*N_a=N_a+\nabla_a\chi \;. \label{dif_N}
\end{eqnarray}

{\em Proof}.
By Lemma 4, for any function $\chi_1$ there exists a vector field $v^a=(\partial/\partial\tau)^a$ so that $N^{\prime}_a = N_a-\delta\tau\nabla_a\chi_1$ under an infinitesimal diffeomorphism generated by $v^a$. Then, at any point $p$ on ${\cal M}$, we have (e.g., \cite{haw73}),
\begin{eqnarray}
	\left.\phi_{\tau*}N_a\right|_p &=& \left.N_a\right|_p-\int_0^\tau\phi_{\tau^\prime*}\left(\left.\pounds_vN_a\right|_{\phi_{-\tau^\prime(p)}}\right)d\tau^\prime = \left.N_a\right|_p-\int_0^\tau\phi_{\tau^\prime*}\left(\left.\nabla_a\chi_1\right|_{\phi_{-\tau^\prime(p)}}\right)d\tau^\prime \nonumber\\
        &=& \left.N_a\right|_p-\int_0^\tau\left.\nabla_a\right|_p\phi_{\tau^\prime*}\left(\left.\chi_1\right|_{\phi_{-\tau^\prime(p)}}\right)d\tau^\prime = \left.N_a\right|_p+\left.\nabla_a\right|_p\chi_p \;,
\end{eqnarray}
where
\begin{eqnarray}
	\chi_p\equiv -\int_0^\tau\phi_{\tau^\prime*}\left(\left.\chi_1\right|_{\phi_{-\tau^\prime(p)}}\right)d\tau^\prime \;. \label{chi_chi1}
\end{eqnarray}
Equation (\ref{chi_chi1}) indicates that $\chi_1=-\phi^{-1}_*(d\chi/d\tau)$.
{\em End of Proof}.

{\sc Theorem 2}. For any closed two-form ${\cal F}_{ab}$ on ${\cal M}$, there exists a symmetry transformation. That is, there exists a vector field $v^a$ on ${\cal M}$ so that
\begin{eqnarray}
	\pounds_v{\cal F}_{ab}=0 \;. \label{pFab}
\end{eqnarray}

{\em Proof.} By the converse of the Poincar\'e lemma, any closed two-form ${\cal F}_{ab}$ can be expressed as in equation (\ref{cF_ab}), i.e., ${\bf F}=d{\bf N}$. Since the exterior derivative operator $d$ commutes with the Lie derivative operator \cite{haw73}, we have $\pounds_v{\bf F}=d\pounds_v{\bf N}$. By Lemma 4, there exists a vector field $v^a$ so that $\pounds_v{\bf N}=d\chi$ (eq.~\ref{pNa}). Then, by the Poincar\'e lemma, we get $\pounds_v{\bf F}=d^2\chi=0$. {\em End of Proof.}

Theorems 1 and 2 indicate that, in a spacetime there always exists a diffeomorphism which gives rise to the gauge transformation in equation (\ref{gauge0}) for electromagnetic fields, where $\chi$ is any smooth function. The electromagnetic field tensor $F_{ab}$ is invariant under the diffeomorphism. Since the derivative operator in the definition of $F_{ab}$, i.e., equation (\ref{F_A}), can be any derivative operator, we can take it to be the $\nabla_a$ before the diffeomorphism. If we use the $g^{ab}$ before the diffeomorphism to raise the indexes of $A_a$, $\nabla_a\chi$, and $F_{ab}$, then we get that $A^a\rightarrow A^a+\nabla^a\chi$, and $F^{ab}$ is also invariant under the gauge transformation. Then, if in the field equation we take the metric tensor to be the $g_{ab}$ and $g^{ab}$ before the diffeomorphism, the derivative operator to be the $\nabla_a$ associated with the $g_{ab}$, we find that the Einstein-Maxwell equation (\ref{meq1g}) is invariant under the gauge transformation. This way we have successfully fitted the electromagnetic gauge symmetry into diffeomorphisms in the background spacetime. 

Of course, the electromagnetic field equation (\ref{meq0}) is not invariant under the gauge transformation if we stay on the original spacetime background, unless $R_{ab}=0$. However, if we apply the diffeomorphic transformation to all tensor variables appearing in the equation, including the electromagnetic field vector and tensor, the current density vector, the metric tensor and quantities derived from it, equation (\ref{meq0}) must be invariant according to Lemma 3.

\section{The Cosmological Constant}
\label{cosmo}

We have interpreted $\dot{g}_{ab}$ as representation of a matter field on ${\cal M}$, but we have not specified the nature of the matter. It may represent dark matter that has been observed to exist in galaxies and clusters of galaxies and interact with ordinary matter only through the gravitational interaction \cite{bin87}, or dark energy (equivalent to a cosmological constant in some models) that is uniformly distributed in the universe and responsible for the observed accelerating expansion of the universe \cite{hin13,pla15}. It is also possible that $\dot{g}_{ab}$ represents some other kind of unknown matter. In this section we investigate if $\dot{g}_{ab}$ can behave as a cosmological constant in some situations.\footnote{The idea of interpreting the extra geometric terms in a four-dimensional Einstein field equation derived from 5D gravity as representing induced matter in a four-dimensional spacetime has been extensively investigated by Wesson and his collaborators (\cite{ove97,wes15}, and references therein). They proposed that the extra geometric terms are the stress-energy tensors of the induced matter and regarded the fifth dimension as being associated with the rest mass of particles instead of a real space dimension. However, in their theory, they did not derive the field equations of matter and electromagnetic fields.}
 
We consider a special case that in the neighbor of the hypersurface ${\cal M}$ defined by $w=0$, the metric tensors $g_{ab}$ and $g^{ab}$ can be approximated by
\begin{eqnarray}
	g_{ab}(w)=g_{ab}(1+\lambda w) \;, \hspace{1cm} g^{ab}(w)=g^{ab}(1-\lambda w) \;, 
\end{eqnarray}
where $g_{ab}=g_{ab}(w=0)$, $g^{ab}=g^{ab}(w=0)$, $\lambda$ is a constant and $\lambda w\ll 1$. Then, we have
\begin{eqnarray}
	\dot{g}_{ab}=\lambda g_{ab} \;, \hspace{1cm}
        \dot{g}^{ab}=-\lambda g^{ab} \;, 
\end{eqnarray}
and
\begin{eqnarray}
	\dot{g}_{a}^{\;\;b}=\ddot{g}_{ab}=\ddot{g}^{ab}=0 \;.
\end{eqnarray}

Then, by the definition of $\Phi^{ab}$, we have
\begin{eqnarray}
	\Phi_{ab}=\frac{1}{2}(n-1)N^{-1}\lambda g_{ab} \;, \hspace{1cm}\Phi=\frac{1}{2}n(n-1)N^{-1}\lambda\;, \label{Phi_ab_g}
\end{eqnarray}
and
\begin{eqnarray}
	\dot{\Phi}^{ab} &=& -\frac{1}{2}N^{-1}\left(\dot{g}^{ac}g^{bd}+g^{ac}\dot{g}^{bd}-\dot{g}^{ab}g^{cd}-g^{ab}\dot{g}^{cd}\right)\dot{g}_{cd} -\frac{1}{2}N^{-2}\dot{N}\left(g^{ac}g^{bd}-g^{ab}g^{cd}\right)\dot{g}_{cd} \nonumber\\
        &=&-(n-1)\lambda\left(\lambda+\frac{1}{2}N^{-1}\dot{N}\right)N^{-1}g^{ab} \;, \label{dotPhi^ab}
\end{eqnarray}
where $\dot{N}=\partial N/\partial w$. 

By equation (\ref{dw_sqrtg}), we get
\begin{eqnarray}
	g_{ac}g_{bd}\frac{1}{\sqrt{-g}}\frac{\partial}{\partial w}\left(\sqrt{-g}\Phi^{cd}\right)=\frac{1}{n-1}N\Phi\Phi_{ab}+g_{ac}g_{bd}\dot{\Phi}^{cd} =(n-1)\lambda N^{-1}\left[\left(\frac{n}{4}-1\right)\lambda-\frac{1}{2}N^{-1}\dot{N}\right]g_{ab}\;. \label{dw_sqrtg_Phi}
\end{eqnarray}
By equation (\ref{Phi_ab_g}), we have
\begin{eqnarray}
	\Phi_{cd}\Phi^{cd}-\frac{1}{n-1}\Phi^2=-\frac{1}{4}n(n-1)N^{-2}\lambda^2 \;, \label{PhiPhi1}
\end{eqnarray}
and
\begin{eqnarray}
	\Phi_{ac}\Phi_b^{\;\;c}-\frac{1}{n-1}\Phi\Phi_{ab}=-\frac{1}{4}(n-1)N^{-2}\lambda^2g_{ab} \;. \label{PhiPhi2}
\end{eqnarray}
Substituting equations (\ref{dw_sqrtg_Phi})--(\ref{PhiPhi2}) into equation (\ref{Tab_m_M}), we get
\begin{eqnarray}
	T_{m,ab} = -\frac{1}{2\kappa}(n-1)\lambda N^{-2}\left[\left(\frac{n}{4}-1\right)\lambda-N^{-1}\dot{N}\right]g_{ab} \;. 
\end{eqnarray}

Letting $n=4$, we have
\begin{eqnarray}
	T_{m,ab} = \frac{3}{2\kappa}\lambda N^{-3}\dot{N}g_{ab} \;, \label{Tab_l}
\end{eqnarray}
which corresponds to a cosmological constant
\begin{eqnarray}
	\Lambda = -\frac{3}{2}\lambda N^{-3}\dot{N} \label{lamda}
\end{eqnarray}
in the four-dimensional spacetime $({\cal M},g_{ab})$.

By equation (\ref{Phi_ab_g}), we have
\begin{eqnarray}
	\Phi_{cd}\Psi^{cd}-\frac{1}{n-1}\Phi\Psi=-\frac{1}{2}N^{-1}\lambda\Psi \;, \label{PhiPsi1}
\end{eqnarray}
and
\begin{eqnarray}
	2\Phi_{c(a}\Psi_{b)}^{\;\;c}-\frac{1}{n-1}(\Phi\Psi_{ab}+\Psi\Phi_{ab})= \left(\frac{n}{2}-1\right)N^{-1}\lambda\Psi_{ab}-\frac{1}{2}N^{-1}\lambda\Psi g_{ab} \;. 
\end{eqnarray}

By equations (\ref{dw_sqrtg}) and (\ref{Phi_ab_g}), we derive that
\begin{eqnarray}
	g_{ac}g_{bd}\frac{1}{\kappa N\sqrt{-g}}\frac{\partial}{\partial w}\left(\sqrt{-g}\Psi^{cd}\right) =\frac{1}{\kappa N}g_{ac}g_{bd}\dot{\Psi}^{cd}+\frac{1}{2\kappa}nN^{-1}\lambda\Psi_{ab}\;. \label{dw_sqrtg_Psi}
\end{eqnarray}
By equation (\ref{Phi_ab_g}) and and the convention of $\nabla_aN=0$, we have $\nabla_a\Phi_{bc}=0$. Hence we get
\begin{eqnarray}
	\frac{1}{\kappa N}\nabla^c\left[2N_{(a}\Phi_{b)c}-N_c\Phi_{ab}\right] =\frac{1}{\kappa}N^{-1}\lambda\left[(n-1)\Psi_{ab}-\frac{1}{2}\Psi g_{ab}\right] \;. \label{dNPhi}
\end{eqnarray}

Substituting equations (\ref{PhiPsi1})--(\ref{dNPhi}) into equation (\ref{Tab_int_M}), we get
\begin{eqnarray}
	T_{\intt,ab} = -\frac{1}{\kappa N}\left[\left(\frac{n}{2}+1\right)\lambda\Psi_{ab}+g_{ac}g_{bd}\dot{\Psi}^{cd}\right] \;. 
\end{eqnarray}
Setting $n=4$, we get
\begin{eqnarray}
	T_{\intt,ab} = -\frac{1}{\kappa N}\left(3\lambda\Psi_{ab}+g_{ac}g_{bd}\dot{\Psi}^{cd}\right) =-\frac{1}{\kappa N}\left(\lambda\Psi_{ab}+\dot{\Psi}_{ab}\right) \;. 
\end{eqnarray}

By equations (\ref{cal_J}) and (\ref{Phi_ab_g}), we have $4\pi J^b=\nabla_a\Phi^{ab}=0$ (since $\nabla_aN=0$). Then, the electromagnetic field equation (\ref{emfield_eq}) becomes a source-free equation
\begin{eqnarray}
	\nabla_a\Psi^{ab}=0 \;.
\end{eqnarray}

Equations (\ref{Tab_l}) and (\ref{lamda}) indicate that in appropriate conditions the matter represented by $\dot{g}_{ab}$ behaves like a cosmological constant $\Lambda$ on ${\cal M}$. However, by equation (\ref{lamda}), $\Lambda\neq 0$ only if $\dot{N}=\partial N/\partial w\neq 0$, and the sign of $\Lambda$ is determined by the sign of $\dot{N}$. To get a positive cosmological constant on ${\cal M}$, we need to have $\dot{N}<0$. Let us write $\dot{N}\sim -\lambda N$, by equation (\ref{lamda}) we have $\Lambda\sim N^{-2}\lambda^2$, and the mass density corresponding to the $\Lambda$ is $\rho_\Lambda=\Lambda/\kappa\sim \kappa^{-1}N^{-2}\lambda^2$. Hence,
\begin{eqnarray}
	\frac{\rho_\Lambda}{\rho_\P}\sim \kappa^{-1}N^{-2}\lambda^2 l_\P^2 \;, \label{rhol_rhop}
\end{eqnarray}
where $\rho_\P=l_\P^{-2}=5.2\times 10^{93}\,\g\,\cm^{-3}$ is the Planck mass density. Observations of WMAP and Planck satellites have indicated that there may exist a positive cosmological constant in the universe with an equivalent mass density $\approx 0.7$ times of the critical mass density \cite{hin13,pla15}, which leads to $\rho_\Lambda/\rho_\P\approx 10^{-123}$. Then, if we set the dimensionless function $N\sim 1$ and $\kappa=8\pi$, equation (\ref{rhol_rhop}) leads to
\begin{eqnarray}
	\lambda^{-1}\sim 6.3\times 10^{60}l_\P\sim 10^{28}\,\cm \;, 
\end{eqnarray}
the same order of the Hubble distance.

On the other hand, if $\lambda^{-1}$ is much smaller than the Hubble distance, equation (\ref{rhol_rhop}) indicates a cosmological constant that is much larger than that we have observed, with an order of
\begin{eqnarray}
	\frac{\rho_\Lambda}{\rho_\P}\sim 10^{-65}\left(\frac{\lambda^{-1}}{1{\rm mm}}\right)^{-2} \;.
\end{eqnarray}
Such an unrealistically large cosmological constant may be canceled by a native cosmological constant in the five-dimensional bulk spacetime, as having been assumed for the brane world theory \cite{shi00}.

The equations derived in Secs.~\ref{field_eqs_I} and \ref{field_eqs_II} can be extended to the case when there is a cosmological constant $\t{\Lambda}$ in the $(n+1)$-spacetime $(\t{\cal M},\t{g}_{ab})$, i.e., when the vacuum Einstein field equation on $\t{\cal M}$ is $\t{G}_{ab}+\t{\Lambda}\t{g}_{ab}=0$, corresponding to an action of gravity on $\t{\cal M}$
\begin{eqnarray}
	\t{S}_G = \int \sqrt{-\t{g}}(\t{R}-2\t{\Lambda})\t{\bf e} = \int \sqrt{-\t{g}}\left(R-K_{ab}K^{ab}+K^2-2\t{\Lambda}\right)\t{\bf e} \;. \label{SGl}
\end{eqnarray} 
The presence of $\t{\Lambda}$ in the action leads only to modification of the ${\cal L}_G$ in equation (\ref{LG0}), and the modified ${\cal L}_G$ is
\begin{eqnarray}
	{\cal L}_G\equiv \sqrt{-g}N(R-2\t{\Lambda}) \;. \label{LG0l}
\end{eqnarray}
The ${\cal L}_\emm$, ${\cal L}_m$, and ${\cal L}_\intt$ defined in equations (\ref{Lem_M2})--(\ref{Lint_M2}) are not changed.

It is easy to derive that, when $\t{\Lambda}$ is present, equations (\ref{scalar_eq_L}) and (\ref{scalar_eq_LK}) become
\begin{eqnarray}
	R+\Pi_{ab}\Pi^{ab}-\frac{1}{n-1}\Pi^2 = 2\t{\Lambda} \;, \label{scalar_eq_Ll}
\end{eqnarray}
and
\begin{eqnarray}
	R+K_{ab}K^{ab}-K^2 = 2\t{\Lambda} \;, \label{scalar_eq_LKl}
\end{eqnarray}
Equations (\ref{emfield_eq}) and (\ref{emfield_eqK}), which are interpreted as the electromagnetic field equations, are not changed since in the Lagrangian density the $\t{\Lambda}$ is not coupled to $N_a$. The gravitational field equation on ${\cal M}$, i.e., equation (\ref{ein_eq_n}), becomes
\begin{eqnarray}
	G_{ab} +\t{\Lambda} g_{ab} = \kappa T_{ab} \;, \label{ein_eq_nl}
\end{eqnarray}
where $T_{ab}$ is unchanged (still given by eqs.~\ref{T_hT}, \ref{Tab_em_M}, \ref{Tab_m_M}, and \ref{Tab_int_M}).

Correspondingly, equation (\ref{scalar_eq_LKb}) becomes
\begin{eqnarray}
	g^{cd}\t{\pounds}_nK_{cd}-K_{cd}K^{cd}-\nabla_ca^c+a_ca^c+\frac{2}{n-1}\t{\Lambda}=0 \;, \label{TT2ax}
\end{eqnarray}
equation (\ref{scalar_eq_La}) becomes
\begin{eqnarray}
	g_{cd}\frac{1}{\sqrt{-g}}\frac{\partial}{\partial w}\left(\sqrt{-g}\Pi^{cd}\right) &=& (3-n)N\left(\Pi_{cd}\Pi^{cd}-\frac{1}{n-1}\Pi^2\right)-(n-1)\nabla_c\nabla^c N \nonumber\\
        &&-\nabla^c\left(2N^a\Pi_{ac}-N_c\Pi\right)-2N\t{\Lambda} \;, \label{TT2x}
\end{eqnarray}
and equation (\ref{scalar_eq_Lb}) becomes
\begin{eqnarray}
	g_{cd}\t{\cal L}_n\Pi^{cd} = -(n-3)\Pi_{cd}\Pi^{cd}+\frac{n-4}{n-1}\Pi^2 -(n-1)\frac{1}{N}\nabla_c\nabla^c N-2\t{\Lambda} \;.  \label{scalar_eq_Lbx}
\end{eqnarray}

The effective cosmological constant on ${\cal M}$ is $\Lambda_\eff=\t{\Lambda}+\Lambda$, with $\Lambda$ being determined by equation (\ref{lamda}). With some unknown fine-tune mechanism (as in the brane world theory), the $\Lambda_\eff$ may become zero or small enough to be compatible with the observations in cosmology.

\section{Discussions on the new electromagnetic field equation}
\label{dis}

In this section we give some discussions on the new electromagnetic field equation (\ref{meq0}), which is the most important equation derived in the paper. Although $\xi=-2$ is preferred since then the equation can be derived from the Einstein field equation in a five-dimensional spacetime, here for generality we treat $\xi$ as an undetermined number of order unity, either positive or negative. Solutions to the equation (\ref{meq0}) will not be provided here, except for a very simple case involving a Killing vector field as a solution. So, we will not attempt to explore phenomenological signals explicitly and quantitatively predicted by solutions to the field equation (\ref{meq0}). Instead, we will only give some general discussions and comments relevant to experimental tests of the new field equation in a four-dimensional spacetime. 

In terms of $A^a$, the electromagnetic field equation (\ref{meq0}) can be written as
\begin{eqnarray}
        \nabla^a\nabla_aA^b-\nabla^b\nabla_aA^a-(\xi+1)R^b_{\;\;a}A^a=-4\pi J^b \;. \label{meq0_A}
\end{eqnarray}
We know that, a Killing vector $\psi^a$ in a spacetime satisfies the equations \cite{wal84}
\begin{eqnarray}
        \nabla_a\psi^a=0 \;, \hspace{1cm} \nabla^a\nabla_a\psi^b+R^b_{\;\;a}\psi^a=0 \;. \label{kill}
\end{eqnarray}
Comparison of equations (\ref{meq0_A}) and (\ref{kill}) leads to the result that {\em when $\xi=-2$, any Killing vector field in a spacetime $({\cal M}, g_{ab})$ is a solution of the source-free electromagnetic field equation (\ref{meq0})}. This is in contrast to the case of the Maxwell equation (\ref{meq1g}), where a Killing vector field solves the source-free equation only if the spacetime is Ricci-flat \cite{wal84,wal74}.

The above result can be more easily derived from equation (\ref{meq4}), which is equivalent to the field equation (\ref{meq0}) with $\xi=-2$. If we set $A^a=\psi^a$, by the Killing equation $\nabla_a\psi_b+\nabla_b\psi_a=0$ we get $H_{ab}=0$ and $H=0$. Hence a Killing vector $\psi^a$ solves the electromagnetic field equation (\ref{meq4}) when $J^a=0$.

The result can be generalized a little bit. Let $\psi^a$ be a conformal Killing vector field, i.e., $\psi^a$ satisfies \cite{wal84}
\begin{eqnarray}
        \nabla_a\psi_b+\nabla_b\psi_a=\alpha g_{ab} \;. \label{kill_eq2}
\end{eqnarray}
Setting $A^a=\psi^a$, we get $H_{ab}=\alpha g_{ab}$ and $H=n\alpha$, where $n=\dim\,{\cal M}$. Substituting them into the electromagnetic field equation (\ref{meq4}), we find that equation (\ref{meq4}) is solved when $J^a=0$ if and only if $\nabla_a\alpha=0$. Hence, {\em any conformal Killing vector field with a constant $\alpha$ solves the source-free electromagnetic field equation (\ref{meq0}) with $\xi=-2$}.

The above conclusion applies to a spacetime of any dimensions. However, in the rest part of this section we assume that $n=4$ since the discussions in the rest part focus on experimental tests of the electromagnetic field equation (\ref{meq0}) in a four-dimensional spacetime.

The stress-energy tensor of electromagnetic fields described by equation (\ref{meq0}) was derived in \cite{li15}, which becomes the equation (\ref{Tem_ab2}) when $\xi=-2$. For any value of $\xi$, the divergence of the stress-energy tensor of electromagnetic fields is evaluated to be \cite{li15}
\begin{eqnarray}
        \nabla^aT_{\emm,ab}=-F_{ba}J^a+A_b\nabla_aJ^a \;, \label{dTem1}
\end{eqnarray}
where the electromagnetic field equation (\ref{meq0}) has been applied. When the electric current is conserved (i.e., $\nabla_aJ^a=0$), equation (\ref{dTem1}) becomes
\begin{eqnarray}
        \nabla^aT_{\emm,ab}=-F_{ba}J^a \;. \label{dTem2}
\end{eqnarray}
If we denote the total stress-energy tensor by $T_{ab}=T_{\emm,ab}+T_{\ot,ab}$, where $T_{\ot,ab}$ represents the stress-energy tensor of other matter fields, by the conservation equation $\nabla^aT_{ab}=0$ we get
\begin{eqnarray}
        \nabla^aT_{\ot,ab}=F_{ba}J^a \;, \label{L_force}
\end{eqnarray}
which is just the Lorentz force law. 

Hence, the curvature-coupled term in the electromagnetic field equation does not affect the Lorentz force law. The force of a charge particle or an electric current in an electromagnetic field is determined by the antisymmetric tensor $F_{ab}$ (i.e., by the electric field ${\bf E}$ and the magnetic field ${\bf B}$) solving the electromagnetic field equation (\ref{meq0}). Although the potential vector $A^a$ explicitly appears in the electromagnetic field equation through the curvature-coupled term, it interacts with charges and currents only through the antisymmetric tensor $F_{ab}$.

When the spacetime is Ricci-flat ($R_{ab}=0$, e.g., outside of a black hole or a star in a vacuum environment\footnote{Strictly speaking, when there is present an electromagnetic field the spacetime cannot be exactly Ricci-flat, since the stress-energy tensor of the electromagnetic field will make $R_{ab}\neq 0$. However, if the electromagnetic field is weak its effect on the spacetime curvature can be ignored and the spacetime can be approximately Ricci-flat if the mass density of other matter is low.}), equation (\ref{meq0}) is equivalent to the Einstein-Maxwell equation (\ref{meq1g}). Hence, in a flat spacetime or in a curved but Ricci-flat spacetime, the new electromagnetic field equation is identical to the Einstein-Maxwell equation. As a result, the curvature-coupled term in the new equation has no effect on all experiments in the lab condition.

In a spacetime with $R_{ab}\neq 0$, the curvature-coupled term can have an effect on the electromagnetic field solution (e.g., inside a star or in a cosmological environment). To estimate the effect, we write
\begin{eqnarray}
	\left|\nabla_a F^{ab}\right|\sim\frac{|A^b|}{l_e^2} \;, \hspace{1cm}
        \left|\xi R^b_{\;\;a}A^a\right|\sim\frac{|A^b|}{r_c^2} \;,
\end{eqnarray}
where $l_e$ is the spacetime scale on which the electromagnetic field varies (e.g., the wavelength of an electromagnetic wave, and the size of the source), and $r_c$ is the spacetime curvature radius defined by the Ricci tensor, i.e., $r_c\equiv\left(R_{ab}R^{ab}\right)^{-1/4}$. So, we have
\begin{eqnarray}
	\frac{\left|\xi R^b_{\;\;a}A^a\right|}{\left|\nabla_a F^{ab}\right|}\sim\frac{l_e^2}{r_c^2} \;.
\end{eqnarray}
When $r_c\gg l_e$, we expect that the term $\xi R^b_{\;\;a}A^a$ is not important. By the Einstein field equation, we can estimate the order of $r_c$ by
\begin{eqnarray}
	r_c\sim\frac{l_\P}{\sqrt{8\pi}}\left(\frac{\rho}{\rho_\P}\right)^{-1/2} \;, \label{rc}
\end{eqnarray}
where $\rho$ is the mass density at the place where the electromagnetic field is present.

\begin{table}
\caption{\label{tab1}Density $\rho$, radius (height) $r$, and the curvature radius $r_c$ of some objects. The $r_c$ is estimated by equation (\ref{rc}), except for the universe where both $r$ and $r_c$ are taken to be the Hubble distance $d_H$.\protect\footnote{In the case of the universe, the $r_c$ estimated by equation (\ref{rc}) is of the same order as $d_H$.}}
\begin{ruledtabular}
\begin{tabular}{llll}
Object & $\rho$ (g/cm$^3$) & $r$ (cm) & $r_c$ (cm)\\[1mm] \hline\\[-3mm]
Atmosphere\footnote{The density is measured at sea level and $15\,^\circ$\!C. The $r$ refers to the approximate height above sea level.} & 0.001225 & $\sim 10^6$ & $6.6\times 10^{14}$\\
Earth\footnote{The density $\rho$ and radius $r$ are averaged values.} & 5.5 & $6.4\times 10^8$ & $1.0\times 10^{13}$ \\
Jupiter$^{\rm c}$ & 1.3 & $7.0\times 10^{9}$ & $2.0\times 10^{13}$ \\
Sun$^{\rm c}$ & 1.4 & $7.0\times 10^{10}$ & $2.0\times 10^{13}$ \\
White Dwarf\footnote{The $\rho$ is the averaged density, and $r$ is the radius of a typical white dwarf.} & $10^6$ & $7\times 10^{8}$ & $2\times 10^{10}$ \\
Neutron Star\footnote{The $\rho$ is the core density, and $r$ is the radius of a typical neutron star.} & $5\times 10^{14}$ & $10^6$ & $1.0\times 10^6$ \\
Universe\footnote{The $\rho$ is the critical density of the universe, where $h$ is the Hubble constant in units of 100 km s$^{-1}$ Mpc$^{-1}$.} & $2\times 10^{-29}h^2$ & $9\times 10^{27}h^{-1}$ & $9\times 10^{27}h^{-1}$\\
\end{tabular}
\end{ruledtabular}
\end{table}

The density and curvature radius of some objects are listed in Table~\ref{tab1}. For comparison with the curvature radius, the sizes (radius or height) of the selected objects are also listed. It can be imagined that the curvature-coupled term is important if the size of an object is larger or at least comparable to the curvature radius given by equation (\ref{rc}). So, according to the results in Table~\ref{tab1}, we can expect that the curvature-coupled term in the new electromagnetic field equation (\ref{meq0}) will give rise to detectable effects for white dwarfs, neutron stars, and the universe. It is worth to note that, \citet{tur88} have shown that an electromagnetic field equation of the form in equation (\ref{meq0}) with a negative $\xi$ can lead to fast growth of a primordial magnetic field during the inflation epoch through a mechanism akin to ``superadiabatic amplification'', which is not possible for the Maxwell equation without a curvature-coupled term. This can be regarded as an evidence that the curvature-coupled term can lead to detectable effects in the early universe.

In Table~\ref{tab1} we also list the atmosphere, which is relevant for the experiments of electromagnetism in laboratory conditions, and determination of the magnetic field of the earth. For the density of the atmosphere, the curvature radius is $\sim 10^6$ times the radius of the earth and the correction from $\xi R^b_{\;\;a}A^a$ should be of order $\lesssim 10^{-12}$ in fraction of $\nabla_a F^{ab}$. According to \cite{gol10}, the best constraint from laboratory tests on the reduced Compton wavelength associated with the rest mass of photons is $\lambdabar\gtrsim 2\times 10^9$ cm. Although the limit is distant from the $r_c$ for the atmosphere by five orders of magnitudes, testing the effect of $\xi R^b_{\;\;a}A^a$ in laboratory conditions may become possible in future with more advanced techniques.

For a vacuum spacetime with a cosmological constant $\Lambda$ (e.g., a de~Sitter spacetime or an anti-de~Sitter spacetime), by the Einstein field equation we have $R_{ab}=\Lambda g_{ab}$ and $R_{ab}A^b=\Lambda A_a$. Then, the electromagnetic field equation (\ref{meq0}) becomes
\begin{eqnarray}
        \nabla_a F^{ab}-\xi\Lambda A^b=-4\pi J^b \;. \label{meq0a}
\end{eqnarray}
In this case, the curvature-coupled term $\xi R_{ab}A^b$ is equivalent to a photon mass term $m^2A_a$, with
\begin{eqnarray}
        m^2=\xi\Lambda \;.
\end{eqnarray}
When $\xi\Lambda<0$, we have $m^2<0$, i.e., the photon has an imaginary mass. Then a photon will travel with a speed faster than $c$. For instance, for a de~Sitter spacetime ($\Lambda>0$) with an electromagnetic field equation derived from five-dimensional gravity we have $\xi=-2$ and $m^2=-2\Lambda<0$. We can then expect that photons can travel superluminally in the de~Sitter spacetime. For a $\Lambda$ comparable to the observed value in the current universe, $m^2$ is sufficiently small and hence the strength of signal off the lightcone should be very weak. However, in the very early universe, e.g., during the inflationary phase, $\Lambda$ is not small so the signal off the lightcone may be appreciable.

\section{Summary and Conclusions}
\label{sum}

A new unified theory of electromagnetic and gravitational interactions is presented in this paper. A four-dimensional spacetime is assumed to be a hypersurface embedded in a five-dimensional bulk spacetime. Then, the field equations in the four-dimensional spacetime are determined by the projection of the Einstein field equation in the five-dimensional spacetime onto the hypersurface, and the contraction with the normal to the hypersurface. This way, three independent equations are obtained, which form a complete set of field equations in the four-dimensional spacetime, including determination of the metric tensor off the hypersurface. 

The first is a scalar constraint equation given by equation (\ref{scalar_eq_L}) (or, equivalently, eq.~\ref{scalar_eq_LK}), which relates the scalar curvature to the extrinsic curvature tensor of the hypersurface. The second is a vector constraint equation given by equation (\ref{dPi_ab_0}) (or, equivalently, eq.~\ref{emfield_eqK}), which can be interpreted as the electromagnetic field equation in a four-dimensional spacetime (eq.~\ref{emfield_eq}). The third is a tensor equation, i.e., equation (\ref{ein_eq_n}) with $T_{ab}$ given by equations (\ref{T_hT}), (\ref{Tab_em_M}), (\ref{Tab_m_M}), and (\ref{Tab_int_M}), which can be interpreted as the Einstein field equation with the stress-energy tensor of electromagnetic fields and other matter as the source. The constraint equation (\ref{scalar_eq_L}) (or, equivalently, eq.~\ref{scalar_eq_LK}) can also be replaced by the equation (\ref{scalar_eq_La}) (or, equivalently, eq.~\ref{scalar_eq_LKb}), since the latter is derived from the combination of equations (\ref{scalar_eq_L}), (\ref{dPi_ab_0}), and (\ref{ein_eq_n}).

The most important result of the new unified theory is that a new electromagnetic field equation in a four-dimensional spacetime is derived from the five-dimensional vacuum Einstein field equation. The new electromagnetic field equation is given by equation (\ref{emfield_eq}), which is equivalent to the equation (\ref{meq0}) with $\xi=-2$ with the assumption that $\nabla_aN=0$ (i.e., $N$ is constant in the four-dimensional spacetime). The new field equation differs from the Einstein-Maxwell equation (\ref{meq1g}) by a curvature-coupled term $\xi R^b_{\;\;a}A^a$, which vanishes in a Ricci-flat spacetime but can be important in an environment with a high mass density. Although practical solutions to the new electromagnetic field equation are not studied, we have shown that a conformal Killing vector field with a constant $\alpha$ (or, a Killing vector field when $\alpha=0$) solves the source-free field equation with $\xi=-2$. We have also argued that the effect of the curvature-coupled term may be detectable in electromagnetic processes inside a neutron star or a white dwarf, and in the early epoch of the universe (see Table \ref{tab1} and relevant discussions in the text).

The electromagnetic field equation (\ref{meq0}) with an undetermined $\xi$ was originally proposed by \citet{li15} to address the incompatibility problem in application of the Einstein-Maxwell equation to a universe with a uniformly distributed net charge. The fact that it can be derived from the five-dimensional Einstein field equation with a determined $\xi=-2$ supports the proposal of the equation (\ref{meq0}) as a solution to the inconsistency problem. Another support for equation (\ref{meq0}) comes from the notice that the Maxwell equation in a flat spacetime can be expressed in terms of a symmetric tensor instead of an antisymmetric tensor. When the Maxwell equation expressed in a symmetric tensor is extended to a curved spacetime via the ``minimal substitution rule'', the field equation (\ref{meq0}) with $\xi=-2$ is naturally obtained. So, we believe that the electromagnetic field equation (\ref{meq0}) and the unified theory of gravity and electromagnetism proposed in this paper deserve studies, although whether they describe the real physical world can only be ultimately determined by experiments and observations.

The theory studied in this paper provides a new way for geometrzing the electromagnetic field and unifying the electromagnetic and gravitational interactions, which has been the ultimate objective of Einstein that has cost his later half life. In the theory, electromagnetic fields are contained in the extrinsic curvature tensor of a four-dimensional spacetime as a hypersurface in a high-dimensional spacetime. The idea of interpreting a four-dimensional spacetime as a hypersurface embedded in a five-dimensional bulk spacetime has also been explored in the brane world theory. Unlike in the brane world theory, where electromagnetic fields are assumed to be confined in a four-dimensional membrane {\it a priori}, here electromagnetic fields and the electromagnetic field equation are {\it derived} from the five-dimensional Einstein field equation and present themselves on the four-dimensional hypersurface. The theory is also different from the Kaluza-Klein theory, since in the Kaluza-Klein theory the Einstein-Maxwell equation was derived.

Besides electromagnetic fields, the extrinsic curvature of a four-dimensional spacetime hypersurface also contains a term proportional to the derivative of the four-dimensional spacetime metric with respect to the fifth dimension ($\dot{g}_{ab}$), i.e., a term depending on the evolution of the metric off the hypersurface. We have tried to interpret it as representing some unidentified matter in the four-dimensional spacetime. The stress-energy tensors of electromagnetic fields and the unidentified matter, and their interaction, are derived. We have shown that, under some conditions, the stress-energy tensor of the unidentified matter may behave like a cosmological constant.

It is well known that the five-dimensional Kaluza-Klein theory can be generalized to a higher-dimensional theory to include the weak and strong interactions (\cite{bai87}, and references therein). It would be interesting to extend the theory presented in this paper to the case of a four-dimensional spacetime embedded in an $n>5$ dimensional bulk spacetime and see if non-Abelian gauge interactions can be derived from an $n$-dimensional Einstein field equation.

\begin{acknowledgments}
The author thanks reviewers Chao-Guang Huang and Lijing Shao for careful reading of the manuscript and for very good reports which have helped to improve the presentation of the contents. This work was supported by the National Basic Research Program (973 Program) of China (Grant No.~2014CB845800) and the NSFC grant (No.~11373012).
\end{acknowledgments}

\appendix

\section{5D Metric in Kaluza-Klein's Representation}
\label{KK}

In this Appendix we study the geometry of the Kaluza-Klein theory and derive the relation between the Kaluza-Klein decomposition of the 5D metric and that adopted in this paper. We use the same coordinate system in Sec.~\ref{4+1}, i.e., the $\{x^0,x^1,x^2,x^3,w\}$ for the 5D spacetime $(\tilde{\cal M},\tilde{g}_{ab})$. The coordinates $\{x^0,x^1,x^2,x^3\}$ are defined on the hypersurface manifold ${\cal M}$ (defined by $w=\mbox{const}$) and carried to the neighbor of ${\cal M}$ in the five-dimensional $\tilde{M}$ with the map generated by the coordinate lines of $w$. 

As stated in Sec.~\ref{4+1}, the strategy of the Kaluza-Klein (KK) theory is to decompose the five-dimensional metric $\t{g}_{ab}$ in the form in equation (\ref{ktg_ab}), or, equivalently,
\begin{eqnarray}
	\t{g}_{ab} = \!\!~_kg_{\mu\nu}dx^\mu_adx^\nu_b+\phi^2\left(A_\mu dx^\mu_a+dw_a\right)\left(A_\nu dx^\nu_b+dw_b\right) \;. \label{ktg_ab1}
\end{eqnarray}

Let us define
\begin{eqnarray}
	\t{g}^{AB} = \left(\begin{array}{cc}
          \!\!\!~_kg^{\mu\nu} & -A^\mu\\[2mm]
          -A^\nu & \frac{1}{\phi^2}+A_\rho A^\rho
          \end{array}\right)
        \;, \label{ktg^ab}
\end{eqnarray}
where the 4-matrix $\!\!\!~_kg^{\mu\nu}$ is the inverse of $\!\!\!~_kg_{\mu\nu}$, i.e., $\!\!\!~_kg_{\mu\nu}\!\!~_kg^{\nu\rho}=\delta_\mu^{\;\,\rho}$; and
\begin{eqnarray}
	A^\mu\equiv \!\!~_kg^{\mu\nu} A_\nu \;. \label{g_A_u}
\end{eqnarray}
It can be checked that the matrix in equation (\ref{ktg^ab}) is the inverse of the matrix in equation (\ref{ktg_ab}). Hence, we have
\begin{eqnarray}
	\t{g}^{ab} &=& \!\!~_kg^{\mu\nu}\left(\frac{\partial}{\partial x^\mu}\right)^a\left(\frac{\partial}{\partial x^\nu}\right)^b 
        -2A^\mu\left(\frac{\partial}{\partial x^\mu}\right)^{(a}\left(\frac{\partial}{\partial w}\right)^{b)} +\left(\frac{1}{\phi^2}+A_\rho A^\rho\right)\left(\frac{\partial}{\partial w}\right)^a\left(\frac{\partial}{\partial w}\right)^b\;. \label{ktg^ab1}
\end{eqnarray}

Equation (\ref{g_A_u}) and the fact that $\!\!\!~_kg^{\mu\nu}$ and $\!\!\!~_kg_{\mu\nu}$ are inverse to each other automatically lead to
\begin{eqnarray}
	\!\!\!~_kg_{\mu\nu}A^\nu=A_\mu \;. \label{g_A_d}
\end{eqnarray}

To understand the geometric nature of the KK variables defined above, we need to find out the KK metric tensor $\!\!\!~_kg_{ab}$ associated with the $4\times 4$ matrix $\!\!~_kg_{\mu\nu}$, and the vector $A_a$ associated with the $4\times 1$ matrix $A_\mu$. Since $\!\!\!~_kg_{ab}$ and $A_a$ are tensors and vectors on $\t{\cal M}$, in general they can be written as
\begin{eqnarray}
	\!\!\!~_kg_{ab} = \!\!~_kg_{\mu\nu}dx^\mu_adx^\nu_b+2\!~_kg_{\mu 4}dx^\mu_{(a} dw_{b)} +\!\!~_kg_{44}dw_adw_b \;, \label{kgab_w}
\end{eqnarray}
and
\begin{eqnarray}
	A_a = A_\mu dx^\mu_a+A_4dw_a \;, \label{Aa_w}
\end{eqnarray}
although some of the tensor components may turn out to be zero.

{\sc Proposition 1.} The vectors $A_a$ and $A^a$ are expressed in coordinate components by
\begin{eqnarray}
	A_a=A_\mu dx^\mu_a \;, \label{A_a}
\end{eqnarray}
and
\begin{eqnarray}
	A^a=A^\mu\left(\frac{\partial}{\partial x^\mu}\right)^a-A_\rho A^\rho\left(\frac{\partial}{\partial w}\right)^a \;, \label{A^a}
\end{eqnarray}
where $\mu,\rho=0,1,2,3$.

{\em Proof}. By equations (\ref{ktg^ab1}) and (\ref{Aa_w}), we have
\begin{eqnarray}
	A^a = \t{g}^{ab}A_b=\left(\!\!~_kg^{\mu\nu}A_\nu-A_4A^\mu\right)\left(\frac{\partial}{\partial x^\mu}\right)^a -\left[A^\rho A_\rho-A_4\left(\frac{1}{\phi^2}+A_\rho A^\rho\right)\right]\left(\frac{\partial}{\partial w}\right)^a \;, \label{A^ax}
\end{eqnarray}
from which we get
\begin{eqnarray}
	A^\mu\equiv A^adx^\mu_a=\!\!~_kg^{\mu\nu}A_\nu-A_4A^\mu \;,
\end{eqnarray}
then by equation (\ref{g_A_u}) we must have $A_4=0$. Hence equation (\ref{A_a}).
 
Setting $A_4=0$, by equations (\ref{A^ax}) and (\ref{g_A_u}) we get the equation (\ref{A^a}). {\em End of Proof}.

{\sc Proposition 2.} 
\begin{eqnarray}
	A_aA^a=A_\rho A^\rho \;, \label{AA_AA}
\end{eqnarray}
which is directly derived from equations (\ref{A_a}) and (\ref{A^a}).

{\sc Proposition 3.} The $\phi$, $A_\mu$, and $A^\mu$ are related to $N$, $N_\mu$, and $N^\mu$ by
\begin{eqnarray}
	\phi^2=N^2+N_\rho N^\rho =\t{g}_{ab}w^aw^b\;, \label{phi}
\end{eqnarray}
\begin{eqnarray}
	A_\mu=\frac{N_\mu}{\phi^2}=\frac{N_\mu}{N^2+N_\rho N^\rho} \;, \label{A_mu}
\end{eqnarray}
and
\begin{eqnarray}
	A^\mu=\frac{N^\mu}{N^2} \;. \label{A^mu}
\end{eqnarray}

{\em Proof.} Comparison of equations (\ref{tg_ab2}) and (\ref{ktg_ab}) leads to equations (\ref{phi}) and (\ref{A_mu}). 
The inverse of the matrix in equation (\ref{tg_ab2}) is
\begin{eqnarray}
	\t{g}^{AB} &=& \left(\begin{array}{cc}
          g^{\mu\nu}+\frac{1}{N^2}N^\mu N^\nu & -\frac{1}{N^2}N^\mu\\[2mm]
          -\frac{1}{N^2}N^\nu & \frac{1}{N^2}
          \end{array}\right)
        \;. \label{tg^ab2}
\end{eqnarray}
Comparison of equations (\ref{tg^ab2}) and (\ref{ktg^ab}) leads to equation (\ref{A^mu}). The second equality in equation (\ref{phi}) follows from equation (\ref{w^a}). {\em End of Proof.}

{\sc Proposition 4.} 
\begin{eqnarray}
	A_\rho A^\rho=\frac{N_\rho N^\rho}{N^2\left(N^2+N_\rho N^\rho\right)} \;, \label{AA}
\end{eqnarray}
\begin{eqnarray}
	A_\rho A^\rho+\frac{1}{\phi^2}=\frac{1}{N^2} \;, \label{AA_pp}
\end{eqnarray}
which are directly derived from equations (\ref{phi})---(\ref{A^mu}).

{\sc Proposition 5.} The $A^a$ is related to the $N^a$ by
\begin{eqnarray}
	A^a=\frac{1}{N^2+N_\rho N^\rho}\left(N^a-\frac{N_\rho N^\rho}{N}n^a\right) \;. \label{Aa_Na}
\end{eqnarray}

{\em Proof.} By equations (\ref{A^a}), (\ref{A^mu}), and (\ref{w^a}), we get
\begin{eqnarray}
	A^a=\frac{N^\mu}{N^2}\left(\frac{\partial}{\partial x^\mu}\right)^a-A_\rho A^\rho\left(Nn^a+N^a\right) \;.
\end{eqnarray}
Then, since $N^a=N^\mu (\partial/\partial x^\mu)^a$, by equation (\ref{AA}) we get
\begin{eqnarray}
	A^a=\frac{N^a}{N^2}-\frac{N_\rho N^\rho}{N^2\left(N^2+N_\rho N^\rho\right)}\left(Nn^a+N^a\right) \;,
\end{eqnarray}
which then leads to equation (\ref{Aa_Na}). {\em End of Proof.}

{\sc Proposition 6.} 
\begin{eqnarray}
	A^an_a=-\frac{N_\rho N^\rho}{N\left(N^2+N_\rho N^\rho\right)} =-NA_\rho A^\rho\;. \label{Aa_na}
\end{eqnarray}
The first identity follows from equation (\ref{Aa_Na}) since $n_a N^a=0$ and $n_an^a=1$. The second follows from equation (\ref{AA}).

Since $N\neq 0$, equation (\ref{Aa_na}) indicates that $A^an_a=0$ (i.e., $A^a$ is a vector tangent to the 4-manifold ${\cal M}$) if and only if (iff) $N_\rho N^\rho=0$, or equivalently, $A_\rho A^\rho=0$. Then, by equation (\ref{AA_AA}), we have

{\sc Proposition 7.}  The $A^a$ is a vector tangent to ${\cal M}$ iff $A^a$ is null, i.e., iff
\begin{eqnarray}
	A^aA_a=0 \;. \label{null2}
\end{eqnarray}
or equivalently,
\begin{eqnarray}
	N^aN_a=0 \;, \label{null1}
\end{eqnarray}

{\sc Proposition 8.}
\begin{eqnarray}
	\!\!\!~_kg_{ab}=\!\!~_kg_{\mu\nu}dx^\mu_a dx^\nu_b\;, \label{g_ab}
\end{eqnarray}
where $\mu,\nu=0,1,2,3$,
\begin{eqnarray}
	\!\!\!~_kg^{ab} = \t{g}^{ac}\t{g}^{bd}\!\!~_kg_{cd} = \!\!~_kg^{\mu\nu}\left(\frac{\partial}{\partial x^\mu}\right)^a\left(\frac{\partial}{\partial x^\nu}\right)^b-2A^\mu\left(\frac{\partial}{\partial x^\mu}\right)^{(a}\left(\frac{\partial}{\partial w}\right)^{b)} +A_\rho A^\rho\left(\frac{\partial}{\partial w}\right)^a\left(\frac{\partial}{\partial w}\right)^b\;, \label{g^ab}
\end{eqnarray}
and
\begin{eqnarray}
	\!\!~_kg_a^{\;\,c} &\equiv& \!\!~_kg_{ab}\!\!~_kg^{bc} = \t{g}^{cb}\!\!~_kg_{ab}= dx^\mu_a\left(\frac{\partial}{\partial x^\mu}\right)^c-A_\mu dx^\mu_a\left(\frac{\partial}{\partial w}\right)^c \;. \label{g_a^b}
\end{eqnarray}

{\em Proof.} By equations (\ref{kgab_w}), (\ref{A^a}), and (\ref{g_A_d}), we get
\begin{eqnarray}
	A_a = \!\!~_kg_{ab}A^b = \left(A_\mu-A_\rho A^\rho\!\!\!~_kg_{\mu 4}\right)dx^\mu_a +\left(A^\rho\!\!\!~_kg_{4\rho}-A_\rho A^\rho\!\!\!~_kg_{44}\right)dw_a \;.
\end{eqnarray}
Comparison with equation (\ref{A_a}) leads to
\begin{eqnarray}
	\!\!~_kg_{\mu4}=\!\!~_kg_{4\mu}=0 \;, \hspace{1cm} \!\!\!~_kg_{44}=0 \;.
\end{eqnarray}
Then equation (\ref{g_ab}) is proved. Equations (\ref{g^ab}) and (\ref{g_a^b}) are then derived by the application of equation (\ref{ktg^ab1}).
{\em End of Proof.}

{\sc Proposition 9.}
\begin{eqnarray}
	\!\!~_kg_{ab} n^b=-NA_a \;. \label{gab_n_k}
\end{eqnarray}

{\em Proof.} By equations (\ref{w^a}) and (\ref{g_ab}), we have
\begin{eqnarray}
	\!\!~_kg_{ab} n^b =\!\!~_kg_{\mu\nu}dx^\mu_a dx^\nu_b \frac{1}{N}\left[\left(\frac{\partial}{\partial w}\right)^b-N^\rho\left(\frac{\partial}{\partial x^\rho}\right)^b\right] = -\frac{1}{N}\!\!~_kg_{\mu\nu}N^\nu dx^\mu_a \;.
\end{eqnarray}
Then, by equations (\ref{A^mu}) and (\ref{A_a}), we get
\begin{eqnarray}
	\!\!~_kg_{ab} n^b=-N\!\!~_kg_{\mu\nu}A^\nu dx^\mu_a =-NA_\mu dx^\mu_a \;.
\end{eqnarray}
{\em End of Proof.}

Equation (\ref{gab_n_k}) leads to the conclusion that $\!\!~_kg_{ab}$ is not a 4-metric tensor field on the hypersurface ${\cal M}$, unless $A_a=0$ (or, equivalently, $N^a=0$). In fact, we have:

{\sc Proposition 10.} The $\!\!~_kg_{ab}$ is a 4-metric tensor on a hypersurface ${\cal M}_k$ orthogonal to $w^a$ (if $w^a$ is hypersurface orthogonal), induced from the 5-metric tensor in the bulk spacetime. The $A^a$ is a vector tangent to ${\cal M}_k$. That is, if we define $\hat{w}^a=w^a/\left(w_cw^c\right)^{1/2}$ (so that $\hat{w}_a\hat{w}^a=1$), we have
\begin{eqnarray}
	\!\!~_kg_{ab}=\t{g}_{ab}-\hat{w}_a\hat{w}_b \;, \label{kgab_uu}
\end{eqnarray}
and
\begin{eqnarray}
	A^a \hat{w}_a=0 \;. \label{Aa_ua}
\end{eqnarray}

{\em Proof.} By equations (\ref{g^ab}), (\ref{ktg^ab1}), and (\ref{phi}), we have
\begin{eqnarray}
	\!\!~_kg^{ab} = \t{g}^{ab}-\frac{1}{\phi^2}\left(\frac{\partial}{\partial w}\right)^a\left(\frac{\partial}{\partial w}\right)^b = \t{g}^{ab}-\hat{w}^a\hat{w}^b\;, 
\end{eqnarray}
since $\hat{w}^a=w^a/\phi$ by equation (\ref{phi}). By equation (\ref{A_a}), we have $A_aw^a=0$ and hence the equation (\ref{Aa_ua}).
{\em End of Proof.}

The metric $g_{ab}$ on ${\cal M}$ is induced from the five-dimensional metric $\t{g}_{ab}$ by $g_{ab}=\t{g}_{ab}-n_an_b$, as discussed in Sec.~\ref{4+1}. The two four-dimensional metrics $\!\!\!~_kg_{ab}$ and $g_{ab}$ are related by 
\begin{eqnarray}
	\!\!\!~_kg_{ab}=g_{ab}+n_an_b-\hat{w}_a\hat{w}_b \;. \label{g_g}
\end{eqnarray}
By equation (\ref{g_g}), we have $g_{\mu\nu}=\!\!~_kg_{\mu\nu}+\hat{w}_\mu\hat{w}_\nu$ for $n_\mu=0$. Since $w_\mu=\t{g}_{\mu w}=N_\mu$ by equation (\ref{tg_ab2}), and $w_aw^a=N^2+N_\rho N^\rho$, by Proposition 3 we have $\hat{w}_\mu\hat{w}_\nu=\phi^2A_\mu A_\nu$. Hence, $g_{\mu\nu}=\!\!~_kg_{\mu\nu}+\phi^2A_\mu A_\nu$, which confirms equation (\ref{ds_4d}).

The $A^a$ is a vector on ${\cal M}_k$. The $N^a$ is a vector on ${\cal M}$. The ${\cal M}_k$ and ${\cal M}$ are respectively orthogonal to $w^a$ and $n^a$. They intersect at a three-dimensional manifold, which we denote by $\Sigma$. The $\Sigma$ is a hypersurface in ${\cal M}$ (and ${\cal M}_k$). The above results are illustrated in Fig.~\ref{kk2}.

{\sc Proposition 11.} When $N_aN^a\neq 0$ (equivalent to $A_aA^a\neq 0$ by eq.~\ref{AA}), $N^a$ and $A^a$ are two independent normals to $\Sigma$.

{\em Proof.} For any $v^a\in{\cal T}(\Sigma)$, we have $N_av^a=(w_a-Nn_a)v^a=w_av^a=0$, where the last equality follows from the fact that $\Sigma$ is also a hypersurface in ${\cal M}_k$ hence $v^a\in{\cal T}({\cal M}_k)$. Then, by equation (\ref{Aa_Na}), we have
\begin{eqnarray}
	A_av^a=-\frac{N_\rho N^\rho}{N\left(N^2+N_\rho N^\rho\right)}n_av^a=0 \;,
\end{eqnarray}
since $v^a\in\tm$. Hence, both $N^a$ and $A^a$ are normals to $\Sigma$. By equation (\ref{Aa_Na}), $A^a$ and $N^a$ are independent iff $N_aN^a\neq 0$. {\em End of Proof.}

{\sc Proposition 12.} When $N_aN^a= 0$ (equivalent to $A_aA^a=0$ by eq.~\ref{AA}), $N^a\in{\cal T}(\Sigma)$ and $A^a\in{\cal T}(\Sigma)$.

{\em Proof.} When $N_aN^a=0$, by equation (\ref{Aa_na}) we have $A^an_a=0$ so $A^a\in\tm$. Since $A^a\in{\cal M}_k$ also, we must have $A^a\in{\cal T}(\Sigma)$. By equation (\ref{Aa_Na}) we must also have $N^a\in{\cal T}(\Sigma)$. (See also Proposition 7.) {\em End of Proof.}

\section{$n+1$ Decomposition of Einstein's Field Equations}
\label{decom}

In this Appendix we discuss direct decomposition of the $(n+1)$-dimensional Einstein field equation, and derive the equivalent equations in an $n$-dimensional spacetime. To make the results general, we assume that the $(n+1)$-dimensional spacetime contains a matter field represented by an $(n+1)$-dimensional stress-energy tensor $\t{T}_{ab}$.

The Einstein field equation on an $(n+1)$-dimensional spacetime $(\t{\cal M},\t{g}_{ab})$ is
\begin{eqnarray}
	\t{G}_{ab}\equiv\t{R}_{ab}-\frac{1}{2}\t{R}\t{g}_{ab}=\t{\kappa} \t{T}_{ab} \;, \label{tEin_eq}
\end{eqnarray}
where $\t{\kappa}$ is the gravitational coupling constant.

Contraction of equation (\ref{tEin_eq}) with $\t{g}^{ab}$ leads to
\begin{eqnarray}
	\t{R}=-\frac{2\t{\kappa}}{n-1}\t{T} \;, \label{tR_eq}
\end{eqnarray}
where $\t{T}\equiv \t{g}^{ab}\t{T}_{ab}$ is the trace of the stress-energy tensor. Hence, the Einstein field equation (\ref{tEin_eq}) can be written in an equivalent form
\begin{eqnarray}
	\t{R}_{ab}=\t{\kappa}\left(\t{T}_{ab}-\frac{1}{n-1}\t{T}\t{g}_{ab}\right) \;. \label{tRab_eq}
\end{eqnarray}

By equation (\ref{g_tg}), the $(n+1)$-dimensional metric tensor $\t{g}_{ab}$ can be decomposed into components tangent and orthogonal to ${\cal M}$ by
\begin{eqnarray}
	\t{g}_{ab}=g_{ab}+n_an_b \;,
\end{eqnarray}
where $g_{ab}n^a=0$, and $\t{g}_{ab}n^an^b=1$. An $(n+1)$-dimensional tensor $\t{\cal T}_{ab}$ can be decomposed into components tangent and orthogonal to ${\cal M}$ by
\begin{eqnarray}
	\t{\cal T}_{ab} = \t{g}_a^{\;\;c}\t{g}_b^{\;\;d}\t{\cal T}_{cd}=g_a^{\;\;c}g_b^{\;\;d}\t{\cal T}_{cd}+n_a\left(g_b^{\;\;d}n^c\t{\cal T}_{cd}\right) +n_b\left(g_a^{\;\;c}n^d\t{\cal T}_{cd}\right) +n_an_b\left(n^cn^d\t{\cal T}_{cd}\right) \;. \label{cT_decom}
\end{eqnarray}

Applying the above decomposition mechanism to the $(n+1)$-dimensional Einstein field equation (\ref{tEin_eq}), we get the following three independent equations on ${\cal M}$:
\begin{eqnarray}
	g_a^{\;\;c}g_b^{\;\;d}\t{G}_{cd}=\t{\kappa}\,g_a^{\;\;c}g_b^{\;\;d}\t{T}_{cd} \;, \label{proj_1G}
\end{eqnarray}
a tensor equation obtained by full projection of the $(n+1)$-dimensional Einstein field equation onto ${\cal M}$;
\begin{eqnarray}
        g_a^{\;\;d}n^c\t{G}_{cd} = \t{\kappa}\,g_a^{\;\;d}n^c\t{T}_{cd} \;, \label{proj_2G}
\end{eqnarray}
a vector equation obtained from the $(n+1)$-dimensional Einstein field equation with one index projected onto $n^a$ and the other index projected onto ${\cal M}$; and
\begin{eqnarray}
        \t{G}_{cd}n^cn^d=\t{\kappa}\,\t{T}_{cd}n^cn^d \;, \label{proj_3G}
\end{eqnarray}
a scalar equation obtained by full projection of the $(n+1)$-dimensional Einstein field equation onto $n^a$.

Alternatively, application of the decomposition mechanism to the $(n+1)$-dimensional Einstein field equation (\ref{tRab_eq}) leads to the following three independent equations on ${\cal M}$:
\begin{eqnarray}
	g_a^{\;\;c}g_b^{\;\;d}\t{R}_{cd}=\t{\kappa}\left(g_a^{\;\;c}g_b^{\;\;d}\t{T}_{cd}-\frac{1}{n-1}\t{T}g_{ab}\right) \;, \label{proj_1}
\end{eqnarray}
\begin{eqnarray}
        g_a^{\;\;d}n^c\t{R}_{cd} = \t{\kappa}g_a^{\;\;d}n^c\t{T}_{cd} \;, \label{proj_2}
\end{eqnarray}
and
\begin{eqnarray}
        \t{R}_{cd}n^cn^d=\t{\kappa}\left(\t{T}_{cd}n^cn^d-\frac{1}{n-1}\t{T}\right) \;. \label{proj_3}
\end{eqnarray}

The complete set of field equations on ${\cal M}$ must contain a scalar equation (eq.~\ref{proj_3G} or \ref{proj_3}), a vector equation (eq.~\ref{proj_2G} or \ref{proj_2}), and a tensor equation (eq.~\ref{proj_1G} or \ref{proj_1}). In fact, equations (\ref{proj_2G}) and (\ref{proj_2}) are equivalent, since $g_a^{\;\;d}n^c\t{G}_{cd}=g_a^{\;\;d}n^c\t{R}_{cd}$.

\subsection{The scalar equation}

For $n_an^a=1$, the Riemann tensor on ${\cal M}$ is related to that on $\t{\cal M}$ by \cite{wal84,mis73}
\begin{eqnarray}
	R_{abc}^{\;\;\;\;\;\;d} = g_a^{\;\;e}g_b^{\;\;f}g_c^{\;\;g}g^d_{\;\;h}\t{R}_{efg}^{\;\;\;\;\;\;h} +\left(K_{ac}K_b^{\;\;d}-K_{bc}K_a^{\;\;d}\right) \;. \label{Rc_Rc}
\end{eqnarray}
From equations (\ref{g_tg}) and (\ref{Rc_Rc}), we get
\begin{eqnarray}
	R_{ac} = g_a^{\;\;b}g_c^{\;\;d}\left(\t{R}_{bd}-n^en^f\t{R}_{bedf}\right) -\left(K_a^{\;\;b}K_{bc}-KK_{ac}\right) \;, \label{Rr_Rr}
\end{eqnarray}
and
\begin{eqnarray}
	R = \t{R}-2\t{R}_{ab}n^an^b -\left(K_{ab}K^{ab}-K^2\right) \;. \label{R_R}
\end{eqnarray}
Equation (\ref{R_R}) is equivalent to
\begin{eqnarray}
	\t{G}_{ab}n^an^b=-\frac{1}{2}\left(R +K_{ab}K^{ab}-K^2\right) \;. \label{G_R}
\end{eqnarray}

Substituting equation (\ref{G_R}) into equation (\ref{proj_3G}), we get the scalar equation on ${\cal M}$
\begin{eqnarray}
        R +K_{ab}K^{ab}-K^2 = -2\t{\kappa}\t{T}_{ab}n^an^b \;. \label{scalar_eq}
\end{eqnarray}
When $\t{T}_{ab}=0$, equation (\ref{scalar_eq}) is equivalent to equation (\ref{scalar_eq_LK}).

By the definition of Riemann tensor and the definition of $K_{ab}$, we have
\begin{eqnarray}
	\t{R}_{ab}n^an^b = -n^a\t{g}^{cd}\left(\t{\nabla}_a\t{\nabla}_c-\t{\nabla}_c\t{\nabla}_a\right)n_d = \t{\nabla}_an^a\t{\nabla}_cn^c -\t{\nabla}_cn^a\t{\nabla}_an^c-\t{\nabla}_av^a = -K_{ab}K^{ab}+K^2-\t{\nabla}_av^a \;, \label{R_K}
\end{eqnarray}
where $v^a$ is defined by equation (\ref{v^a}). Substituting equation (\ref{v^a}) into equation (\ref{R_K}), we get
\begin{eqnarray}
	\t{R}_{ab}n^an^b = -K_{ab}K^{ab}-\t{\pounds}_nK+\t{\nabla}_aa^a \;, \label{Rnn_x1}
\end{eqnarray}
where $a^a$ is defined by equation (\ref{aa_def}). Since
\begin{eqnarray}
	\t{\pounds}_nK=g^{ab}\t{\pounds}_nK_{ab}-2K_{ab}K^{ab} \;, \label{Rnn_x2}
\end{eqnarray}
and
\begin{eqnarray}
	\t{\nabla}_aa^a=\nabla_aa^a-a_aa^a \;,
\end{eqnarray}
from equation (\ref{Rnn_x1}) we get
\begin{eqnarray}
	\t{R}_{ab}n^an^b = -g^{ab}\t{\pounds}_nK_{ab}+K_{ab}K^{ab}+\nabla_aa^a-a_aa^a \;. \label{Rnn_K}
\end{eqnarray}

Substituting equation (\ref{Rnn_K}) into equation (\ref{proj_3}), we get another scalar equation on ${\cal M}$
\begin{eqnarray}
	g^{ab}\t{\pounds}_nK_{ab}-K_{ab}K^{ab}-\nabla_aa^a+a_aa^a =-\t{\kappa}\left(\t{T}_{ab}n^an^b-\frac{1}{n-1}\t{T}\right) \;. \label{scalar_eq2}
\end{eqnarray}
When $\t{T}_{ab}=0$, equation (\ref{scalar_eq2}) is equivalent to equation (\ref{scalar_eq_LKb}).

By equations (\ref{G_R}) and (\ref{R_K}), we get
\begin{eqnarray}
	\frac{1}{2}\t{R} = \left(\t{R}_{ab}-\t{G}_{ab}\right)n^an^b =\frac{1}{2}\left(R-K_{ab}K^{ab}+K^2\right)-\t{\nabla}_av^a \;,
\end{eqnarray}
which agrees with the equation (\ref{R_R2}).

\subsection{The vector equation}

By the definition of $K_{ab}$ (eq.~\ref{Kab}), we get
\begin{eqnarray}
	\nabla_aK^a_{\;\;b}-\nabla_bK = g_b^{\;\;d}g^{ce}\left(\t{\nabla}_c\t{\nabla}_d-\t{\nabla}_d\t{\nabla}_c\right)n_e = g_b^{\;\;d}g^{ce}\t{R}_{cde}^{\;\;\;\;\;\;f}n_f \;.
\end{eqnarray}
Hence, we have
\begin{eqnarray}
	\nabla_aK^a_{\;\;b}-\nabla_bK = g_b^{\;\;c}\t{R}_{cd}n^d \;, \label{D_Kab}
\end{eqnarray}
since $\t{R}_{cde}^{\;\;\;\;\;\;f}n_fn^e=0$ according to properties of the Riemann tensor. Equations (\ref{Rc_Rc}) and (\ref{D_Kab}) are called the Gauss-Codacci relations \cite{wal84}.

Substituting equation (\ref{D_Kab}) into equation (\ref{proj_2}), we get the vector equation on ${\cal M}$
\begin{eqnarray}
        \nabla_aK^a_{\;\;b}-\nabla_bK = \t{\kappa}g_b^{\;\;a}\t{T}_{ac}n^c \;. \label{vector_eq}
\end{eqnarray}
When $\t{T}_{ab}=0$, equation (\ref{vector_eq}) is equivalent to equation (\ref{emfield_eqK}).

Since $g_b^{\;\;c}\t{G}_{cd}n^d=g_b^{\;\;c}\t{R}_{cd}n^d$, equation (\ref{proj_2G}) leads to the same vector equation (\ref{vector_eq}).

\subsection{The tensor equation}

From equations (\ref{Rr_Rr}) and (\ref{R_R}) we get
\begin{eqnarray}
	G_{ac} = g_a^{\;\;b}g_c^{\;\;d}\t{G}_{bd} +n^bn^d\t{R}_{bd}g_{ac} -g_a^{\;\;b}g_c^{\;\;d}n^en^f\t{R}_{bedf} -\left(K_a^{\;\;b}K_{bc}-KK_{ac}\right)+\frac{1}{2}\left(K_{bd}K^{bd}-K^2\right)g_{ac} \;. 
         \label{RRg}
\end{eqnarray}
The Riemann tensor on the $(n+1)$-dimensional $\t{\cal M}$ can be written as ($n+1\ge3$, \cite{wal84})
\begin{eqnarray}
	\t{R}_{abcd} = \t{C}_{abcd}+\frac{2}{n-1}\left(\t{g}_{a[c}\t{R}_{d]b}-\t{g}_{b[c}\t{R}_{d]a}\right) -\frac{2}{n(n-1)}\t{R}\t{g}_{a[c}\t{g}_{d]b} \;, \label{weyl}
\end{eqnarray}
where $\t{C}_{abcd}$ is the traceless Weyl tensor on $\t{\cal M}$. Substituting equation (\ref{weyl}) into equation (\ref{RRg}), we get
\begin{eqnarray}
	G_{ab} = \frac{n-2}{n-1}\left[g_a^{\;\;c}g_b^{\;\;d}\t{G}_{cd}+\left(n^cn^d\t{R}_{cd}-\frac{1}{2n}\t{R}\right)g_{ab}\right] -\left(K_a^{\;\;c}K_{cb}-KK_{ab}\right)+\frac{1}{2}\left(K_{cd}K^{cd}-K^2\right)g_{ab} -E_{ab}\;, \label{RRg2}
\end{eqnarray}
where 
\begin{eqnarray}
	E_{ab}\equiv g_a^{\;\;c}g_b^{\;\;d}n^en^f\t{C}_{cedf} \;, \hspace{1cm}g^{ab}E_{ab}=0 \;. \label{Eab}
\end{eqnarray}

If we substitute equations (\ref{tEin_eq})--(\ref{tRab_eq}) into equation (\ref{RRg2}), we get a tensor equation that agrees with the eq.~8 in \cite{shi00}. However, before doing the substitution, we should express $E_{ab}$ in $K_{ab}$ and its derivatives. This is necessary since, as we will see, the expression for $E_{ab}$ contains $\t{R}_{ab}$ and hence $\t{T}_{ab}$.

By the definition of Riemann tensor, we have
\begin{eqnarray}
	\t{R}_{cedf}n^f=\left(\t{\nabla}_c\t{\nabla}_e-\t{\nabla}_e\t{\nabla}_c\right)n_d \;.
\end{eqnarray}
Then, by equation (\ref{nab_n}), we get
\begin{eqnarray}
	\t{R}_{cedf}n^en^f = -K_{ed}K_c^{\;\;e}-K_{ed}n_ca^e+\t{\nabla}_ca_d-n^e\t{\nabla}_eK_{cd} -n_cn^e\t{\nabla}_ea_d-a_ca_d \;. \label{riem_nn}
\end{eqnarray}
Since
\begin{eqnarray}
	n^c\t{\nabla}_cK_{ab} = \t{\pounds}_nK_{ab}-2K_{ac}K_b^{\;\;c}-K_{ac}n_ba^c-K_{bc}n_aa^c \;, 
\end{eqnarray}
from equation (\ref{riem_nn}) we get
\begin{eqnarray}
	g_a^{\;\;c}g_b^{\;\;d}n^en^f\t{R}_{cedf} = -g_a^{\;\;c}g_b^{\;\;d}\t{\pounds}_nK_{cd}+K_{ac}K_b^{\;\;c} +\nabla_{(a}a_{b)}-a_aa_b \;, \label{Riemann_nn}
\end{eqnarray}
where the indexes $a$ and $b$ of $\nabla_aa_b$ are symmetrized since the left-hand side of the equation is symmetric with respect to them. In fact, it can be verified that $\nabla_aa_b=\nabla_ba_a$.

Then, by equations (\ref{weyl}), (\ref{Eab}), and (\ref{Riemann_nn}), we get
\begin{eqnarray}
	E_{ab} = -g_a^{\;\;c}g_b^{\;\;d}\t{\pounds}_nK_{cd}+K_{ac}K_b^{\;\;c}+\nabla_{(a}a_{b)}-a_aa_b -\frac{1}{n-1}\left(g_a^{\;\;c}g_b^{\;\;d}\t{R}_{cd}+n^cn^d\t{R}_{cd}g_{ab}\right) +\frac{1}{n(n-1)}\t{R}g_{ab} \;. \label{Eab_K}
\end{eqnarray}
Equation (\ref{Eab_K}) and the identity $g^{ab}E_{ab}=0$ leads to equation (\ref{Rnn_K}).

Substituting equation (\ref{Eab_K}) into equation (\ref{RRg2}), we get
\begin{eqnarray}
	G_{ab} &=& g_a^{\;\;c}g_b^{\;\;d}\t{G}_{cd}+n^cn^d\t{R}_{cd}g_{ab}-\left(2K_a^{\;\;c}K_{cb}-KK_{ab}\right) +\frac{1}{2}\left(K_{cd}K^{cd}-K^2\right)g_{ab} \nonumber\\
        && +g_a^{\;\;c}g_b^{\;\;d}\t{\pounds}_nK_{cd} -\nabla_{(a}a_{b)}+a_aa_b \;. \label{RRg3}
\end{eqnarray}
Substituting equation (\ref{Rnn_K}) into equation (\ref{RRg3}), we get
\begin{eqnarray}
	G_{ab} &=& g_a^{\;\;c}g_b^{\;\;d}\t{G}_{cd}-\left(2K_a^{\;\;c}K_{cb}-KK_{ab}\right) +\frac{1}{2}\left(3K_{cd}K^{cd}-K^2\right)g_{ab} +\left(g_a^{\;\;c}g_b^{\;\;d}-g_{ab}g^{cd}\right)\t{\pounds}_nK_{cd} \nonumber\\
        && -\nabla_{(a}a_{b)}+a_aa_b +\left(\nabla_ca^c-a_ca^c\right)g_{ab} \;, \label{RRg4}
\end{eqnarray}
from which we get the full projection of $\t{G}_{ab}$ onto ${\cal M}$
\begin{eqnarray}
	g_a^{\;\;c}g_b^{\;\;d}\t{G}_{cd} &=& G_{ab} +\left(2K_a^{\;\;c}K_{cb}-KK_{ab}\right) -\frac{1}{2}\left(3K_{cd}K^{cd}-K^2\right)g_{ab} -\left(g_a^{\;\;c}g_b^{\;\;d}-g_{ab}g^{cd}\right)\t{\pounds}_nK_{cd} \nonumber\\
        &&+\nabla_{(a}a_{b)}-a_aa_b -\left(\nabla_ca^c-a_ca^c\right)g_{ab} \;. \label{RRg4a}
\end{eqnarray}

Substituting the Einstein field equation (\ref{tEin_eq}) into equation (\ref{RRg4}), we get the tensor field equation on ${\cal M}$
\begin{eqnarray}
	G_{ab} &=& \t{\kappa} g_a^{\;\;c}g_b^{\;\;d}\t{T}_{cd}-\left(2K_a^{\;\;c}K_{cb}-KK_{ab}\right) +\frac{1}{2}\left(3K_{cd}K^{cd}-K^2\right)g_{ab} +\left(g_a^{\;\;c}g_b^{\;\;d}-g_{ab}g^{cd}\right)\t{\pounds}_nK_{cd} \nonumber\\
        && -\nabla_{(a}a_{b)}+a_aa_b +\left(\nabla_ca^c-a_ca^c\right)g_{ab} \;. \label{tensor_eq}
\end{eqnarray}
The right-hand side of equation (\ref{tensor_eq}) can be interpreted as representing the stress-energy tensor of matter on ${\cal M}$. When $\t{T}_{ab}=0$, equation (\ref{tensor_eq}) is equivalent to the $n$-dimensional Einstein field equation (\ref{ein_eq_n}) with the $T_{ab}$ given by equation (\ref{Tm+Tint+Tem5s}).

The tensor equation (\ref{tensor_eq}) is obtained by full projection of the $(n+1)$-dimensional Einstein field equation onto ${\cal M}$.

Substituting equation (\ref{scalar_eq2}) into equation (\ref{tensor_eq}), we get another version of the tensor equation on ${\cal M}$
\begin{eqnarray}
	G_{ab} &=& \t{\kappa}\left[g_a^{\;\;c}g_b^{\;\;d}\t{T}_{cd}+\left(\t{T}_{cd}n^cn^d-\frac{1}{n-1}\t{T}\right)g_{ab}\right] -\left(2K_a^{\;\;c}K_{cb}-KK_{ab}\right)+\frac{1}{2}\left(K_{cd}K^{cd}-K^2\right)g_{ab} \nonumber\\
        &&+g_a^{\;\;c}g_b^{\;\;d}\t{\pounds}_nK_{cd}-\nabla_{(a}a_{b)}+a_aa_b \;. \label{tensor_eq2}
\end{eqnarray}
Equation (\ref{tensor_eq2}) can also be obtained by directly substituting equations (\ref{tEin_eq}) and (\ref{tRab_eq}) into equation (\ref{RRg3}).

\section{The Pseudo-Hamiltonian Formulation}
\label{hamilton}

The procedure that we have used to derive field equations on ${\cal M}$ is very similar to that in the Hamiltonian formulation of general relativity: define a hypersurface in a $(n+1)$-dimensional spacetime by the coordinate $w=\mbox{const}$ then decompose the metric and curvature tensors in terms of $N$, $N_a$, and the metric $g_{ab}$ on the hypersurface induced from the metric tensor in the $(n+1)$-dimensional bulk spacetime \cite{wal84,arn62,mis73}. However, unlike in the Hamiltonian formulation, the $w$-coordinate used in this paper is a space coordinate instead of a time coordinate. Even so, it does not prohibit us from formulating the problem in a pseudo-Hamiltonian way. In this Appendix we derive the field equations by the pseudo-Hamiltonian formulation and confirm the results that we have obtained in Secs.~\ref{field_eqs_I} and \ref{field_eqs_II}.

The Lagrangian density in equation (\ref{LG}) contains independent variables of $g_{ab}$, $N$, and $N_a$. However, only $g_{ab}$ appears as a ``dynamical'' variable, since $\t{L}_G$ contains the $w$-derivative of $g_{ab}$ only. The momentum canonically conjugate to $g_{ab}$ is
\begin{eqnarray}
	\pi^{ab} \equiv \frac{\partial\t{L}_G}{\partial\dot{g}_{ab}} = -\sqrt{-g}\left(K^{ab}-Kg^{ab}\right)= \sqrt{-g}\Pi^{ab} \;, \label{pi_ab}
\end{eqnarray}
where $\Pi_{ab}$ is defined by equation (\ref{Pi_ab_K}). 

Since $\t{L}_G$ does not contain any $w$-derivative of $N$ and $N_a$, their conjugate momenta vanish identically. Hence, we get the pseudo-Hamiltonian density
\begin{eqnarray}
	\t{\cal H}_G = \pi^{ab}\dot{g}_{ab}-\t{\cal L}_G = -\sqrt{-g}NR-\frac{N}{\sqrt{-g}}\left(\pi_{ab}\pi^{ab}-\frac{1}{n-1}\pi^2\right) +\pi^{ab}M_{ab} \;, \label{HG}
\end{eqnarray}
where
\begin{eqnarray}
	\pi\equiv g_{ab}\pi^{ab}=\sqrt{-g}(n-1)K \;.
\end{eqnarray}

The Hamiltonian is defined by
\begin{eqnarray}
	\t{H}_G\equiv\int_{\cal M} \t{\cal H}_G\left(g_{ab},\pi^{ab},N,N_a\right)\,{\bf e} \;. \label{HHG}
\end{eqnarray}
Variation of $\t{H}_G$ with respect to $N$ and $N_a$ leads to two constraint equations:
\begin{eqnarray}
	0=\frac{\delta\t{H}_G}{\delta N} \;,\hspace{1cm}
        0=\frac{\delta\t{H}_G}{\delta N_a} \;. \label{con_eq}
\end{eqnarray}
Variation of $\t{H}_G$ with respect to $g_{ab}$ and $\pi^{ab}$ leads to two ``dynamical'' equations:
\begin{eqnarray}
	\dot{g}_{ab}=\frac{\delta\t{H}_G}{\delta\pi^{ab}} \;,\hspace{1cm}
        \dot{\pi}^{ab}=-\frac{\delta\t{H}_G}{\delta\g_{ab}} \;. \label{dyn_eq}
\end{eqnarray}

\subsection{Variation with respect to $N$ and $N_a$}

Variation of $\t{H}_G$ with respect to $N$ leads to the first constraint equation
\begin{eqnarray}
	R-g^{-1}\left(\pi_{ab}\pi^{ab}-\frac{1}{n-1}\pi^2\right) = 0 \;, \label{con1}
\end{eqnarray}
which agrees with the scalar equation (\ref{scalar_eq_L}).

By equation (\ref{HG}), for variation with respect to $N_a$ we get
\begin{eqnarray}
	\delta\t{\cal H}_G = \pi^{ab}\delta M_{ab}\doteq -2\sqrt{-g}\left(\nabla_a\Pi^{ab}\right)\delta N_b \;.
\end{eqnarray}
Then, by equation (\ref{con_eq}), variation of $\t{H}_G$ with respect to $N$ leads to the second constraint equation
\begin{eqnarray}
	 \nabla_a\left(\frac{1}{\sqrt{-g}}\pi^{ab}\right) = 0 \;, \label{con2}
\end{eqnarray}
which agrees with the vector equation (\ref{dPi_ab_0}).

\subsection{Variation with respect to $\pi^{ab}$ and $g_{ab}$}

By equation (\ref{HG}), for variation with respect to $\pi^{ab}$ we get
\begin{eqnarray}
	\delta\t{\cal H}_G &=&  -\frac{2N}{\sqrt{-g}}\left(\pi_{ab}-\frac{\pi}{n-1}g_{ab}\right)\delta\pi^{ab}+M_{ab}\delta\pi^{ab} \;. 
\end{eqnarray}
Then, by equation (\ref{dyn_eq}), we get
\begin{eqnarray}
	\dot{g}_{ab}=-\frac{2N}{\sqrt{-g}}\left(\pi_{ab}-\frac{1}{n-1}\pi g_{ab}\right)+M_{ab} \;. \label{dyn1}
\end{eqnarray}
This is just the $\dot{g}_{ab}$ derived from equation (\ref{Kab_M}) and the definition of $\pi^{ab}$.

For variation with respect to $g_{ab}$, by equation (\ref{HG}) we get
\begin{eqnarray}
	\delta\t{\cal H}_G = -N\delta\left(\sqrt{-g}R\right)+\pi^{ab}\delta M_{ab} -N\delta\left[\frac{1}{\sqrt{-g}}\left(\pi_{ab}\pi^{ab}-\frac{1}{n-1}\pi^2\right)\right] \;. \label{dH}
\end{eqnarray}
Evaluation of the first term of variation leads to
\begin{eqnarray}
	-N\delta\left(\sqrt{-g}R\right) \doteq \sqrt{-g}\left(NG^{ab} -\nabla^a\nabla^bN+g^{ab}\nabla_c\nabla^c N\right)\delta g_{ab} \;. \label{dH1}
\end{eqnarray}
By the definition of $M_{ab}$, we have
\begin{eqnarray}
	\pi^{ab}\delta M_{ab} \doteq \sqrt{-g}\nabla_c\left\{\frac{1}{\sqrt{-g}}\left[2N^{(a}\pi^{b)c}-N^c\pi^{ab}\right]\right\}\delta g_{ab} \;. \label{dH2}
\end{eqnarray}

Since $\delta\pi^{ab}=0$ when the variation is taken with respect to $g_{ab}$, we have
\begin{eqnarray}
	\delta\left(\pi_{ab}\pi^{ab}\right)=\pi^{ab}\delta\pi_{ab} =2\pi^a_{\;\;c}\pi^{bc}\delta g_{ab} \;, \hspace{1cm} \delta\pi^2=2\pi\delta\pi=2\pi\pi^{ab}\delta g_{ab} \;.
\end{eqnarray}
Hence,
\begin{eqnarray}
	\delta\left[\frac{1}{\sqrt{-g}}\left(\pi_{ab}\pi^{ab}-\frac{1}{n-1}\pi^2\right)\right] = \frac{2}{\sqrt{-g}}\left(\pi^a_{\;\;c}\pi^{bc}-\frac{1}{n-1}\pi\pi^{ab}\right)\delta g_{ab} -\frac{1}{2\sqrt{-g}}\left(\pi_{cd}\pi^{cd}-\frac{1}{n-1}\pi^2\right)g^{ab}\delta g_{ab} \;. \label{dH3}
\end{eqnarray}

Substituting equations (\ref{dH1}), (\ref{dH2}), and (\ref{dH3}) into equation (\ref{dH}), we get
\begin{eqnarray}
	\delta\t{\cal H}_G &\doteq& \sqrt{-g}\left(NG^{ab} -\nabla^a\nabla^bN +g^{ab}\nabla_c\nabla^c N\right)\delta g_{ab} +\sqrt{-g}\nabla_c\left\{\frac{1}{\sqrt{-g}}\left[2N^{(a}\pi^{b)c}-N^c\pi^{ab}\right]\right\}\delta g_{ab} \nonumber\\
        && -\frac{N}{\sqrt{-g}}\left[2\left(\pi^a_{\;\;c}\pi^{bc}-\frac{1}{n-1}\pi\pi^{ab}\right)-\frac{1}{2}\left(\pi_{cd}\pi^{cd}-\frac{1}{n-1}\pi^2\right)g^{ab}\right]\delta g_{ab} \;. \label{dHa}
\end{eqnarray}
Therefore, by equation (\ref{dyn_eq}), we get
\begin{eqnarray}
	\dot{\pi}^{ab} &=& -N\sqrt{-g}G^{ab}+\frac{2N}{\sqrt{-g}}\left(\pi^a_{\;\;c}\pi^{bc}-\frac{1}{n-1}\pi\pi^{ab}\right) -\frac{N}{2\sqrt{-g}}\left(\pi_{cd}\pi^{cd}-\frac{1}{n-1}\pi^2\right)g^{ab} \nonumber\\
        && +\sqrt{-g}\left(\nabla^a\nabla^bN -g^{ab}\nabla_c\nabla^c N\right) -\sqrt{-g}\nabla_c\left\{\frac{1}{\sqrt{-g}}\left[2N^{(a}\pi^{b)c}-N^c\pi^{ab}\right]\right\}\;. \label{dyn2}
\end{eqnarray}

It can be checked that equation (\ref{dyn2}) is identical to the Einstein field equation on ${\cal M}$ with a stress-energy tensor given by equation (\ref{Tm+Tint+Tem}).


\begin{thebibliography}{99}

\bibitem[Einstein(1915a)]{ein15a}
    A. Einstein, Seitsber. Preuss. Akad. Wiss. Berlin, 778 (1915a)
\bibitem[Einstein(1915b)]{ein15b}
    A. Einstein, Seitsber. Preuss. Akad. Wiss. Berlin, 844 (1915b)
\bibitem[Goenner(2004)]{goe04}
    H. F. M. Goenner, Living Rev. Relativity {\bf 7}, 2 (2004)
\bibitem[Einstein(1948)]{ein48}
    A. Einstein, Rev. Mod. Phys. {\bf 20}, 35 (1948)
\bibitem[Einstein(1955)]{ein55}
    A. Einstein, The Meaning of Relativity, Fifth Edition: Including the Relativistic Theory of the Non-Symmetric Field. Princeton University Press, Princeton (1955)
\bibitem[Weyl(1918)]{wey18}
    H. Weyl, Seitsber. Preuss. Akad. Wiss. Berlin, 465 (1918)
\bibitem[Eddington(1921)]{edd21}
    A. Eddington, Proc. R. Soc. Ser. A {\bf 99}, 104 (1921)
\bibitem[Schr\"odinger(1947)]{sch47}
    E. Schr\"odinger, Proc. Royal Irish Acad. A {\bf 51}, 163 (1947)
\bibitem[Nordstr\"om(1914)]{nor14}
    G. Nordstr\"om, Phys. Z. {\bf 15}, 504 (1914)
\bibitem[Kaluza(1921)]{kal21}
    T. Kaluza, Sitzungsber. Preuss. Akad. Wiss, 966 (1921)
\bibitem[Klein(1926)]{kle26a}
    O. Klein, Zeit. Phys. {\bf 37}, 895 (1926a)
\bibitem[Klein(1926)]{kle26b}
    O. Klein, Nature {\bf 118}, 516 (1926b)
\bibitem[Bailin \& Love(1987)]{bai87}
    D. Bailin \& A. Love, Rep. Prog. Phys. {\bf 50}, 1087 (1987)
\bibitem[Overduin \& Wesson(1997)]{ove97}
    J. M. Overduin \& P. S. Wesson, Phys. Rep. {\bf 283}, 303 (1997)
\bibitem[Arkani-Hamed, Dimopoulos, \& Dvali(1998)]{ark98} 
    N. Arkani-Hamed, S. Dimopoulos, \& G. Dvali, Phys. Lett. B {\bf 429}, 263 (1998)
\bibitem[Antoniadis et al.(1998)]{ant98} 
    I. Antoniadis, N. Arkani-Hamed, S. Dimopoulos, \& G. Dvali, Phys. Lett. B {\bf 436}, 257 (1998)
\bibitem[Randall \& Sundrum(1999a)]{ran99a}
    L. Randall \& R. Sundrum, Phys. Rev. Lett. {\bf 83}, 3370 (1999a)
\bibitem[Randall \& Sundrum(1999)b]{ran99b}
    L. Randall \& R. Sundrum, Phys. Rev. Lett. {\bf 83}, 4690 (1999b)
\bibitem[Wald(1984)]{wal84}
    R. M. Wald, General Relativity. The University of Chicago Press, Chicago (1984)
\bibitem[Hawking \& Ellis(1973)]{haw73}
    S. W. Hawking \& G. F. R. Ellis, The Large Scale Structure of Space-Time. The Cambridge University Press, Cambridge (1973)
\bibitem[Carroll(2003)]{car03}
    S. Carroll, Spacetime and Geometry: An Introduction to General Relativity. Addison-Wesley, New York (2003)
\bibitem[Li(2015)]{li15}
    L.-X. Li, Gen. Relativ. Gravit. {\bf 48}, 28 (2016)
\bibitem[Weinberg(1972)]{wei72}
    S. Weinberg, Gravitation and Cosmology. John Wiley \& Sons, New York (1972)
\bibitem[Misner, Thorne, \& Wheeler(1973)]{mis73}
    C. W. Misner, K.S. Thorne, \& J. A. Wheeler, Gravitation. W. H. Freeman, New York (1973)
\bibitem[Einstein(1916)]{ein16}
    A. Einstein, Ann. Phys. {\bf 354}, 769 (1916)
\bibitem[Campbell(1926)]{cam26}
    J. E. Campbell, A Course of Differential Geometry. Clarendon Press, Oxford (1926)
\bibitem[Romero(1996)]{rom96}
    C. Romero, R. Tavakol, \& R. Zalaletdinov, Gen. Relativ. Gravit. {\bf 28}, 365 (1996)
\bibitem[Arnowitt, Deser, \& Misne(1962)]{arn62}
    R. L. Arnowitt, S. Deser, \& C. W. Misner, in Gravitation: An Introduction to Current Research, ed. L. Witten. John Wiley \& Sons, Inc., New York, p. 227 (1962)
\bibitem[Stueckelberg(1957)]{stu57}
    E. C. G. Stueckelberg, Helv. Phys. Acta {\bf 30}, 209 (1957)
\bibitem[Peskin \& Schroeder(1995)]{pes95}
    M. E. Peskin \& D. V. Schroeder, An Introduction To Quantum Field Theory. Westview Press, New York (1995)  
\bibitem[Liang \& Zhou(2006)]{lia06}
    C. Liang \& B. Zhou, Introduction to Differential Geometry and General Relativity, Vol. I. Science Press, Beijing (2006)
\bibitem[Binney \& Tremaine(1987)]{bin87}
    J. Binney \& S. Tremaine, Galactic Dynamics. Princeton University Press, Princeton (1987)
\bibitem[Hinshaw et al.(2013)]{hin13}
    G. Hinshaw et al. 2013,  Astrophys. J. Supp. {\bf 208}, 19 (2013)
\bibitem[Planck(2015)]{pla15}
    Planck Collaboration, P. A. R. Ade, N. Aghanim, M. Arnaud et al., arXiv:1502.01589 (2015)
\bibitem[Wesson(2015)]{wes15}
    P. S. Wesson, Int. J. Mod. Phys. D {\bf 24}, 1530001 (2015)
\bibitem[Shiromizu, Maeda, \& Sasaki(2000)]{shi00}
    T. Shiromizu, K. Maeda, \& M. Sasaki, Phys. Rev. D {\bf 62}, 024012 (2000)
\bibitem[Wald(1974)]{wal74}
    R. M. Wald, Phys. Rev. D {\bf 10}, 1680 (1974)
\bibitem[Turner \& Widrow(1974)]{tur88}
    M. S. Turner \& L. M. Widrow, Phys. Rev. D {\bf 37}, 2743 (1988)
\bibitem[Goldhaber \& Nieto(2010)]{gol10}
    A. S. Goldhaber \& M. M. Nieto, Rev. Mod. Phys. {\bf 82}, 939 (2010)

\end{thebibliography}
\end{document}